\documentclass[onecolumn,apj,numberedappendix]{emulateapj}
\usepackage{srcltx}
\usepackage{graphicx}
\usepackage{multirow}
\usepackage{amsmath}
\newcommand\Tstrut{\rule{0pt}{4.6ex}}         
\newcommand\Bstrut{\rule[-2.9ex]{0pt}{0pt}}   

\renewcommand{\vec}[1]{\mathbf{#1}}
\newcommand{\be}{\begin{equation}}
\newcommand{\ee}{\end{equation}}
\newcommand{\ba}{\begin{eqnarray}}
\newcommand{\ea}{\end{eqnarray}}

\begin{document}
\title{Spectrum and Anisotropy of Turbulence from Multi-Frequency Measurement of
Synchrotron Polarization}

\author{A. Lazarian}
\affil{Department of Astronomy, University of Wisconsin,
Madison, US}
\author{D. Pogosyan}
\affil{Physics Department, University of
Alberta, Edmonton, Canada}

\begin{abstract}
We consider turbulent synchrotron emitting media that also exhibits Faraday
rotation and provide a statistical description of synchrotron polarization
fluctuations.  In particular, we consider these fluctuations as a function of
the spatial separation of the direction of measurements and as a function of
wavelength for the same line-of-sight. On the basis of our general analytical
approach, we introduce several measures that can be used to obtain the spectral
slopes and correlation scales of both the underlying magnetic turbulence
responsible for emission and the spectrum of the Faraday rotation fluctuations.
We show the synergetic nature of these measures and discuss how the study can
be performed using sparsely sampled interferometric data. We also discuss how
additional characteristics of turbulence can be obtained, including the
turbulence anisotropy, the three dimensional direction of the mean magnetic
field. We consider both cases when the synchrotron emission and Faraday
rotation regions coincide and when they are spatially separated. Appealing to
our earlier study in Lazarian \& Pogosyan (2012) we explain that our new
results are applicable to a wide range of spectral indexes of relativistic
electrons responsible for synchrotron emission. We expect wide application of
our techniques both with existing synchrotron data sets as well as with big
forthcoming data sets from LOFAR and SKA.
\end{abstract}


\section{Introduction}
\label{sec:introduction}

Radio observations of synchrotron emission is an important source of information about astrophysical magnetic fields
(see Ginzburg 1981). Diffuse synchrotron emission is observed throughout the ISM and the ICM, as well as in the lobes of radio galaxies (e.g. Westerhout et al. 1962,  Carilli et al. 1994, Reich et al. 2001,  Wolleben et al. 2006, Haverkorn et al. 2006,  Clarke \& En{\ss}lin 2006, Schnitzeler et al. 2007, Laing et al. 2008). Observations testify that turbulence is ubiquitous in astrophysics (see
Armstrong et al. 1994, Lazarian 2009, Chepurnov \& Lazarian 2010). As most
astrophysical environments are magnetized and relativistic electrons are in most cases are present, the turbulence results in
synchrotron fluctuations, which carry important information, but
at the same time, interfere with attempts to measure Cosmic Microwave Background
(CMB) with high precision. In addition, synchrotron fluctuations present an impediment
for studying fluctuations of atomic hydrogen distribution in the early Universe. The latter
has become a direction of intensive discussion recently (see Loeb \& Zaldarriaga 2004, Pen et al. 2008,
Loeb \& Wyithe 2008, Liu, et al. 2009, Ferdinandez et al. 2014). If we know the spectrum of underlying turbulence, these fluctuations can be separated from the CMB signal (see Cho \& Lazarian 2010). Better cleaning of the CMB maps is particularly important while analyzing polarized radiation in the search of enigmatic B-modes produced by gravitational waves in the Early Universe. The polarized synchrotron present a important foreground that such studies have to deal with.

A number of earlier studies tried to utilize synchrotron intensity fluctuation to obtain the spectrum
and anisotropies of underlying magnetic turbulence (see Getmantsev 1959, Chibisov \& Ptuskin 1981, Lazarian \& Shutenkov 1990, Lazarian \& Chibisov 1991, Chepurnov 1998). In addition, polarization fluctuations were
proposed to address the complex issues of measuring magnetic field helicity (Waelkens et al. 2009, Junklewitz \& En{\ss}lin 2011). The
serious limitation of all the above studies was that it was done for a single spectral index of relativistic electrons that allowed
to write the synchrotron emissivity not as generally applicable $i\sim H_{\bot}^{\alpha}$, where $H_{\bot}$ is a perpendicular component
of magnetic field and $\alpha$ depends on the spectrum of emitting electrons, but only for $\alpha=2$. In a way, the studies were limited
to a single point of a parameter space.

The above deficiency was addressed in our recent study (Lazarian \& Pogosyan 2012, henceforth LP12) where we provided the statistical description
synchrotron fluctuations for an arbitrary index $\alpha$ corresponding to the actual energy distribution of relativistic electrons. Very importantly, rather than taking a usual ad hoc and incorrect assumption that magnetic field in turbulent media can be presented as $\textbf{H}_{total}= \textbf{H}_{regular} + \textbf{h}_{turbulent}$, i.e. as a superposition of a regular magnetic field $\textbf{H}_{regular}$ and \textit{isotropic} stochastic magnetic field we used the model of the turbulence for realistic anisotropic magnetic turbulence which corresponds to theoretical expectations (Goldreich \& Sridhar 1995, see Brandenburg \& Lazarian 2013 for a review) and supported by numerical simulations in Cho \& Lazarian (2003) and Kowal \& Lazarian (2010). These two advances brought the studies of magnetic turbulence using synchrotron to a new stage. Testing of the expressions obtained in LP12 has been performed with synthetic data in
Heron et al. (2015).

The study in LP12 was mostly dealing with synchrotron intensities. Present day telescopes present opportunities to get detailed maps of polarization. In fact, the Position-Position Frequency (PPF) data cubes are getting
available with high spatial and spectral resolution. Such data cubes present a good opportunity for studying magnetic turbulence, provided that the description of the relation of the synchrotron polarization statistics and the statistics of the underlying magnetic fields is available.

We also derived
correlations of synchrotron polarization but did not deal with the important effect of
Faraday rotation of the polarized radiation that arises as radiation propagates in the magnetized plasmas. The angle of polarized radiation
rotation is proportional to the $\lambda^2\int n_e \bf{H}_{total, \|} dz$, where $\lambda$ is the wavelength of the radiation, the integration is done along the line-of-sight, while $\bf{H}_{total, \|}$ is the component of magnetic field along the line-of-sight and $n_e$ is the density of electrons in thermal plasmas. In terms of synchrotron
polarization the effects of Faraday rotation decrease the polarization and introduce additional fluctuations arising from both fluctuations of parallel component of magnetic field as well as electron density. Therefore ignoring the Faraday rotation while dealing with polarized intensity can only be justified for sufficiently short wavelengths.

Faraday rotation measurements have been extensively used for studying regular and fluctuating components of magnetic fields using radio emission of external sources, e.g. point radio sources. In addition, the effect of Faraday depolarization was used to probe magnetic field at different distances from the observer. Indeed,  by changing the wavelength of the radiation one can vary the contribution of polarized synchrotron emission from the regions at different distances along the line of sight. Indeed, using longer the wavelengths one can sample emission from closer
emitting volumes. In fact, our present study shows that the criterium for sampling the turbulence with synchrotron polarization is different from the one for intensity studies.

More
recently there have been renewed interest
to getting detailed maps of diffuse synchrotron emission that experiences Faraday rotation within the emitting volume (Beck et al. 2013).
 These new studies provide Position-Position-Frequency (PPF) data cubes which exhibit an intricate structure of fluctuations that arise from both
the fluctuations of magnetic field and the fluctuations in the Faraday measure. Our paper opens new ways of using these PPF data cubes for studying turbulence by
 providing the analytical description of fluctuations in these data cubes.
In particular, we below we describe techniques for studying
polarization fluctuations at a given wavelength as a function of spatial separation. We also explore the potential
of the dispersion of the polarized signal when it is studies as
 a function of frequency. The first technique with separated lines-of-sight some has similarities to the Velocity Channel Analysis (VCA) technique
that employs spectral Doppler-shifted lines to study velocity turbulence introduced by us some time ago (Lazarian \& Pogosyan 2000, 2004), while the studies of the frequency dependence of the dispersion has some similarities to the
Velocity Correlation Spectrum (VCS) technique that was suggested by us later, i.e. in
Lazarian \& Pogosyan (2006, 2008). Both VCA and VCS make use of Position Position Velocity (PPV) spectral data which is an analog of PPF in the present analysis. Both techniques have been successfully employed to study velocity turbulence data (see Lazarian 2009 for a review).
 In analogy with these techniques we term the technique based on the
analysis of
spatial fluctuations of polarization Polarization Spatial Analysis (PSA), which is
an analog of VCA for velocity data cubes,
 and on the analysis of frequency dependence of the polarization variance,
Polarization Variance Analysis (PVA), which is an analog of VCS. In view of the revival of interest to the Faraday rotation synthesis technique (Brentjens \& Bruyn 2005)\footnote{The original version of the technique was formulated by Burn (1966).} we discuss how to use this technique within the PVA approach.

We would like to stress that there are two major advantages of using different techniques for studying turbulence. First of
all, they measure different components of turbulent cascade. For instance, it is advantageous to measure independently both the spectrum of velocity and the spectrum of magnetic field. This, for instance, is possible combining VCA and PSA measurements for the same media. Second, combining different techniques it is possible to study whether properties of magnetic turbulence in different media, e.g. to explore the continuity the turbulent cascade in different phases of the ISM and test whether the cascade is these phases corresponds to the Big Power Law in the Sky (Armstrong et al. 1994, Chepurnov \& Lazarian 2010).

The present paper follows the pattern of our earlier publications on studying spectrum of turbulence from observations (see Lazarian \& Pogosyan 2000, 2004, 2006). We obtain general expressions, but are focused on obtaining the asymptotic regimes for turbulence statistics. While, as we discuss in the paper, these asymptotic expressions are informative, the full expressions may have advantages for the analysis of
observational data as was shown in Chepurnov et al. (2010). Indeed, in the latter paper, apart from the spectral slope, the injection scale of turbulence and the turbulence Mach number were obtained. We expect that additional measures, e.g. injection scale and variations of turbulence intensities along different directions, can be available.

In what follows, we discuss the basic statistics of MHD turbulence that we seek
to obtain using synchrotron polarization fluctuations in \S\ref{sec:MHDstat},
introduce the measures and explore the properties of synchrotron statistics in
\S\ref{sec:Syncstat}, introduce the correlations of polarization at different
spatial points, i.e. PSA technique in \S\ref{sec:samelambda} and discuss the
statistics of measures along the same line of sight in \S\ref{sec:lineofsight}.
In \S\ref{sec:additionalways} we discuss additional measures, including
spatial correlations of the derivatives of polarization \textit{wrt}
to wavelength, and application to interferometers.
On the basis of our formalism we formulate new techniques of turbulence study
with synchrotron polarization in \S\ref{sec:results}  and
provide the discussion of our results and comparison of the different ways
to study magnetic turbulence in \S\ref{sec:discussion}.
The latter section may be the most useful for researchers interested in
practical application of the techniques.
Our findings are summarized in \S\ref{sec:summary}.

\section{Spectrum of MHD turbulence}
\label{sec:MHDstat}

\subsection{Importance}
\label{2.1}

This paper deals with developing the technique for obtaining properties of magnetic turbulence from observations.
Turbulence in magnetized plasmas plays a crucial role for the processes of cosmic ray propagation (see Schlickeiser 2003, Longair 2011), star formation (see Elmegreen \& Scalo 2005, McKee \& Ostriker 2007), heat transfer in magnetized plasmas (see Narayan \& Medvedev 2001, Lazarian 2006), magnetic reconnection (see Lazarian \& Vishniac 1999, Kowal et al. 2009,
Eyink et al. 2011, see review Lazarian et al. 2015 and ref. therein). The advantage of
statistical description of turbulence is that it allows to reveal regular features within chaotic picture of turbulent fluctuations. Recent reviews on MHD turbulence include Brandenburg \& Lazarian (2013) 
and Beresnyak \& Lazarian (2015). 

In this paper we use the statistical description of turbulence and claim that it is
 an adequate and concise way to characterize many essential properties of interstellar turbulence. 
Indeed, while turbulence is an extremely complex chaotic non-linear phenomenon, it
allows for a remarkably simple statistical description (see Biskamp 2003).
If the injections and sinks of the energy are correctly identified, we can
describe turbulence for \textit{arbitrary} $Re$ and $Rm$. The simplest
description
of the complex spatial variations of any physical variable, $X({\bf r})$, is
related to the amount of change of $X$ between points separated by a chosen
displacement ${\bf l}$, averaged over the entire volume of interest.  Usually
the result is given in terms of the Fourier transform of this average, with the
displacement ${\bf l}$ being replaced by the wavenumber ${\bf k}$ parallel to
${\bf l}$ and $|{\bf k}|=1/|{\bf l}|$.  For example, for isotropic turbulence
the kinetic energy spectrum, $E(k)dk$, characterizes how much energy resides at
the interval $k, k+dk$.  At some large scale $L$ (i.e., small $k$), one expects
to observe features reflecting energy injection.  At small scales, energy
dissipation should be seen.  Between these two scales we expect to see a
self-similar power-law scaling reflecting the process of non-linear energy
transfer, which for Kolmogorov turbulence results in the famous
$E(k)\sim k^{-5/3}$ relation. However, from the point of view of astrophysics
both the injection scale or multiple injection scales (see Yoo \& Cho 2014), which are also 
expected in interstellar medium, as well as dissipation scale are of great
interest.

For our statistical description, we need to know what are expected properties of
MHD turbulence, e.g. to know which range of spectral indexes we should consider deriving
our asymptotic solutions and what other effects, e.g. related to anisotropy we should consider.
 Apparently,
MHD turbulence is more complex than the hydrodynamical one. Magnetic field
defines the chosen direction of anisotropy (Montgomery \& Turner 1981, 
Shebalin, Matthaeus \& Montgomery 1983, Higdon 1984).
For small scale motions this is true even in the absence of
the mean magnetic field in the system. In this situation the magnetic field of
large eddies defines the direction of anisotropy for smaller eddies. This
observation brings us to the notion of \textit{local system of reference},
which is one of the major pillows of the modern theory of MHD turbulence
\footnote{The
theory by Goldreich \& Sridhar (1995) did not have the notion of local
system of reference or local direction of magnetic field. This notion appeared
later (Lazarian \& Vishniac 1999, Cho \& Vishniac 2000, Goldreich \& Maron 2001).}. Therefore a
correct formulation of the theory requires wavelet description (see Kowal \&
Lazarian 2010). Indeed, a customary description of anisotropic turbulence using
parallel and perpendicular wavenumbers assumes that the direction is fixed in
space. In this situation, however, the turbulence loses its universality in the
sense that, for instance, the critical balance condition of the widely accepted
model incompressible MHD turbulence (Goldreich \& Sridhar 1995, henceforth GS95) 
which expresses the equality of the time of wave
transfer along the magnetic field lines and the eddy turnover time is not
satisfied.

The existence of the local scale dependent anisotropy does not mean that describing synchrotron fluctuations
as they are seen by the observer one should use the GS95 description. In fact, the local system of reference in most cases is not available to the observer who measures turbulence in the system of mean magnetic field. In such a system, the anisotropy is also present, but it is independent of scale (see the discussion in Cho et al. 2002) and the
perpendicular Alfv\'enic perturbations absolutely dominate the spectrum of fluctuations from
Alfv\'enic turbulence\footnote{The study by Vestuto et al. (2003) reported difference in spectra and scale dependent
anisotropy in the global frame. However, the reported effect was not related to the GS95 predictions, but due to
numerical set up the authors used. The erroneous idea that the scale dependent anisotropy of turbulence are available in global system of reference available through observations influenced some further studies (e.g. Heyer et al. 2008).}. Therefore the expected spectrum from GS95 is Kolmogorov-type with $E_{Alf}(k)\sim k^{-5/3}$ (see Cho, Lazarian \& Vishniac 2002). The tensor for turbulence with the scale-independent anisotropy in the global system of reference is given in LP12 and discussed for realizations of compressible MHD turbulence. 

It is important to understand that, contrary to the entrenched in the community notion, MHD \textit{compressible} turbulence can be described in simple terms. 
Numerical studies support the notion that the turbulence of fast modes develops mostly on its
own and corresponds to the spectrum of acoustic turbulence, i.e. $E_{fast}\sim k^{-3/2}$ (Cho \& Lazarian
2002, 2003, Kowal \& Lazarian 2010). The slow mode fluctuations are expected to follow the spectrum of
the Alfv\'en mode (GS95, Lithwick \& Goldreich 2001, Cho \& Lazarian 2002). Shocks, which are inevitable
for highly supersonic turbulence, are expected to induce steeper turbulence with spectrum $E_{shock}\sim k^{-2}$. Steepening of observed velocity fluctuations was reported in Padoan et al. (2009) and Chepurnov et al. (2010). These groups used for their studies the VCS technique (
Lazarian \& Pogosyan 2000, 2006)\footnote{Our studies in Esquivel \& Lazarian (2005) and Esquivel et al. (2006)
showed that the traditionally used velocity centroids get unreliable for 
studying spectra of
supersonic turbulence.}. It is interesting to know whether magnetic field fluctuations will also demonstrate the corresponding steepening. Thus the development of the corresponding techniques is important from the point of view of establishing the correct
spectral slope of magnetic fluctuations in MHD turbulence.

We should mention that the theory of MHD turbulence is a developing field with its ongoing debates\footnote{Recent debates, for instance, were centered on the role of dynamical alignment or
polarization intermittency that could change the slope of incompressible MHD turbulence from
the Kolmogorov slope predicted in the GS95 to a more shallow slope observed in
numerical simulations (see Boldyrev 2005, 2006, Beresnyak \& Lazarian 2006).
Other studies (see Beresnyak \& Lazarian 2009, 2010) indicated that
numerical simulations may not have enough dynamical range to test the actual
spectrum of turbulence and the flattening of the spectrum measured in the
numerical simulations is expected due to a bottleneck, arising from MHD turbulence being less local than
its hydro counterpart. This explanation seems consistent with more recent numerical simulations in Beresnyak (2014) (see also a discussion in Beresnyak \& Lazarian 2015).}. While we feel that among all the existing models the
GS95 provides the best correspondence to the existing numerical and
observational data (see Beresnyak \& Lazarian 2010, Chepurnov \& Lazarian 2010, 
Beresnyak 2011, 2013), the issue of the actual nature of MHD turbulence requires further research for
which observational techniques will play important role. Indeed, the inertial range provided by astrophysical turbulence is much larger than that of numerical
simulations. Thus the synchrotron studies may be very useful for getting insight into the nature of MHD turbulence.

The issues of the spectrum of magnetic fluctuations are important for turbulence theory and its implications, e.g.
cosmic ray and heat propagation (see Brandenburg \& Lazarian 2013). However, for
describing many astrophysical processes, the issues of intensity of turbulent fluctuations,
degree of turbulence compressibility, degree of magnetization of turbulence, injection and dissipation scales of turbulence are essential. In particular, interstellar medium is a very complex system and therefore one cannot be a priori sure that simple models of isothermal
turbulence can be directly applicable to its parts not to speak about the interstellar medium in general. Additional physics, as well as multiple sources of energy injection can affect
the shape of the turbulence spectrum and this makes obtaining the 
turbulence spectrum from observations essential. Within our treatment we do not assume that the spectral index of 
magnetic fluctuations is $-5/3$, but treat it as a parameter that should be established from observations.

All in all, the techniques that we propose in this paper are important for establishing (a) sources and sinks of turbulent energy, (b) the distribution of turbulence in Galaxy and other astrophysical objects, (c) clarification of the properties of compressible MHD turbulence.

\subsection{Shallow and Steep spectra}

Magnetic fields that are sampled by synchrotron polarization are turbulent. The prediction for the magnetic turbulence within the GS95 theory is the Kolmogorov spectrum,
 which is in terms of 3D spectrum corresponds to 
$k^{-11/3}$. Note, when direction averaged, the spectrum gets another $k^2$, which provides the usual value $E(k)\sim k^{-5/3}$, which is a more common reference to the Kolmogorov spectrum.
We, however,
will use in this paper, similar to our other publications (see LP00) the 3D spectra.

At the same time, in some cases the spectrum of turbulence may be more shallow. This, for instance, corresponds to the magnetic field in the viscosity-damped turbulence, which is the high $k$ regime of turbulence in the fluid where the ratio of viscosity to resistivity is much larger than one (Cho, Lazarian \& Vishniac 2002, 2003, Lazarian, Vishniac \& Cho 2004). There it was shown that the one-dimensional spectrum can be $E(k)\sim k^{-1}$ which corresponds to the 3D spectrum of magnetic field of $k^{-3}$. The spectrum of $-3$ corresponded to the border-line spectrum of turbulence between the steep and shallow regimes (see  LP00) with shallow regime corresponding to most of the turbulent energy being at
small scales, while the steep regime corresponds to most of the energy being at large scales. We are not aware of any expectations of the turbulent magnetic spectrum with
the index more shallow than this borderline value. Therefore we shall consider only
\textit{steep} magnetic field spectra.

The fluctuations of synchrotron polarization are affected not only by magnetic perturbations, but also
by Faraday rotation fluctuations that are proportional to the product of the parallel to the line of
sight component of magnetic field and density. Shallow and steep spectra are, however, a confirmed reality of the spectrum of density in MHD turbulence (see Beresnyak, Lazarian \& Cho 2005, Kowal, Lazarian \& Beresnyak 2007). The spectrum gets shallow as turbulence gets supersonic and more density fluctuations are localized in corrugated structures of shock-compressed gas.
This regime in case of interstellar medium is relevant to cold phases of the medium, e.g. to molecular clouds (see Draine \& Lazarian 1998 for the list of the idealized phases), but localized ionization sources may result also
in a shallow spectrum of random density. The warm interstellar medium responsible for the synchrotron radiation corresponds to transonic turbulence with Mach number of the order of unity (see Burkhart, Lazarian \& Gaensler 2010). Nevertheless, Faraday rotation may take place in any media between the observer and the region of region of synchrotron emission, which may include cold high Mach number turbulence. Therefore, in our treatment we consider both shallow and steep spectra of turbulence.

The anisotropy of the density correlations depends on the sonic Mach number.
In MHD turbulence at low Mach numbers density follows the velocity scaling
and exhibit GS95 type anisotropy, while at large Mach numbers
the density gets isotropic (Cho \& Lazarian 2003, Beresnyak et al. 2005,
Kowal et al. 2007). 
In this paper we will use
only very general spectral properties of the RM correlations, while
we continue elsewhere our studies of anisotropies (see Lazarian, Esquivel
\& Pogosyan 2001, Esquivel \& Lazarian 2005, 2009, 
 Lazarian \& Pogosyan 2012).

There can be processes that create correlations between the magnetic field and the density of thermal electrons.
Such correlations are possible for shocked regions. We consider such correlations. However, as we discuss further,
the correlations of the vector, i.e. the line of sight component of magnetic field, and a scalar, i.e. thermal electron density are always zero. Our assumption within the paper is that the squared perpendicular component of magnetic field and the density of cosmic rays are not correlated, as observations are indicative of more isotropic distribution of cosmic electrons. However, this is not a crucial assumption for our work, as we discuss below.

\subsection{Statistical description of magnetic fields}
\label{subsec:Bfieldstat}

MHD turbulence is more complex than the hydrodynamical one. Magnetic field
defines the preferred direction  
(Montgomery \& Turner 1981, Shebalin et al. 1983, Higdon 1984) and
the statistical properties of magnetized turbulence are anisotropic.
For small scale motions this is true even in the absence of
the mean magnetic field in the system.
The local system of reference
that, as we discussed earlier, is fundamental for modern theory of Alfv\'enic turbulence (see Lazarian \& Vishniac 1999, Cho \& Vishniac 2000, Maron \& Godreich 2001)
is not accessible to an observer who deals with
projection of magnetic fields from the volume to the pictorial plane. The
projection effects inevitably mask the actual direction of magnetic field
within individual eddies along the line-of-sight.
 As the observer maps the projected magnetic
field in the global reference frame, e.g. system of reference of the mean
field, the anisotropy of eddies becomes \textit{scale-independent} and the
degree of anisotropy gets determined by the anisotropy of the largest eddies
which projections are mapped
(Cho et al. 2002, Esquivel \& Lazarian 2005).

In the presence of the mean magnetic field in the volume under study, an
observer will see anisotropic turbulence, where statistical properties of magnetic field
differ in the directions orthogonal and parallel to the mean magnetic field
that defines the symmetry axis.
The description of axisymmetric turbulence was
given by Batchelor (1946), Chandrasekhar (1950)
and later 
Matthaeus \& Smith (1981) and Oughton (1997).
This is the description that was employed in our earlier paper (LP12) for the description of anisotropic fluctuations of the synchrotron emission.
The index-symmetric part of the correlation tensor can then be presented in the
following form:
\begin{equation}
\langle H_i({\bf x_1}) H_j({\bf x_2}) \rangle = A_\xi(r, \mu)
\hat r_i \hat r_j +  B_\xi(r, \mu) \delta_{ij} + C_\xi(r, \mu) \hat \lambda_i
\hat \lambda_j + D_\xi(r, \mu) \left( \hat r_i \hat \lambda_j + \hat r_j
\hat \lambda_i \right)
\label{eq:xi_ij}
\end{equation}
where  the separation vector ${\bf r} = {\bf x}_1 - {\bf x}_2$ has the
magnitude $r$ and the direction specified by the unit vector $\hat {\bf r}$.
The direction of the symmetry axis set by the  mean magnetic field is given by
the unit
vector
$\hat {\bf \lambda}$ and $\mu = \hat {\bf r} \cdot \hat {\bf \lambda}$.  The
magnetic field correlation tensor may
also have antisymmetric, helical part (see Appendix~\ref{app:antisym}).
This part was not considered in LP12
as its contribution to synchrotron intensity fluctuations may be shown
to be small.
However, as we will discuss in the present paper, this part provides a very
distinct response within the polarization studies that we discuss. Therefore,
while we do not dwell upon this part within this paper, we
would like to stress, that, as we discuss below, with polarization correlations
it is feasible to detect the helical part of the tensor. Such a detection would be very important understanding of many problems of magnetic dynamo.

The structure function of the field has the same representation
\begin{eqnarray}
\frac{1}{2} \langle \left( H_i({\bf x_1}) - H_i({\bf x_2}) \right) \left(
H_j({\bf x_1}) - H_j({\bf x_2}) \right) \rangle =
A(r, \mu) \hat r_i \hat r_j +  B(r, \mu) \delta_{ij} + C(r, \mu) \hat \lambda_i
\hat \lambda_j + D(r, \mu) \left( \hat r_i \hat \lambda_j + \hat r_j
\hat \lambda_i
\right)
\label{eq:Dij}
\end{eqnarray}
with coefficients $A(r,\mu) = A_\xi(0,\mu) - A_\xi(r,\mu) ~, \ldots$ etc.  In
case of power-law spectra, one can use either the correlation function or the
structure function, depending on the spectral slope. 
For the shallow spectrum it is natural to use the correlation function,
while for steep one the structure function (see e.g. Lazarian \& Pogosyan 2004).
In this case the structure
function coefficients can be thought of as renormalized correlation
coefficients.

Magnetic field spectrum can be obtained by a Fourier transform of correlation and structure functions of
magnetic fields (see Monin \& Yaglom 1975). First we consider the case when statistics of the magnetic field is isotropic, which may, for instance,
correspond to the super-Alfv\'enic turbulence, i.e. for the turbulence with the injection velocity much
in excess of the Alfv\'enic one.
The structure tensor of a Gaussian isotropic vector field, a special case of Eq.~(\ref{eq:Dij}),
is usually written in the form
\begin{equation}
\left\langle \left( H_i ( {\bf x}_1 )- H_i ({\bf x}_2) \right) \left( H_j ( {\bf
x}_1 )- H_j ({\bf x}_2) \right) \right\rangle
= \left( D_{LL} - D_{NN}  \right) \hat r_i \hat r_j + D_{NN} \delta_{ij} \quad,
\end{equation}
where $D_{LL}(r)$ and $D_{NN}(r)$ are structure functions that describe,
respectively, the correlation of the vector components
normal and orthogonal to point separation ${\bf r}$.
In case of solenoidal vector field, in particular the magnetic field,
two structure functions are related by
\begin{equation}
\frac{d}{d r} D_{LL} = - \frac{2}{r} \left( D_{LL} - D_{NN} \right)
\label{eq:solenoid_condition}
\end{equation}
which in the regime of the power-law behaviour $D_{NN} \propto r^m$ leads to
both functions being proportional to each other
$D_{LL} = \frac{2}{2+m} D_{NN}$.

From the point of view of observations, our considerations in \S\ref{2.1} suggest that the
slope of the spectrum is not expected to change when the measurements are done in the system 
of reference of the mean magnetic field, which is the only system of reference available to the 
observer. Therefore, if we are interested only in the crude description of turbulence, which 
includes the spectral slope and approximate measures of the injection/dissipation scales, to use
the isotropic description of turbulence. This was the description that we adopted in our earlier
papers dealing with velocity spectra (Lazarian \& Pogosyan 2000, 2004, 2006, 2008), but it is
different from the description adopted for describing anisotropy of synchrotron fluctuations in 
LP12.

Potentially, our treatment of synchrotron fluctuations may be done for the case of relativistic electrons
correlating with the strength of squared component of the magnetic field perpendicular to the line of sight.
Then the correlated quantities should be not $H_{\bot}^2$, but $n_{CR}H_{\bot}^2$.

\section{Statistical description of the polarization signal from emitting distributed medium}
\label{sec:Syncstat}

\subsection{Basic definitions}
To characterize fluctuations of the synchrotron polarization
one can use different combinations of Stokes parameters 
(see our discussion in LV12). In this paper we shall focus on the complex
measure of the linear polarization
\begin{equation}
P \equiv Q + i U ~ .
\end{equation}
Other combinations may have their advantages and should be discussed
elsewhere.

In case of an extended synchrotron sources, the polarization of the
synchrotron emission at the source is 
characterized by the polarized intensity density $P_i({\bf X},z)$,
where ${\bf X}$ marks the two-dimensional position of the source on a sky
and $z$ is a line-of-sight distance. The polarized intensity detected by an
observer
in the direction ${\bf X}$ at wavelength $\lambda$
\begin{equation}
P({\bf X},\lambda^2) = \int_0^L dz P_i({\bf X},z) e^{2 i \lambda^2 \Phi({\bf X},z) }
\label{eq:Plambda2}
\end{equation}
is a line-of-sight integral over emission at the sources
modified by the Faraday rotation of
the polarization plane (see Brentjens \& Bruyn 2005).
Here $L$ is the extent of the source along
the line-of-sight and the Faraday rotation measure (RM) is given  by 
\begin{equation}
\Phi(z)=0.81 \int_0^z n_e(z^\prime) H_{z}(z^\prime) dz^\prime \quad {\rm rad} ~ {\rm m}^{-2}~~,
\end{equation}
where $n_e$ is the density of thermal electrons in $\mathrm{cm}^{-3}$,
$H_{z}$ is the strength of the parallel to the 
line-of-sight component of magnetic field in $\mu$Gauss, and
the radial distance is in $parsecs$. 

In general, synchrotron emission intensity depends on the wavelength $\lambda$,
as discussed in Appendix~\ref{app:basics}.
In this study we consider the polarization measure $P$ in which
this dependence has been scaled out. This can be accomplished, for instance,
by determining the mean wavelength scaling from the total intensity
measurements.  With such rescaling,
polarization at the source $P_i(\mathbf{x})$ is treated as wavelength
independent, while the observed $P(\mathbf{x},\lambda^2)$
contains the residual wavelength dependence 
due to Faraday rotation only.

While Faraday rotation reflects the line-of-sight $H_{\|}=H_z$ 
component of the magnetic field, the intrinsic synchrotron emission at the
source is determined locally by its transverse, $H_{\bot}=H_{\bf X}$, part
(see \S~\ref{app:basics}).
Clearly both polarization at the source and Faraday rotation
influence the observed signal. This makes the analysis of polarization fluctuations
much more complicated compared to pure intensity that we studied in LP12.

\subsection{Correlations of the Faraday rotation measure}

Let us write the Faraday rotation measure as
\begin{equation}
\Phi({\bf X},z) = \kappa \int_0^z  n_e H_z dz^\prime =
\int_0^z \phi({\bf X},z^\prime) dz^\prime
\label{eq:faradayPhi}
\end{equation}
where $\phi({\bf X}, z^\prime) =\kappa n_e H_z $ denotes the RM per unit length
along the line-of-sight. 

Electron density $n_e$, the line-of-sight component of the magnetic field
$H_z$ and, correspondingly, $\phi({\bf X},z)$ are spatially local quantities
which we assume to be statistically homogeneous. Both density and magnetic field
can be presented as a sum of the mean value and a fluctuation. Therefore, 
the average value of the RM linear density is
\begin{equation}
\bar \phi \propto
\left\langle n_e H_z \right\rangle = \left\langle \left(\overline{n}_e + 
\Delta n_e\right)\left(\overline{H}_z + \Delta H_z\right)\right\rangle = \overline{n}_e \overline{H}_z +
\left\langle \Delta n_e \Delta H_z \right\rangle = \overline{n}_e \overline{H}_z
\label{eq:mean_phi}
\end{equation}
where $\Delta n_e$ and $\Delta H_z$ are zero mean fluctuations of the electron
density and the magnetic field.
Note that these fluctuations taken at the same point are generally uncorrelated due to the vector nature of the magnetic field
and the symmetry under the local reversal of its direction.  Thus, it is just the product of the mean 
$\overline{n}_e \overline{H}_z$ that defines the mean RM density.

The variance of fluctuations in Faraday RM density is
\begin{equation}
\sigma^2_{\phi} \equiv
\left\langle \Delta(n_e H_z)^2 \right\rangle = 
{\overline{H_z}}^2 \left\langle (\Delta n_e)^2\right\rangle
+{\overline{n_e}}^2 \left\langle (\Delta H_z)^2\right\rangle
+\left\langle \Delta n_e^2\right\rangle 
\left\langle \Delta H_z^2\right\rangle
\label{eq:sigma_phi}
\end{equation}

Correlation properties of the RM density at two points in space are described by
the correlation or structure functions
\begin{eqnarray}
\xi_{\phi}({\bf X_1-X_2},z^\prime-z^{\prime\prime}) &\equiv&
\kappa^2 \left\langle \Delta(n_e H_z)({\bf X_1},z^\prime)\Delta(n_e H_z)({\bf X_2},z^{\prime\prime}) \right\rangle \\
D_{\phi}({\bf X_1-X_2},z^\prime-z^{\prime\prime}) &\equiv& 
\kappa^2 \left\langle \left( (n_e H_z)({\bf X_1},z^\prime)
-(n_e H_z)({\bf X_2},z^{\prime\prime}) \right)^2\right\rangle
\end{eqnarray}
Assumption of statistical homogeneity of the medium is reflected in the fact
that $\xi_{\phi}$ and $D_{\phi}$ depend only on the coordinate difference
between the two positions.
In what follows we illustrate our treatment of the problem using 
power-law correlation model
\begin{eqnarray}
\xi_{\phi}(\mathbf{X_1-X_2},z^\prime-z^{\prime\prime}) & = & \sigma_\phi^2 
\frac{ {r_\phi}^{m_\phi}}{ {r_\phi}^{m_\phi} + \left(R^2+\Delta z^2\right)^{m_\phi/2}} \\
D_{\phi}(\mathbf{X_1-X_2},z^\prime-z^{\prime\prime}) & = & 2 \sigma_\phi^2 
\frac{ \left(R^2+\Delta z^2\right)^{m_\phi/2}}{ {r_\phi}^{m_\phi} +
\left(R^2+\Delta z^2\right)^{m_\phi/2}}
\label{eq:RMden_powerlaw}
\end{eqnarray}
where $m_\phi$ is the scaling slope and $r_\phi$ is the correlation length
of RM density, $R=|\mathbf{X}_1-\mathbf{X_2}| $ and 
$\Delta z = z^\prime-z^{\prime\prime}$.
These expressions are not the most 
general expressions for the correlations in the 
magnetic field (compare with Eqs.~(\ref{eq:xi_ij}),(\ref{eq:Dij})),
but they are adequate if we discuss 
measuring the scaling properties.
Similar models were employed e.g. in Lazarian \& Pogosyan (2006). 
In several regimes obtained in this paper, models with $m_\phi > 1$
produce similar asymptotical behaviour as $m_\phi=1$ model, thus we will
frequently use the notation
\begin{equation}
\widetilde{m}_\phi = \textrm{min}(m_\phi,1) ~.
\end{equation}

The total RM $\Phi({\bf X},z)$, in contrast to RM density, is not a local but
an integral quantity.
Its behaviour in $z$ coordinate is not statistically homogeneous, rather it
depends on the length of the integration path, which becomes a critical
feature in our studies. At the same time the behaviour 
transverse to the line-of-sight remains
statistically homogeneous.
In particular, 
the mean RM is proportional to the distance along the line-of-sight
from the emitter at $z$ to the observer
\begin{equation}
\overline{\Phi}(z) \equiv \left\langle \Phi({\bf X},z) \right\rangle
= \alpha \langle n_e H_z \rangle z = \overline{\phi} z
\end{equation}
but does not depend on $\textbf{X}$. 

Due to inhomogeneity in $z$, one has to separate the mean Faraday
RM and its fluctuations
$\Phi(\textbf{X},z) = \overline{\Phi}(z) + \Delta \Phi(\textbf{X},z)$,
even when studying the structure functions (normally, insensitive to the mean).
The variance of the fluctuation is
\begin{equation}
\label{eq:sigmaPhi_def}
\sigma^2_{\Delta \Phi}(z_1) \equiv
\left\langle \Delta \Phi({\bf X}_1,z_1)^2 \right\rangle 
= \int_0^{z_1} \!\!\!\! dz'\! \int_0^{z_1}\!\!\!\! dz'' \xi_{\phi}(0,z'-z'')
~,
\end{equation}
while the structure function for fluctuations in the RM we define as
\begin{eqnarray}
\label{eq:DPhi_def}
D_{\Delta \Phi}({\bf R},z_1,z_2) &\equiv&
\frac{1}{2} \left\langle \left( \Delta \Phi({\bf X}_1,z_1) - \Delta \Phi({\bf X}_2, z_2)
\right)^2 \right\rangle \\
&=& 
\frac{1}{2} \int_0^{z_1} \!\!\!\! dz'\! \int_0^{z_1}\!\!\!\! dz'' \xi_{\phi}(0,z'-z'') + 
\frac{1}{2} \int_0^{z_2} \!\!\!\! dz'\! \int_0^{z_2}\!\!\!\! dz'' \xi_{\phi}(0,z'-z'') 
-\int_0^{z_1}\!\!\!\! dz' \! \int_0^{z_2}\!\!\!\! dz'' \xi_{\phi}({\bf X_1-X_2},z'-z'')~.
\nonumber
\end{eqnarray}
Note the non-standard factor $1/2$ in the definition, which we introduced to
simplify the subsequent equations.

\begin{figure}[ht]
\includegraphics[width=0.45\textwidth]{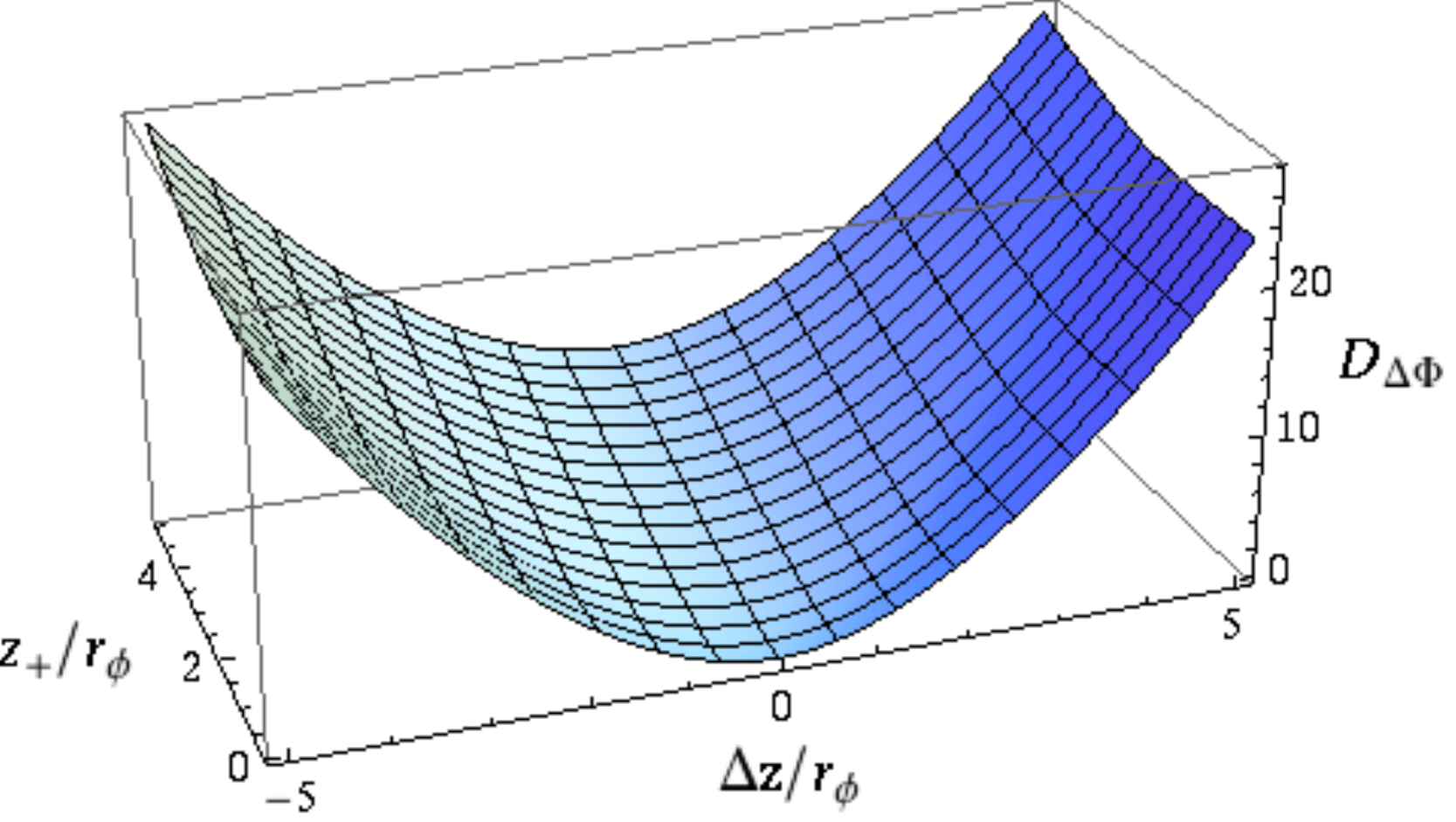} \hfill
\includegraphics[width=0.45\textwidth]{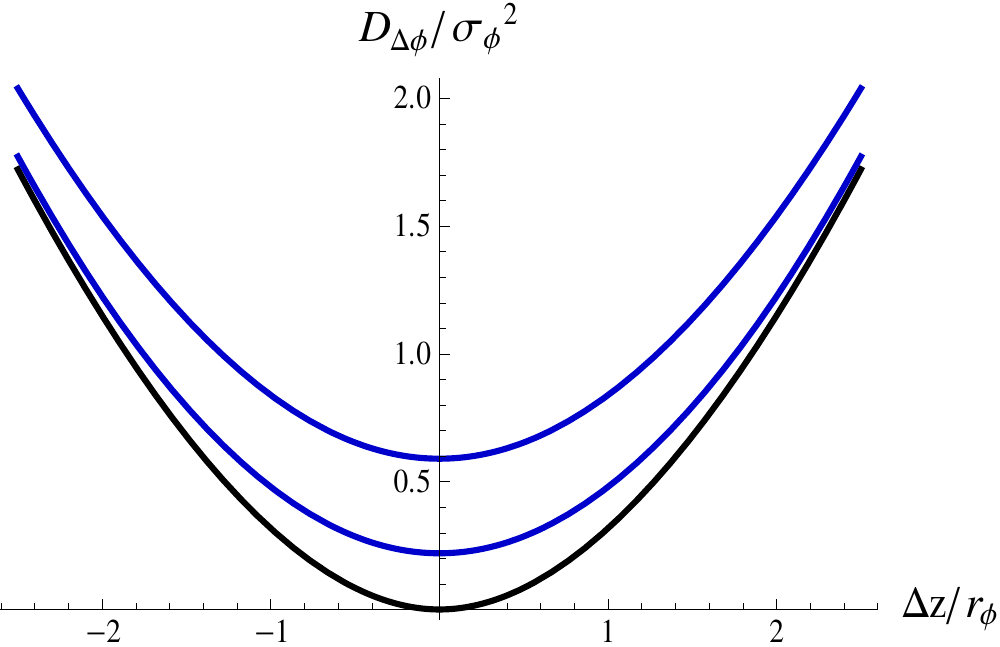} 
\caption{Left: profile of the structure function of RM fluctuations 
$D_{\Delta \Phi}({\bf R},z_1,z_2)/(\sigma_\phi^2 r_\phi^2)$
in $\Delta z = z_1-z_2$ and $z_+ = (z_1+z_2)/2$ space
for $R=r_\phi$, $m_\phi=2/3$.
Right: cross-sections of the left panel surface 
at fixed $z_+/r_\phi=0, 1, 2$ (from lower to upper
curve, respectively).
}
\label{fig:DDFview}
\end{figure}

Let us study geometrical properties of the above structure function in $(z_1,z_2)$
plane. In Figure~\ref{fig:DDFview} its behaviour is demonstrated for the
power law model given by Eq.~(\ref{eq:RMden_powerlaw}) and a fixed $R \ne 0$.
It demonstrates  a valley shape with the bottom at $z_1=z_2$ line that
slowly rises with $z_+=(z_1+z_2)/2$  
and steep walls in $\Delta z=z_1-z_2$ direction.
Main conclusion is that the dependence
in $\Delta z = z_1-z_2$ close to
the local minima of $D_{\Delta\Phi}$ is primarily simply quadratic
due to geometrical reason of different integration lengths.

\begin{figure}[ht]
\includegraphics[width=0.47\textwidth]{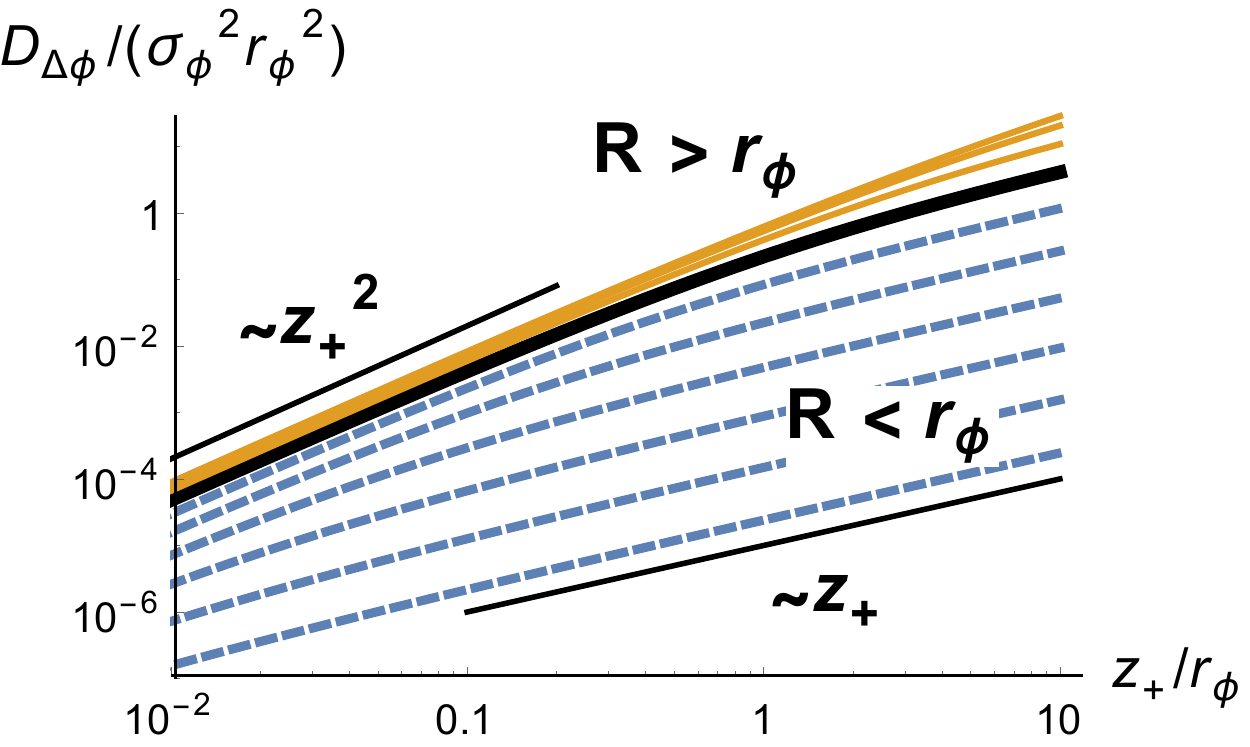} \hfill
\includegraphics[width=0.47\textwidth]{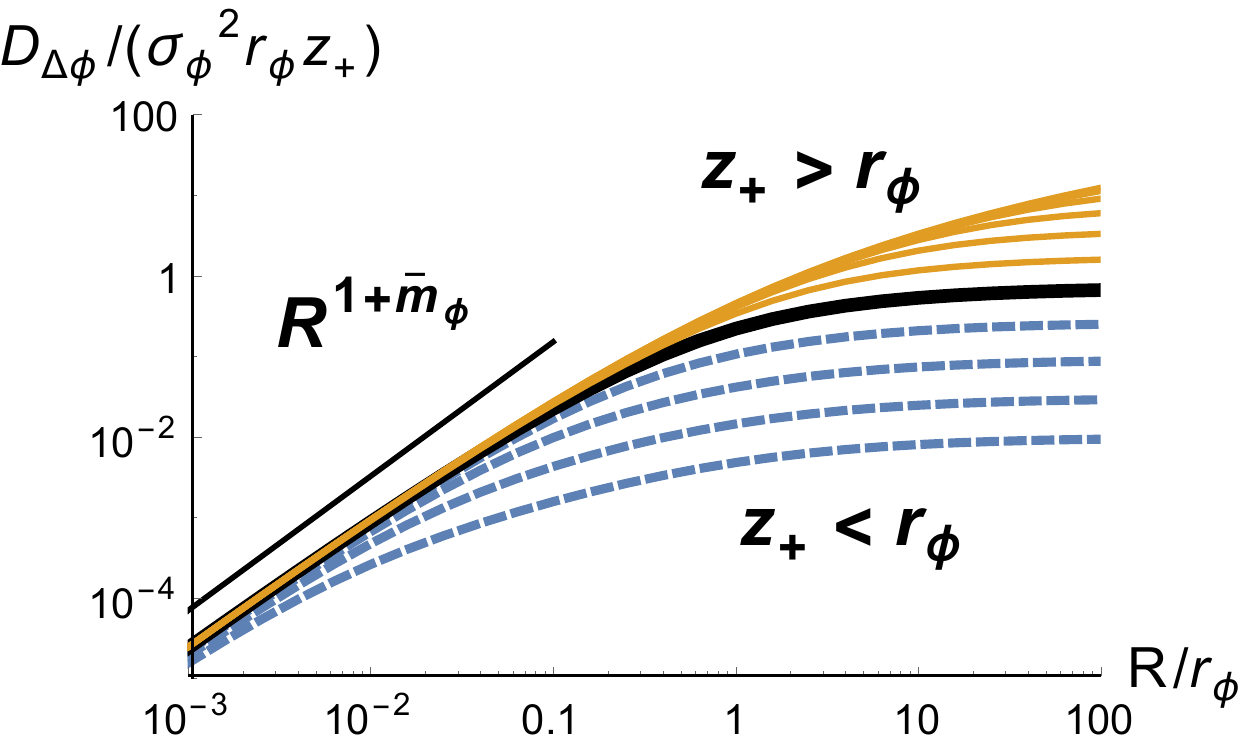}
\caption{
$\Delta z=0$ valley bottom profile $D_{\Delta\Phi}^+(R,z_+)$ 
Left: as the function of $z_+$ 
for several $R$ in the range from $0.001 \; r_\phi$ to $30 \; r_\phi$
(from bottom to the top, dashed $R < r_\phi$, solid $R \ge r_\phi$ with 
the thickest curve corresponding to $R=r_\phi$),
Right: as the function of $R$ 
for several $z_+$ in the range from $0.01 \; r_\phi$ to $1000 \; r_\phi$
(from bottom to the top, dashed $z_+ < r_\phi$, solid $z_+ \ge r_\phi$ with 
the thickest curve corresponding to $z_+=r_\phi$),
}
\label{fig:DDF}
\end{figure}

Analytic considerations in Appendix~\ref{app:DPhi} suggest the following
quadratic approximation
\begin{equation}
D_{\Delta \Phi}({\bf R},z_1,z_2) =
D_{\Delta \Phi}^+({\bf R},z_+) + \frac{1}{4} (\Delta z)^2 
\Lambda_-(R,z_+) 
\label{eq:DPhi_quadratic}
\end{equation}
where along the bottom of the valley $z_1=z_2$
\begin{equation}
D_{\Delta \Phi}^+({\bf R},z_+) = 
\int_0^{z_+} \!\!\!\! dz'\! \int_0^{z_+}\!\!\!\! dz''
\left( \xi_{\phi}(0,z'-z'')  
- \xi_{\phi}({\bf R},z'-z'') \right)
= 2 \int_0^{z_+} dz_- (z_+-z_-) 
\left( \xi_{\phi}(0,z_-)  
- \xi_{\phi}({\bf R},z_-) \right)
\label{eq:DPhi_valley}
\end{equation}
and the curvature in $\Delta z \equiv z_1-z_2$ is 
\begin{equation}
\label{eq:Lambda-}
\Lambda_{-}(R,z_+) =
\xi_{\phi}(0,z_+) -\xi_{\phi}({\bf R},z_+) +2 \xi_{\phi}({\bf R},0) 
\end{equation}
The residual dependence of coefficients on $z_+$ is
one more manifestation of inhomogeneity of statistical measures in
$z$ direction. Figure~\ref{fig:DDF} shows that 
$D_{\Delta \Phi}^+({\bf R},z_+)$ as a function of $z_+$ is initially 
quadratic but then becomes linear at larger $z_+$.
However, for $R < r_\phi$ 
both low and high $z_+$ asymptotics have $R$ dependence
\begin{eqnarray}
D_{\Delta\Phi}^+(\mathbf{R},z_+) &\sim& 
\sigma_\phi^2 \left(R/r_\phi \right)^{\widetilde{m}_\phi} z_+^2
~, \quad\quad\quad\quad\quad ~\; R < r_\phi ~, \quad z_+ \ll R ~,
\label{eq:Dz+_smallRsmallz}
\\
D_{\Delta\Phi}^+(\mathbf{R},z_+) &\sim& 
\sigma_\phi^2 R \left(R/r_\phi \right)^{\widetilde{m}_\phi} z_+
~, \quad\quad\quad\quad\quad R < r_\phi ~, \quad z_+ \gg R ~,
\label{eq:Dz+_smallRlargez}
\end{eqnarray}
while for $R > r_\phi$
at low $z_+$ there is no dependence on $R$ and at high $z_+$ 
$\widetilde{m}_\phi$ scaling is inverted
\begin{eqnarray}
D_{\Delta \Phi}^+({\bf R},z_+) &\sim& \sigma_\phi^2 z_+^2
~, ~ ~\quad\quad\quad\quad\quad\quad\quad\quad\quad\quad
R > r_\phi , \quad
z_+ < r_\phi ~,\\
\label{eq:Dz+_largeRsmallz}
D_{\Delta \Phi}^+({\bf R},z_+) &\sim& \sigma_\phi^2 z_+^2
\left(z_+/r_\phi \right)^{-\widetilde{m}_\phi}
~, \quad\quad\quad\quad\quad
R > r_\phi , \quad
r_\phi < z_+ < R ~,\\
\label{eq:Dz+_largeRmidz}
D_{\Delta \Phi}^+({\bf R},z_+) &\sim& 
\sigma_\phi^2 R \left(R/r_\phi \right)^{-\widetilde{m}_\phi} z_+
~, ~ ~\quad\quad\quad\quad
R > r_\phi , \quad
z_+ \gg R ~.
\label{eq:Dz+_largeRlargez}
\end{eqnarray}
In the subsequent sections we shall use these results extensively to analyze the
asymptotics for synchrotron polarization structure functions.

Important special case is that of a single line-of-sight, ${\bf R} = 0$.
The approximation Eq.~(\ref{eq:DPhi_quadratic}) 
is reduced to $D_{\Delta\Phi}(0,\Delta z) = 
\frac{1}{2}(\Delta z)^2 \sigma^2_{\phi}$,
demonstrating that within its range of validity Faraday effect
is dominated by a purely geometrical factor, insensitive to correlations of
$n H_{\|}$ quantity.
We can study this case in more detail using the exact formula
\begin{equation}
D_{\Delta\Phi}(0,z_1,z_2) \equiv \frac{1}{2}
\int_{z_1}^{z_2}\!\!\!\! dz' \! \int_{z_1}^{z_2}\!\!\!\! dz'' \xi_{\phi}(0,z'-z'') =
\frac{1}{2} \left(
(\Delta z)^2 \sigma_{\phi}^2 
- \int_0^{\Delta z} \!\! dz \; (\Delta z - z) D_{\phi}(0,z)
\right)
\label{eq:DFR0}
\end{equation}
which explicitly shows that the correlated terms are further and further subdominant
to the first geometrical one 
as $\Delta z \equiv |z_1-z_2|$ decreases.  The exact criterium is that the
quadratic geometrical term dominates at $\Delta z < r_\phi$, where
$r_\phi$ is the correlation length of the product of the electron density 
fluctuations and the parallel component of the magnetic field.
At $\Delta z > r_\phi$ the
Faraday structure function tends to another, linear, universal  behaviour $
D_{\Delta\Phi}(\Delta z) \propto
\sigma^2_{\phi} r_\phi \Delta z$ that represents a random walk in the value of the Faraday RM accumulated 
over different intervals of the line-of-sight. This tendency
to random walk at large $\Delta z$ is also seen in Figure~\ref{fig:DDF}
in the general case of separated lines-of-sight of greatly non-equal
lengths. Transition for $\propto\Delta z^2$ to $\propto\Delta z$ behaviour
depends on the details of correlation of the RM density.
Note that statistics of RM fluctuations are homogeneous along a single
line-of-sight.

\subsection{Correlation of the synchrotron polarization at the source}
\label{sec:xi_i}

Magnetic field at the source can be decomposed into regular and random components. 
The regular component provides mean polarization, while the random component
provides fluctuations of polarization. Our study is mostly devoted to the statistical
description of the random component of polarization as it is measured by the observer being averaged
along the line-of-sight and rotated through Faraday rotation,
although the effect of the regular magnetic field is also discussed where
appropriate.

The polarization at the source provides an
initial polarization in our study, which is described by
polarized intensity density denoted as $P_i$ in this paper.
As we discuss in Appendix A, polarized emissivity 
depends on the transverse to the line of observation
magnetic field $\vec{H}_{\perp}$ and the
wavelength $\lambda$, $j(\lambda,\mathbf{x}) \propto \lambda^{\gamma-1} 
\left|\vec{H}_{\perp}\right|^{\gamma}$. In this paper we shall consider  
observational measures in which the underlying dependence on the wavelength 
is scaled out. 
This can be accomplished, for instance,
by measuring the wavelength dependence of the mean intensity of the synchrotron
radiation.
Thus we consider
$P_i(\mathbf{x}) \propto \lambda^{1-\gamma} j(\lambda,\mathbf{x})$
that is wavelength independent. 

Fractional power dependence on the magnetic field 
$\propto\left|\vec{H}_{\perp}\right|^\gamma$ 
is the same for the intensity and polarized intensity.
This allows us
to apply the results of LP12 to the polarized intensity and express the
fluctuation of polarization at the source for an arbitrary index $\gamma$ using
the fluctuations of magnetic field obtained for $\vec{H}_{\perp}^{2}$.
\begin{equation} 
\left[
\left\langle P_i(\mathbf{x_1})P_i^*(\mathbf{x_2}) \right\rangle
- \left\langle P_i\right\rangle \left\langle P_i^*\right\rangle 
\right]_\gamma
\approx
A(\gamma) 
\left[ \left\langle P_i(\mathbf{x_1})P_i^*(\mathbf{x_2}) \right\rangle 
- \left\langle P_i \right\rangle \left\langle P_i^*\right\rangle 
\right]_{\gamma=2}
\end{equation}
Here $A(\gamma)$ is a factor given by the ratio of the variances 
$A(\gamma) = \langle P^2 \rangle_\gamma/ \langle P^2 \rangle_{\gamma=2}$ which
dependence on $\gamma$ is similar to that for the intensity correlations
discussed in LP12. In isotropic turbulence, average polarization is zero,
unless there is a uniform average component to the magnetic field.
If the turbulence is anisotropic, difference between the variances of different
components of the magnetic field may contribute to the mean polarization as well.

In terms of $Q$ and $U$ Stokes parameters in the observers frame,
the correlation between polarizations at two sources is, in general,
\begin{equation}
\xi_i \equiv \left\langle P_i(\mathbf{x_1})P_i^*(\mathbf{x_2}) \right\rangle = 
\left\langle Q(\mathbf{x_1}) Q(\mathbf{x_2}) +
U(\mathbf{x_1}) U(\mathbf{x_2}) \right\rangle 
+ i \left\langle U(\mathbf{x_1}) Q(\mathbf{x_2})
- Q(\mathbf{x_1}) U(\mathbf{x_2}) \right\rangle 
\end{equation}
Two parts of the correlation, real and imaginary, describe correlation
invariants with respect to rotation of the observers frame. Explicit 
expressions via the magnetic field components for $\gamma=2$ are given
in Appendix~\ref{app:antisym}.
The real part is the trace of the polarization correlation
matrix (see LP12) and imaginary part is the antisymmetric contribution to the
correlation. For synchrotron signal, the latter one can be present only
if the magnetic field correlation tensor 
$\left\langle H_i(\mathbf{x_1}) H_j(\mathbf{x_2})\right\rangle$ has index
antisymmetric part, which, in general, is related to the helical correlations
(Oughton et al. 1997, also see Appendix~\ref{app:antisym}).
Although we shall not consider
these antisymmetric correlations in this paper, we stress
that the very detection of helical correlations will be a major discovery.

The main parameters of the correlation function
of the polarization at the source
is the correlation length $r_i$ and the characteristic scaling slope $m$ of
its fluctuations, and the relative contribution from the mean and fluctuating
polarization. While
our subsequent analysis does not rely on a specific shape of $\xi_i$, 
for numerical illustrations we adopt a saturated isotropic power law 
similar to Eq.~(\ref{eq:RMden_powerlaw})
\begin{equation}
\xi_i(\mathbf{X}_1-\mathbf{X}_2,z^\prime-z^{\prime\prime}) = 
\bar P_i^2 +
\sigma_i^2 
\frac{ {r_i}^{m}}{ {r_i}^{m} + \left(R^2+\Delta z^2\right)^{m/2}}
\label{eq:xi_def}
\end{equation}
The mean polarization dominates on all scales if 
$\bar P_i^2 \equiv
\left\langle P_i\right\rangle \left\langle P_i^*\right\rangle > \sigma_i^2 $,
in which case the functional form for intrinsic correlation effectively
corresponds to the infinite correlation length $r_i \to \infty$.
Otherwise, the mean contribution can be neglected for separations 
$R < r_i (\sigma_i^2/\bar P_i^2 -1)^{1/m}$ which covers
all the separations within
the correlation length of intrinsic fluctuations if 
$\bar P_i^2 < \frac{1}{2} \sigma_i^2$.

\subsection{Correlation of the observed polarization}

The observed polarization is subject to both integration along the
line-of-sight and to the Faraday rotation. As a result, 
the invariant over frame rotation measure of the observed
correlation is
\begin{equation}
\left\langle P({\bf X_1},\lambda^2_1)P^*({\bf X_2},\lambda^2_2) \right\rangle
= \int_0^L dz_1 \int_0^L dz_2 \left\langle P_i({\bf X_1},z_1) P_i^*({\bf X_2},z_2) 
e^{2 i \left( \lambda_1^2  \Phi({\bf X_1},z_1)  - \lambda_2^2  \Phi({\bf X_2},z_2) \right)}
\right\rangle
\label{eq:corrP}
\end{equation}
We shall consider all quantities to be statistically homogeneous in real space, however we do not have homogeneity
property in the square-of-wavelength ``direction'' $\lambda^2$.
With the mean effect separated,  Eq.~(\ref{eq:corrP}) becomes
\begin{equation}
\left\langle P({\bf X_1},\lambda^2_1)P^*({\bf X_2},\lambda^2_2) \right\rangle
= \int_0^L \!\! dz_1 \int_0^L \!\! dz_2 \; e^{2 i \overline{\phi}\left(
\lambda_1^2 z_1  - \lambda_2^2 z_2 \right)}
\left\langle P_i({\bf X_1},z_1) P_i^*({\bf X_2},z_2) 
e^{2 i \left( \lambda_1^2  \Delta \Phi({\bf X_1},z_1)  - \lambda_2^2  \Delta \Phi({\bf X_2},z_2) \right)}
\right\rangle
\label{eq:corrPfluct}
\end{equation}
The formula represents the general expression for correlation function 
in PPF (position-position-frequency) data cube and is
the starting point for our further study.


Observable correlation function in terms of the Stokes parameters 
is split again into real and imaginary parts 
that are separately invariant with respect to frame rotation
\begin{equation}
\left\langle P(\mathbf{X_1})P^*(\mathbf{X_2}) \right\rangle = 
\left\langle Q(\mathbf{X_1}) Q(\mathbf{X_2}) +
U(\mathbf{X_1}) U(\mathbf{X_2}) \right\rangle 
+ i \left\langle U(\mathbf{X_1}) Q(\mathbf{X_2}) 
- Q(\mathbf{X_1}) U(\mathbf{X_2}) \right\rangle 
\label{eq:Pcorr_inQU}
\end{equation}
In this paper we focus on the symmetric real part 
$\left\langle Q(\mathbf{X_1}) Q(\mathbf{X_2}) 
+ U(\mathbf{X_1}) U(\mathbf{X_2}) \right\rangle$
which is easier to determine and which carries
the most straightforward information about the magnetized turbulent medium.
Antisymmetric imaginary part 
$\left\langle U(\mathbf{X_1}) Q(\mathbf{X_2}) 
- Q(\mathbf{X_1}) U(\mathbf{X_2}) \right\rangle$
potentially reflects helical correlations
of the magnetic field, but, as will be shown, can be also generated by Faraday
rotation in the anisotropic MHD turbulence. 
Its measurement in data provides valuable observational constraints on
such contributions 
\footnote{We note, that the structure 
function defined in the standard way
\begin{equation}
\left\langle \left|P(\mathbf{X_1})-P(\mathbf{X_2}) \right|^2 \right\rangle = 
\left\langle \left(Q(\mathbf{X_1})-Q(\mathbf{X_2}) \right)^2 \right\rangle +
\left\langle \left( U(\mathbf{X_1})-U(\mathbf{X_2}) \right)^2\right\rangle 
\label{eq:Pstruct_inQU}
\end{equation}
is symmetric and measuring it
cannot provide information about possible antisymmetric correlations.}.

Let us summarize the parameters and scales of the problem that determine
the observed synchrotron polarization correlations, subject to Faraday
rotation. Long list of parameters and notations is summarized in 
Table~\ref{tab:parameters}, however not all of them determine the results
independently. Our problem contains
the correlation length of the rotation
measure $r_\phi$, the correlation length of the transverse magnetic field
$r_i$, the line-of-sight size of the emitting region $L$ and the separation
between two line-of-sight $R$ over which we correlate two polarization 
polarization measurements.
As well we have scaling slopes for RM measure $m_\phi$ and intrinsic
correlations $m$, amplitude of fluctuations in RM $\sigma_\phi$ and intrinsic
correlations $\sigma_i$, possible mean rotation $\bar \phi$ and mean intrinsic
polarization $\bar P_i$,
and the wavelength of observations $\lambda$. Among them,
$\bar P_i$ is trivial
to account for separately,
$\sigma_i$ is 
a simple coefficient the signal is proportional to,
while the 
magnitude of RM, either random $\sigma_\phi$ or mean $\bar\phi$ together
with observation wavelength $\lambda$ determine the characteristic distance 
${\cal L}_{\sigma_\phi,\bar\phi}$ (see next section for exact definition)
over which Faraday effect rotates the polarization by one \textit {radian}.
As the final tally, we have five scales, ${\cal L}_{\sigma_\phi,\bar\phi}$,
$r_\phi$, $r_i$, $L$, $R$ and two scaling slopes $m_\phi$ and $m$.
\begin{table}[h]
\begin{tabular}{ll|l|c} 
Parameter&& Meaning & First appearance \\
Scales:& & & \\
&$\lambda$& wavelength of observations & Eq.\;\;\;\ref{eq:Plambda2} \\
&$ L $ & line-of-sight extent of the emitting region & Eq.\;\;\;\ref{eq:Plambda2} \\
&$ R $ & separation between lines-of-sight & Eq.~\ref{eq:RMden_powerlaw} \\
&$ r_\phi $ & correlation length for Faraday Rotation Measure density & Eq.~\ref{eq:RMden_powerlaw} \\
&$ r_i $ & correlation length for polarization at the source& Eq.~\ref{eq:xi_def} \\
&$ {\cal L}_{\bar\phi} $ & distance of one \textit{rad} revolution by random Faraday rotation &Eq.~\ref{eq:barphi} \\
&$ {\cal L}_{\sigma_\phi} $ & distance of one \textit{rad} revolution by mean Faraday rotation & Eq.~\ref{eq:sigmaphi} \\
&$ {\cal L}_{\sigma_\phi,\bar\phi} $ & the smallest of $ {\cal L}_{\sigma_\phi} $ 
and $ {\cal L}_{\bar\phi} $  & Eq.~\ref{min} \\
Spectral indexes:& & & \\
&$m_\phi$& Correlation index for Faraday RM density & Eq.~\ref{eq:RMden_powerlaw} \\
&$m$& Correlation index for polarization at the source & Eq.~\ref{eq:xi_def} \\
Basic statistical:& & & \\
&$\bar\phi$& Mean Faraday RM density & Eq.~\ref{eq:mean_phi} \\
&$\sigma_\phi$& \textit{rms} Faraday RM density fluctuation & Eq.~\ref{eq:sigma_phi}\\
&$\bar P_i$& Mean polarization at the source & Eq.~\ref{eq:xi_def} \\
&$\sigma_i$& \textit{rms} polarization fluctuation at the source & Eq.~\ref{eq:xi_def} \\
\end{tabular}
\caption{Parameters for correlation studies of the synchrotron polarization 
from an extended emitting region with Faraday rotation}
\label{tab:parameters}
\end{table}

\section{Statistics of the turbulence from single wavelength PPF slice}
\label{sec:samelambda}

In this section we study how spatial correlation properties of the observed 
polarization of synchrotron emission reflect the underlying statistical
properties of magnetic and electron density turbulence. 
Observed polarization correlation properties depend on the separation between
the lines-of-sight and the wavelengths of the observation. 

Let us consider spatial correlations in polarization maps for measurements at
the fixed wavelength. Such approach we shall call Polarization Spatial Analysis
(PSA). The signal is accumulated along pairs of lines-of-sight,
separated by $R$. The main effect of the Faraday rotation in
the sufficiently turbulent (criterium to follow) medium
is to suppress the observed
correlations by establishing an effective narrow line-of-sight depth
over which correlated part of the signal is accumulated.
As we shall show, at small separations
$R < r_\phi$, this depth depends on $R$, resulting in modified scaling
of the polarization correlations that reflects the correlation
of the Faraday RM density. At large separations, the suppression is uniform,
synchrotron correlations are accumulated over an effectively thin slice and
reflect the underlaying correlations of the magnetic field.

We make two approximations in our quantitative treatment. First
we take $\phi$ to be a Gaussian quantity, definitely good approximation
when its fluctuations are dominated by the fluctuations in the magnetic field.
Second, we neglect the correlations between the fluctuations in intrinsic
polarization at the source $P_i$ and the Faraday RM. Here we note that when
both are dominated by fluctuations of magnetic field, which may give the
most of cross-correlation, these are different (perpendicular and parallel to 
line-of-sight) components of the magnetic field 
that define intrinsic polarization and Faraday RM.
At small separations between the lines-of-sight the correlation between them is
suppressed (and is formally zero along coincident lines-of-sight or 
between sources at the same distance when turbulence is isotropic or is 
a strong turbulent mix of Alfv\'en and slow modes as this is the case of nearly incompressible
turbulence (see Goldreich \& Sridhar 1995)).
\footnote{As one sees, then  $\langle H_x(z_1) H_z(z_2) \rangle = 0$,
and, as we discussed, the two-point correlations between the magnetic field
and the electron density are not present either. In general there can be
higher order correlations between the electron density and
the magnetic field vector. Their presence will indicate non-Gaussian nature
of at least electron density distribution, the situation to be studied
in the subsequent papers.}
Whereas at large separations effect of Faraday rotation is, as we'll see below,
mostly amounts to providing a window over which synchrotron polarization
fluctuations are sampled.

Under stated assumptions
\begin{eqnarray}
\left\langle P({\bf X}_1)P^*({\bf X}_2) \right\rangle
&=& \int_0^L dz_1 \int_0^L dz_2\;e^{2 i \overline{\phi}\lambda^2 ( z_1 - z_2)}
\left\langle P_i({\bf X}_1,z_1) P_i^*({\bf X}_2,z_2) \right\rangle
e^{- 2 \lambda^4 \left\langle 
\left( \Delta \Phi({\bf X}_1, z_1) - \Delta \Phi({\bf X}_2, z_2)
\right)^2\right\rangle}
\nonumber \\
&=& \int_0^L dz_1 \int_0^L dz_2\;e^{2 i \overline{\phi}\lambda^2 ( z_1 - z_2)}
\xi_i(\mathbf{R},z_1-z_2) 
e^{- 4 \lambda^4  D_{\Delta \Phi}(\mathbf{R}, z_1,z_2) }
\label{eq:corrP_lambda}
\end{eqnarray}
For observations done at sufficiently long wavelength (criterium to follow),
we can use quadratic approximation of Eq.~(\ref{eq:DPhi_quadratic})
\begin{equation}
\left\langle P({\bf X}_1)P^*({\bf X}_2) \right\rangle
\approx 2 \int_0^{L/2} dz_+ 
e^{-4 \lambda^4 D_{\Delta\Phi}^+(R,z_+)} 
\int_{-2 z_+}^{2 z_+} \!\!\! d\Delta z \;
e^{2 i \overline{\phi}\lambda^2 \Delta z}
\xi_i({\bf R},\Delta z)
e^{- \lambda^4 \Lambda_-({\bf R},z_+) (\Delta z)^2}
\label{eq:corrP_quadratic}
\end{equation}
According to this formulae, both the mean field and the fluctuating, turbulent
Faraday rotation establish an effective width in $\Delta z$ separation
over which the polarization correlations are accumulated over.
For the mean field,
the effective width is
\begin{equation} 
{\cal L}_{\bar\phi} \equiv (\lambda^2 \bar\phi)^{-1} ,
\label{eq:barphi}
\end{equation}
while the one for the fluctuative rotation is 
${\cal L}_{\sigma_\phi} \approx (\lambda^2 \sqrt{\Lambda_-})^{-1}$,
both windows decreasing with the increase in the wavelength of the observations.
These scales have the meaning of a line-of-sight distance over
which polarization direction rotates by approximately a radian.
In what follows we consider the spatial extent of the emitting region
to be much larger that the smallest of these two scales,
$L \gg {\cal L}_{\sigma_\phi, {\bar \phi}}$, where
\begin{equation}
{\cal L}_{\sigma_\phi, {\bar \phi}}\equiv \mathrm{min}({\cal L}_{\bar\phi},{\cal L}_{\sigma_\phi}).
\label{min}
\end{equation}
 In the opposite
case the effect of Faraday decorrelation can be neglected.

Note that the effect of turbulent rotation can be 
more dramatic, leading to Gaussian window in comparison to slower 
oscillatory cutoff from the mean field Faraday rotation.
The quadratic approximation Eq.~(\ref{eq:corrP_quadratic}) 
is sufficient when this effective window produced by 
turbulent component of Faraday rotation ${\cal L}_{\sigma_\phi}$ is narrower than the intrinsic
correlation length $r_i$ of synchrotron fluctuations arising from the magnetic field component 
$H_{\bot}$, i.e. when $r_i \lambda^2 \sqrt{\Lambda_-} > 1$.
Following Eq.~(\ref{eq:Lambda-}), $\Lambda_-$ is bounded from below by
$2 \xi_\phi(R,0)$, i.e $\Lambda_- \approx 2 \sigma^2_\phi$ for $ R < r_e $ and
$\Lambda_- > 2 \sigma^2_\phi (r_\phi/R)^m $ at $R > r_\phi$. Thus
the required criterium is $ \sqrt{2} \lambda^2 r_i \sigma_\phi > 1$ 
with line-of-sight 
separation $ R < (\lambda^2 \sigma_\phi r_i)^\frac{1}{m} r_\phi $.
This criterium can also be written in terms of scales as 
$r_i > {\cal L}_{\sigma_\phi} $ and 
$ R < (r_i/{\cal L}_{\sigma_\phi})^{1/m} r_\phi$ where
we define 
\begin{equation}
{\cal L}_{\sigma_\phi}
\equiv ( \sqrt{2} \lambda^2 \sigma_\phi)^{-1} ,
\label{eq:sigmaphi}
\end{equation}
which is the quantity that will be used through the rest of our paper.

\subsection{Dominance of turbulent rotation, 
${\cal L}_{\sigma_\phi} < {\cal L}_{\bar\phi}$}
\label{subsec:turbdom}
We first consider
the case when $\bar\phi < \sigma_\phi$. This
is the case of either weak regular magnetic field, with respect 
to its fluctuations, 
or of strongly inhomogeneous distribution of electron density, or both.
The problem is complex, having five scales involved, namely the scale for Faraday
rotation ${\cal L}_{\sigma_\phi}$, the correlation length of the rotation
measure $r_\phi$, the correlation length of the transverse magnetic field
$r_i$, the line-of-sight size of the emitting region $L$ and the separation
between two line-of-sight $R$ over which we correlate to polarization signal.
We shall always consider the extent $L$ to exceed the correlation length
of the polarization fluctuations at the source, $ r_i < L$ and 
we limit our studies to $R < L$.
This leaves us with
two parameters ${\cal L}_{\sigma_\phi}/r_i$ and $r_\phi/r_i$ to study
polarization correlation as a function of $R/r_i$. 
We expect $r_\phi \le r_i$ which in case of inequality will give rise to
the intermediate regime $r_\phi < R < r_i$ which can be potentially used to
investigate two correlation lengths separately. In the limiting case
when polarization at the source is dominated by the mean contribution,
we should replace $r_i$ by $L$ in all criteria and 
results that follow.

\begin{figure}[ht]
\includegraphics[width=0.45\textwidth]{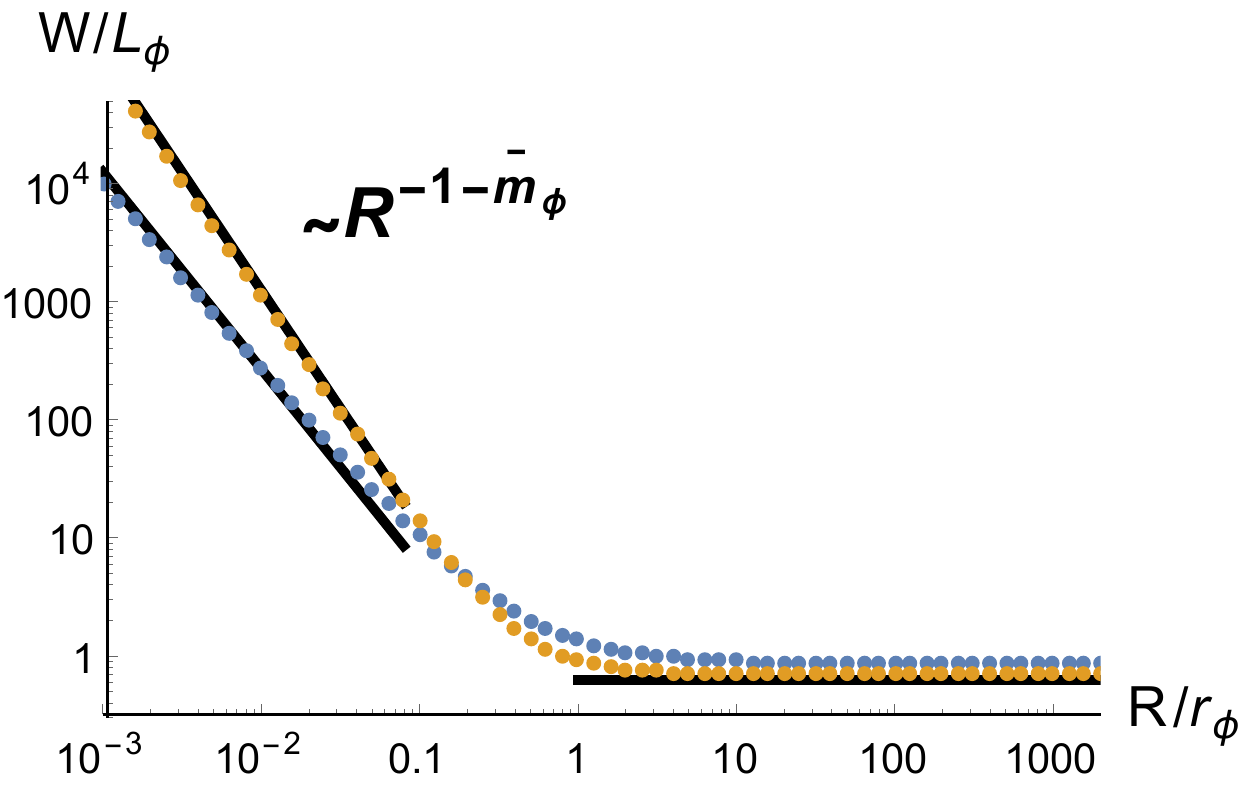}
\caption{Effective Faraday depth $W_\phi(R)$ 
in units of ${\cal L}_{\sigma_\phi}$ as the function of separation
between lines-of-sight $R$ given
in units of the Faraday correlation length $r_\phi$. Dashed lines correspond
to $m_\phi=2/3$ and $m_\phi=3/2$. Solid lines are asymptotical scalings,
$W_\phi/{\cal L}_{\sigma_\phi} \sim \frac{1}{4}
( {\cal L}_{\sigma_\phi}/ r_\phi)
(R/r_\phi)^{-1-\widetilde{m}_\phi}$ at $R \ll r_\phi$ and $W_\phi \sim 
\sqrt{\frac{\pi}{4 m_\phi}} {\cal L}_{\sigma_\phi} $ at $ R > r_\phi$.
Here $\widetilde{m}_\phi = min(1,m_\phi)$.
Note that for $m_\phi=3/2$, the small $R$ asymptotics behaves
as if $m_\phi=1$.
}
\label{fig:Wl}
\end{figure}

Now two basic regimes can be distinguished:

(a) the regime of strong Faraday rotation, $ {\cal L}_{\sigma_\phi} < r_i$.
In this regime, Faraday rotation does not decorrelate the polarization
only from sources with $\Delta z < {\cal L}_{\sigma_\phi} < r_i$.
In the approximation of Eq.~(\ref{eq:corrP_quadratic}), the 
integral over such narrow window of $\Delta z$ gives
\begin{eqnarray}
\left\langle P({\bf X}_1)P^*({\bf X}_2) \right\rangle
&\sim& \sqrt{\pi} \; \xi_i({\bf R},0) \times {\cal L}_{\sigma_\phi}
\int_0^L \frac{dz_+}{\sqrt{\Lambda_-({\bf R},z_+)/(2 \sigma_\phi^2)}}
e^{-4 \lambda^4 D_{\Delta\Phi}^+({\bf R},z_+)} 
\nonumber\\
&=& \sqrt{\pi} \; \xi_i({\bf R},0) \times {\cal L}_{\sigma_\phi}
W_\phi(R)
\label{eq:Pthin}
\end{eqnarray}
The remaining line-of-sight integral provides the effective
depth along the line-of-sight $W_\phi$ over which the signal is accumulated. It
depends on $R$ and warrants a detailed examination.
For $R=0$ it evaluates simply to $L$ as $D_{\Delta\Phi}^+(0,z_+)=0$ and
$\Lambda_-(0,z_+)=2 \sigma_\phi^2$.
At finite $R$, however, it is shortened, since
Faraday rotation decorrelates the signal as we integrate along
two non-coincident lines-of-sight.
Mathematically, $D_{\Delta\Phi}^+(\mathbf{R},z_+)$ increases with $z_+$
with coefficients that increase with $R$ as described by
Eqs.~(\ref{eq:Dz+_smallRsmallz},\ref{eq:Dz+_smallRlargez},
\ref{eq:Dz+_largeRsmallz}).
To compute the Faraday effective depth,  $D_{\Delta\Phi}^+(\mathbf{R},z_+)$ 
is exponentiated and then integrated over $z_+$. Since $D_{\Delta\Phi}^+$ is 
growing in both $R$ and $z_+$, small $R$ behaviour of the $W_\phi(R)$ will
be defined by the functional form of
$D_{\Delta\Phi}^+(\mathbf{R},z_+)$ at large $z_+$, while
large $R$ dependence will be determined by small $z_+$.
As the result the effective depth decreases with  $R$
\begin{equation}
W_\phi(R)
\propto 
{\cal L}_{\sigma_\phi}^2 r_\phi^{\widetilde{m}_\phi} R^{-1-\widetilde{m}_\phi}
~, \quad\quad  R < r_\phi \; \mathrm{min}
\left(\left({\cal L}_{\sigma_\phi}/r_\phi \right)^{\frac{2}{2+\widetilde{m}_\phi}},1\right)
\label{eq:Wphi}
\end{equation}
from it's maximum value of $ L $ as dictated by Eq.~(\ref{eq:Dz+_smallRlargez})
until it becomes effectively constant $ W_\phi \approx {\cal L}_{\sigma_\phi}$  
(with weak dependence on $r_\phi$ and $m_\phi$) at $R \ge r_\phi$ 
as follows from Eq.~(\ref{eq:Dz+_largeRsmallz}).
This behaviour of the Faraday window is summarized in Figure~\ref{fig:Wl}.
If $ r_\phi > {\cal L}_{\sigma_\phi}$
one may detect the intermediate asymptotics 
$W_\phi(R) \propto {\cal L}_{\sigma_\phi} (r_\phi/R)^{m_\phi/2}$,
over the range of scales 
$\left[r_\phi 
\left({\cal L}_{\sigma_\phi}/r_\phi \right)^{\frac{2}{2+\widetilde{m}_\phi}},
r_\phi \right]$,
as governed by Eq.~(\ref{eq:Dz+_smallRsmallz}).

Thus, at $R \ll r_\phi$ we have asymptotic behaviour of polarization correlation
\begin{eqnarray}
\label{eq:xi_sync_smallR_m<1}
\left\langle P({\bf X}_1)P^*({\bf X}_2) \right\rangle &\sim&
{\cal L}_{\sigma_\phi}^2 \xi_i({\bf R},0)
\frac{ {\cal L}_{\sigma_\phi} r_\phi^{\widetilde{m}_\phi}}{R^{1+\widetilde{m}_\phi}}
\quad \quad \sqrt{2} \lambda^2 r_i \sigma_\phi > 1, \quad
R < r_\phi
\label{eq:xi_sync_smallR_m>1}
\end{eqnarray}
that is proportional to the underlying turbulent correlations taken in thin,
$z=0$ slice, but is modified due to Faraday rotation by
$R^{-1-\widetilde{m}_\phi}$. 
Moreover, with $r_\phi \le r_i$ as expected, the underlying correlations
are almost constant at such small separations, $\xi_i(R \ll r_\phi,0) \approx
\bar P_i^2+\sigma_i^2 = const$, and  $R^{-1-\widetilde{m}_\phi}$ scaling is
the dominant one.

Whereas at larger separations, $ R \gg r_\phi $,
the correlation signal
is simply accumulated from a thin slice of depth ${\cal L}_{\sigma_\phi}$
from the observer and the scaling
of the observed polarization reflects that of the intrinsic
polarization unmodified 
\begin{equation}
\left\langle P({\bf X}_1)P^*({\bf X}_2) \right\rangle \sim
{\cal L}_{\sigma_\phi}^2 \xi_i({\bf R},0)
\quad \quad \sqrt{2} \lambda^2 r_i \sigma_\phi > 1, \quad
R > r_\phi ~.
\label{eq:xi_sync_largeR}
\end{equation}

\begin{figure}[ht]
\includegraphics[width=0.45\textwidth]{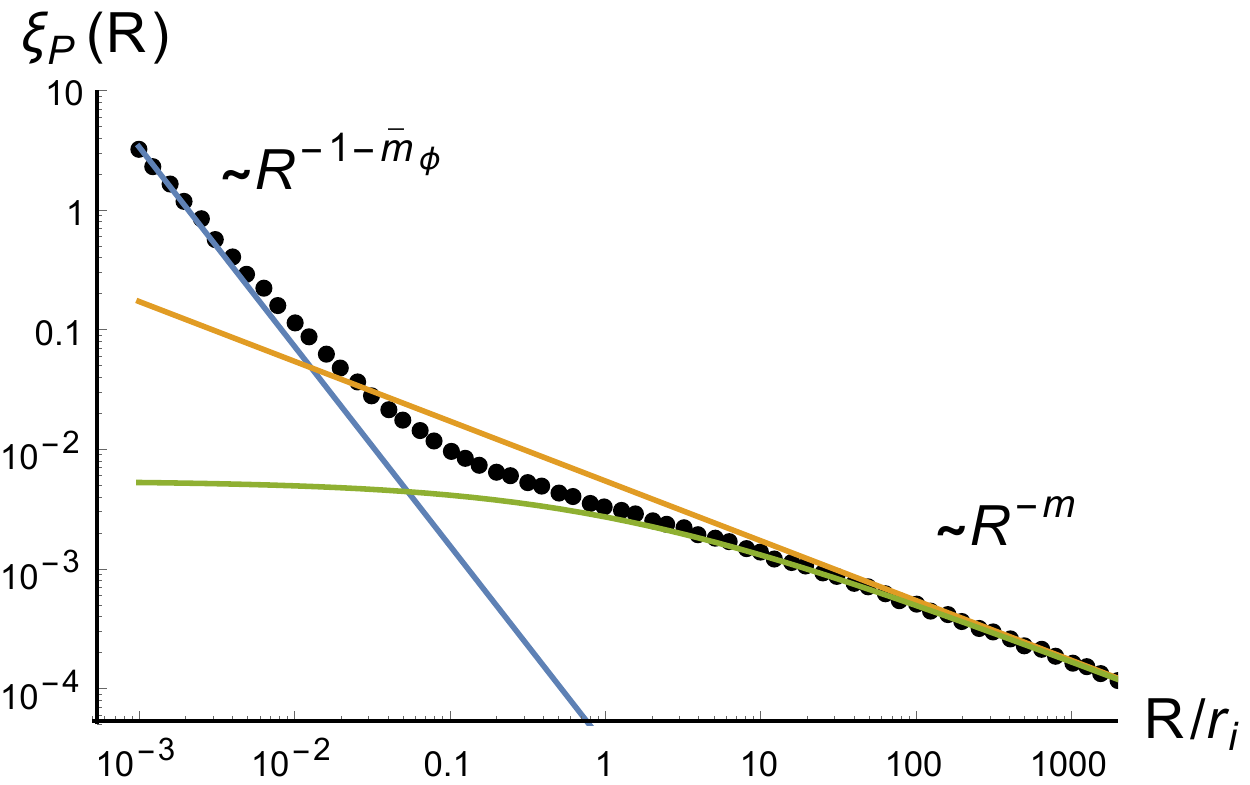}
\includegraphics[width=0.45\textwidth]{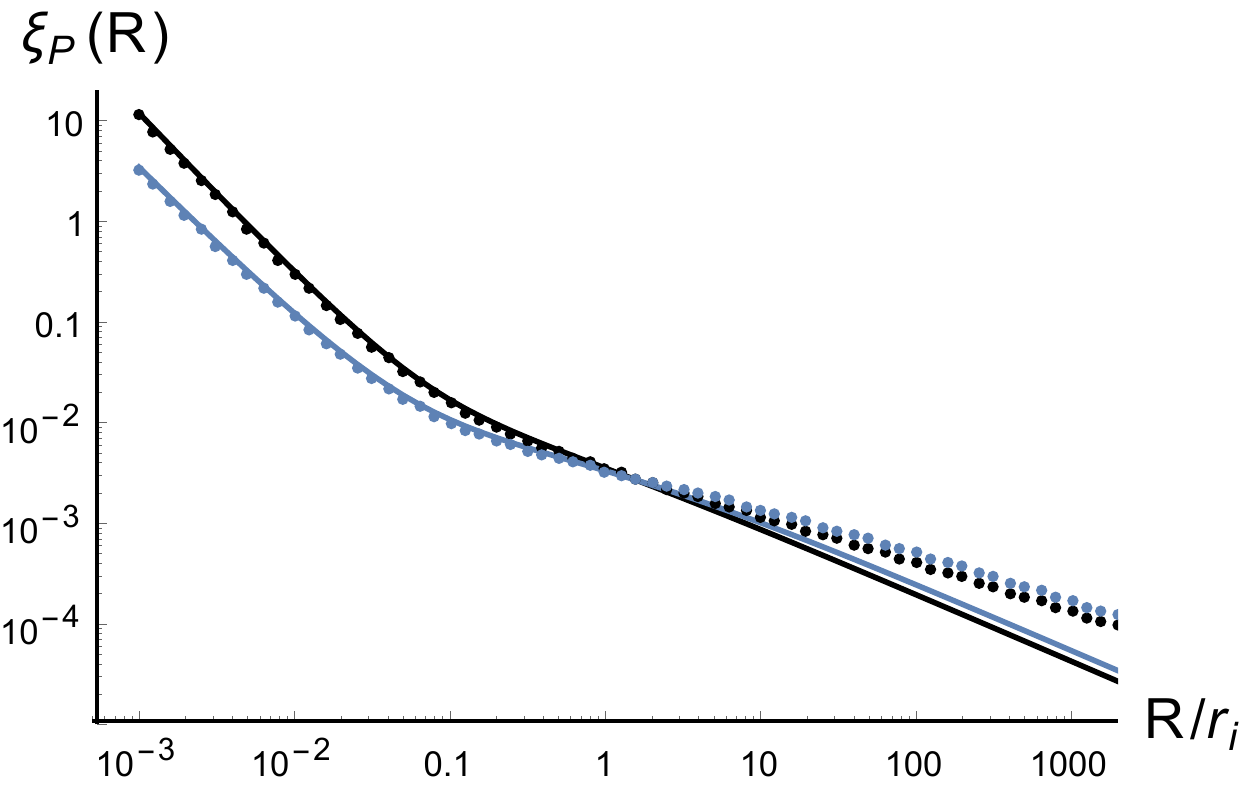}
\caption{The correlation function of polarization
$ \xi_{P}(R) \equiv \left\langle P({\bf X}_1)P^*({\bf X}_2) \right\rangle $
in the limit of strong turbulent Faraday rotation,
${\cal L}_{\sigma_\phi} < r_i$, specifically ${\cal L}_{\sigma_\phi}=0.07\;r_i$.
Left: $\xi_{P}$ asymptotic regimes. Dotted line is the numerical evaluation. 
The correlation follows 
Eq.~(\ref{eq:xi_sync_smallR_m<1})
at $R < r_\phi$ (blue) and Eq.~(\ref{eq:xi_sync_largeR}) at $ R > r_\phi$
(green).
The latter regime reaches a power law behaviour for $R > r_i$ (orange).
Here $r_\phi=0.1\; r_i$ and $\widetilde{m}_\phi=min(m_\phi,1)$. 
Right: dependence of
$\xi_{P}(R)$ on $m$ and $r_\phi$; $r_\phi=r_i$ (black) and 
$r_\phi=0.1\; r_i$ (blue); $m=2/3$ (solid) and $m=1/2$ (dotted).
}
\label{fig:asymp}
\end{figure}

In Figure~\ref{fig:asymp} we show
the results of numerical integration of Eq.~(\ref{eq:corrP_lambda}) 
with $\bar P_i=0$ that demonstrate the discussed regimes. It shows that the
regime Eq.~(\ref{eq:xi_sync_largeR}) remains valid at large $R$ even 
beyond the range of validity
of the quadratic approximation of Eq.~(\ref{eq:corrP_quadratic}).
Importantly, the change of the correlation slope
is expected at $R=r_\phi$ which can be used to determine the RM density 
correlation scale $r_\phi$.

We also see that in the regime of strong random
Faraday rotation, 
3D statistical anisotropy of the turbulence, if present,
is directly mapped into 2D observational statistics via 
$ \xi_i({\bf R},0)$.

\begin{figure}[ht]
\includegraphics[width=0.45\textwidth]{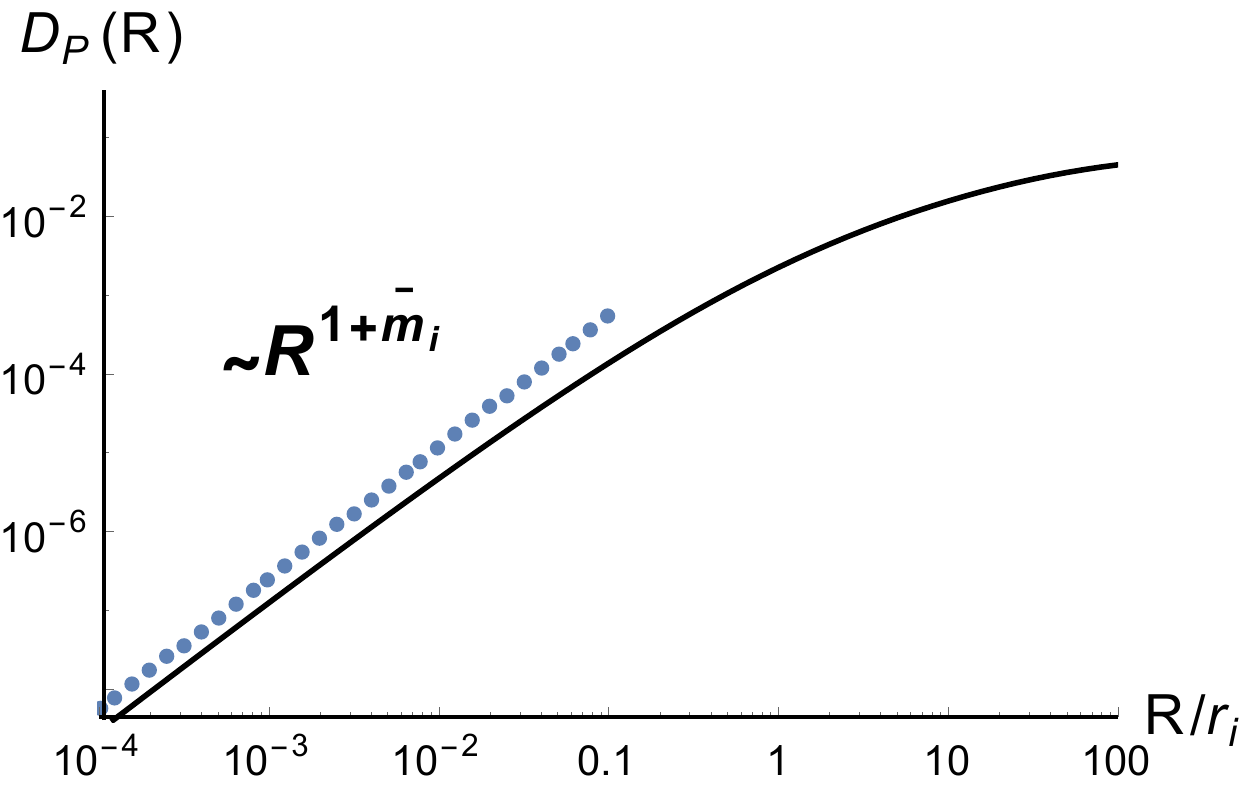}
\includegraphics[width=0.45\textwidth]{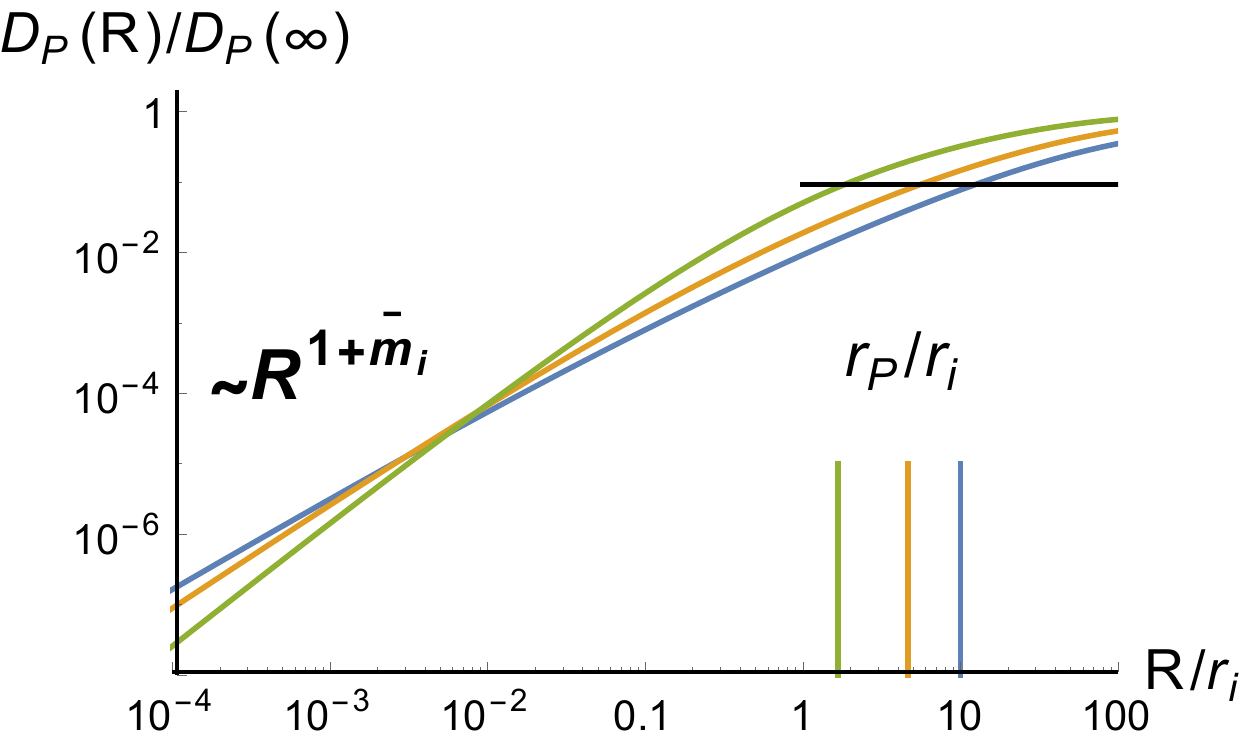}
\caption{The structure function of polarization
$ D_{P}(R) \equiv \left\langle \left| P({\bf X}_1)- P({\bf X}_2)\right|^2 \right\rangle $
in the limit of negligible Faraday rotation. Depth of the emitting region is
taken to be $L = 100 r_i$.
Left: $D_{P}$ asymptotic regime at $R \ll r_i$. Solid line is the
full numerical evaluation, specifically for $m=2/3$. Corresponding asymptotic slope (dotted line) is offset for clarity.
Right: dependence of $D_{P}(R)$ and the observed correlation length $R_P$,
given by Eq.~(\ref{eq:R_P}), on $m$;  
$m=1/3,1/2,4/5$ -- correspondingly from the longest to the shortest correlation
lengths, as marked by vertical lines. According to Eq.~\ref{eq:corrP_weakF_scaling}, normalized structure functions intersect at $R/r_i \approx r_i/L$.
}
\label{fig:DR_weakF}
\end{figure}

(b) the weak rotation case, $r_i < {\cal L}_{\sigma_\phi}$, when
the Faraday rotation is small over the distances on which
intrinsic polarization is correlated. Eq.~(\ref{eq:corrP_lambda})
asymptotically gives for observed polarization correlation
\begin{equation}
\left\langle P({\bf X}_1)P^*({\bf X}_2) \right\rangle
\approx \int_0^L \!\!\! d \Delta z \; \xi_i({\bf R},\Delta z)
\int_{\Delta z/2}^{L-\Delta z/2} \!\!\! dz_+
\left( 1 - 4 \lambda^4 D_{\Delta\Phi}(R,z_1,z_2) \right)
\label{eq:corrP_weakF}
\end{equation}
i.e. simply the intrinsic correlations integrated over the line-of-sight with,
if the required accuracy warrants it, perturbative
correction from Faraday rotation. 
Using the structure function, the leading
behaviour at small scales is
\begin{eqnarray}
\label{eq:corrP_weakF_scaling}
\left\langle \left| P({\bf X}_1) - P({\bf X}_2)\right|^2 \right\rangle
&\sim& \sigma_i^2 L^{2-\bar{m}} r_i^{\bar{m}}  (R/R_P)^{1+\bar{m}}
= \sigma_i^2 L R \left( R/r_i \right)^{\bar{m}} ~,
\quad \bar{m} \equiv \mathrm{min}(1,m)~,
\end{eqnarray}
where for $m < 1$ the observed correlation length $R_P$ (defined as the scale
where structure function reaches one half of its asymptotic limit 
$\approx  L^{2-\bar{m}} r_i^{\bar{m}}$) depends
not only on $r_i$, but also on
the size of the emitting region, 
\begin{equation}
R_P \approx r_i (L/r_i)^\frac{1-\bar{m}}{1+\bar{m}} ~,
\label{eq:R_P}
\end{equation}
and will 
significantly exceed $r_i$ if the emitting volume extend is large, $L \gg r_i $.
The reason is that for such low $m$ observed correlations are accumulated
from pairwise volume correlations at all distance separations up to $L$.
For the fixed $L$ and $r_i$, the larger the $m$ is, the steeper is the slope and
the shorter is the correlation length of the observed correlations.
For $m > 1$ the
observed correlation saturates at quadratic behaviour, which hides the
information about the underlying turbulence.
While the first expression in Eq.~(\ref{eq:corrP_weakF_scaling}) focuses on
the asymptotic value of the structure function and the observed correlation 
length, the second, equivalent, form reminds us that power-law asymptotics is
accurate only for $R < r_i$. This is illustrated in Fig.~\ref{fig:DR_weakF}.

In all the regimes discussed in this section,
the intrinsic correlation function $\xi_i$ is
factorized, thus the effects from the mean intrinsic polarization and
its fluctuations are additive in the observational measures. If fluctuating part
is negligible, $\xi_i \approx \bar P_i^2 $, one should replace $r_i$ by $L$ in
all the above results and criteria. 
The only interesting case for observations is at
short separations $R < r_\phi$ under the strong Faraday rotation, 
where the scaling of the Faraday rotation depth can be determined
\begin{equation}
\label{eq:mean_xi_sync_smallR}
\left\langle P({\bf X}_1)P^*({\bf X}_2) \right\rangle \sim
{\cal L}_{\sigma_\phi}^2  \bar P_i^2
\frac{ {\cal L}_{\sigma_\phi} r_\phi^{\widetilde{m}_\phi}}
{R^{1+\widetilde{m}_\phi}}
\quad \quad \sigma_i^2 \ll \bar P_i^2 , 
\quad  \sqrt{2} \lambda^2 L \sigma_\phi > 1, \quad
R < r_\phi,  
\end{equation}
At large separations or if the Faraday rotation is weak, 
the observed correlations will exhibit the plateau value,
correspondent to the mean observed polarization. It can be subtracted out by
measuring the structure function instead.

\subsection{Dominance of the mean field in Faraday rotation, 
${\cal L}_{\sigma_\phi} > {\cal L}_{\bar\phi}$}

Let us now consider the situation when the RM due to the mean field is
much larger than the turbulent contribution.  
With fluctuations in RM neglected, 
Eq.~(\ref{eq:corrP_lambda}) can be rearranged as 
\begin{equation}
\left\langle P({\bf X}_1)P^*({\bf X}_2) \right\rangle
\approx 2 \int_{0}^{L} \!\!\! d \Delta z (L - \Delta z) 
\left( \cos \left( 2 \overline{\phi}\lambda^2 \Delta z \right)
\xi_i^e({\bf R},\Delta z)
+ i \sin \left( 2 \overline{\phi}\lambda^2 \Delta z \right)
\xi_i^o({\bf R},\Delta z) \right)
\label{eq:xi_meanF}
\end{equation}
where 
$ \xi_i^e({\bf R},\Delta z) = \frac{1}{2} \left( \xi_i({\bf R},\Delta z)
+\xi_i({\bf R},-\Delta z) \right) $ is the even and 
$ \xi_i^o({\bf R},\Delta z) = \frac{1}{2} \left( \xi_i({\bf R},\Delta z)
-\xi_i({\bf R},-\Delta z) \right) $ is the odd part of the intrinsic 
correlations with respect to $\Delta z \to -\Delta z$ change.
Both parts, however, can be complex, with real part symmetric and imaginary
part antisymmetric with respect to $\mathbf{r} \to -\mathbf{r}$ as has been
discussed in \S~\ref{sec:xi_i}. So
for the observable quantities
\begin{eqnarray}
\left\langle Q({\bf X}_1)Q({\bf X}_2) +  U({\bf X}_1)U({\bf X}_2) \right\rangle
&\approx& 2 \int_{0}^{L} \!\!\! d \Delta z (L - \Delta z) 
\left( \cos \left( 2 \overline{\phi}\lambda^2 \Delta z \right)
\Re \xi_i^e({\bf R},\Delta z)
- \sin \left( 2 \overline{\phi}\lambda^2 \Delta z \right)
\Im \xi_i^o({\bf R},\Delta z) \right) \\
\left\langle Q({\bf X}_1)U({\bf X}_2) -  U({\bf X}_1)Q({\bf X}_2) \right\rangle
&\approx& 2 \int_{0}^{L} \!\!\! d \Delta z (L - \Delta z) 
\left( \cos \left( 2 \overline{\phi}\lambda^2 \Delta z \right)
\Im \xi_i^e({\bf R},\Delta z)
+ \sin \left( 2 \overline{\phi}\lambda^2 \Delta z \right)
\Re \xi_i^o({\bf R},\Delta z) \right)
\label{eq:QU_meanF}
\end{eqnarray}

In isotropic MHD turbulence with no helical correlations, 
$ \xi_i({\bf R},\Delta z) =  \xi_i(R^2+\Delta z^2)$ is even and real,
thus $\Im\xi_i^e=\xi_i^o=0$,
and no antisymmetric correlations should be observed
\begin{eqnarray}
\label{eq:QU_meanF_iso}
\left\langle Q({\bf X}_1)Q({\bf X}_2) +  U({\bf X}_1)U({\bf X}_2) \right\rangle
&\approx& 2 \int_{0}^{L} \!\!\! d \Delta z (L - \Delta z) 
 \cos \left( 2 \overline{\phi}\lambda^2 \Delta z \right)
\xi_i({\bf R},\Delta z) \\
\left\langle Q({\bf X}_1)U({\bf X}_2) -  U({\bf X}_1)Q({\bf X}_2) \right\rangle
&=& 0 \quad .
\end{eqnarray}

Helicity in isotropic turbulence contributes only the purely odd
$\Im\xi_i^o$ imaginary part and a correction to $\Re\xi_i^e$
(see Appendix~\ref{app:antisym}).
Thus, despite leading to antisymmetric correlations between two emitters,
it does not give rise to antisymmetric 
$\left\langle Q({\bf X}_1)U({\bf X}_2)-U({\bf X}_1)Q({\bf X}_2) \right\rangle$
in the observable polarization from an extended emitting region,
whether Faraday rotation is present or not. It modifies the symmetric trace
of the correlations only, namely
\begin{eqnarray}
\left\langle Q({\bf X}_1)Q({\bf X}_2) +  U({\bf X}_1)U({\bf X}_2) \right\rangle
&\approx& 2 \int_{0}^{L} \!\!\! d \Delta z (L - \Delta z) 
\left( \cos \left( 2 \overline{\phi}\lambda^2 \Delta z \right)
\Re \xi_i^e({\bf R},\Delta z)
- \sin \left( 2 \overline{\phi}\lambda^2 \Delta z \right)
\Im \xi_i^o({\bf R},\Delta z) \right) \\
\left\langle Q({\bf X}_1)U({\bf X}_2) -  U({\bf X}_1)Q({\bf X}_2) \right\rangle
&\approx& 0
\label{eq:QU_isohel}
\end{eqnarray}

However, in contrast to the random Faraday rotation, the regular rotation can
generate the observable antisymmetric correlations from anisotropy
of the turbulence. 
For instance, for axisymmetric turbulence, the 3D correlation functions
are real functions of separation magnitude $r=\sqrt{R^2+\Delta z^2}$
and the modulus of its angle with the
symmetry axis $\hat \lambda$, 
$\xi_i = \xi_i(r, |\mathbf{r} \cdot \hat \lambda|)$.
Unless the preferred direction is strictly perpendicular, 
$\hat\lambda_z=0$, or parallel, $\hat\lambda_x=\hat\lambda_y=0$, to the 
line-of-sight, such correlations, generally, contain the odd in $\Delta z$ 
contribution, and since $\Im\xi_i=0$, lead to
\begin{eqnarray}
\left\langle Q({\bf X}_1)Q({\bf X}_2) +  U({\bf X}_1)U({\bf X}_2) \right\rangle
&\approx& 2 \int_{0}^{L} \!\!\! d \Delta z (L - \Delta z) 
\cos \left( 2 \overline{\phi}\lambda^2 \Delta z \right)
\xi_i^e({\bf R},\Delta z) \\
\left\langle Q({\bf X}_1)U({\bf X}_2) -  U({\bf X}_1)Q({\bf X}_2) \right\rangle
&\approx& 2 \int_{0}^{L} \!\!\! d \Delta z (L - \Delta z) 
\sin \left( 2 \overline{\phi}\lambda^2 \Delta z \right)
\xi_i^o({\bf R},\Delta z) 
\label{eq:QU_meanF_ani}
\end{eqnarray}
for the correlations in $Q,U$ observables.

\begin{figure}[ht]
\includegraphics[width=0.45\textwidth]{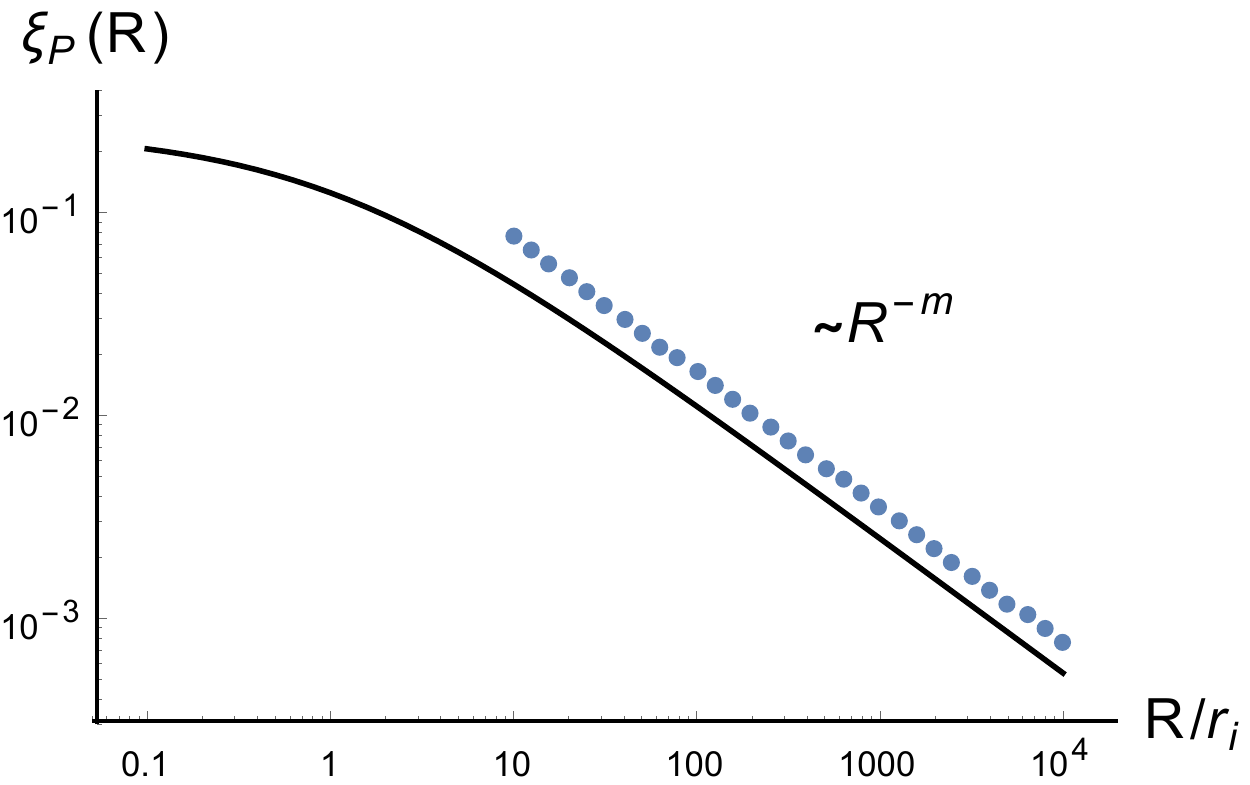}
\caption{The correlation function of polarization
$ \xi_{P}(R)=\left\langle P({\bf X}_1)P^*({\bf X}_2) \right\rangle  $
in the regime of strong mean Faraday rotation, 
${\cal L}_{\bar\phi} < r_i$, ${\cal L}_{\bar\phi} < {\cal L}_{\sigma_\phi}$.
In this plot ${\cal L}_{\bar\phi} = 0.1\; r_i$.
$ \xi_{P}(R) \propto \xi_i(R,0)$ over all range of R and is plotted 
with $\bar P_i=0$.
}
\label{fig:asymp_meanF}
\end{figure}

Following the main line of the paper, let us restrict our considerations to the
real, symmetric, part of the correlations as given in Eq.~(\ref{eq:QU_meanF_iso}).
In this problem we have four scales, ${\cal L}_{\bar\phi}$, $r_i$, $L$ and $R$.
We still assume $L \gg r_i$, leaving us with  
${\cal L}_{\bar\phi}/r_i$ parameter and $R/r_i$ variable.
The only new regime is when Faraday rotation is strong again,
${\cal L}_{\bar\phi} < r_i$, in which case the integral is estimated as
\begin{equation}
\left\langle P({\bf X}_1)P^*({\bf X}_2) \right\rangle \sim
L \; {\cal L}_{\bar\phi} \; \xi_i({\bf R},0)~,
\quad \quad \lambda^2 r_i \bar\phi > 1~.
\label{eq:xi_sync_meanF}
\end{equation}
This represents the behaviour similar to
Eq.~(\ref{eq:xi_sync_largeR}) for random RM.
The difference is the absence of the
effect of scale dependent RM density correlations at $R < r_\phi$ and
the enhanced overall amplitude by $L/{\cal L}_{\bar\phi}$ factor.
The amplitude that is $\propto (L \; {\cal L}_{\bar\phi})$ 
tells us that the observed
correlation is dominated by pairs of emitters at the same 
(within the window ${\cal L}_{\bar\phi}$) line-of-sight distance, 
but all such pairs throughout the emitting region contribute, not just 
the ``thin'' layer closest to the observer as in the case of random Faraday
rotation. 
To neglect any fluctuating component in the RM density is of course, an
idealization. Presence of any even small, 
${\cal L}_{\sigma_\phi} > {\cal L}_{\bar\phi}$,  random contribution will
limit the depth from which polarization correlations are coming
to ${\cal L}_{\sigma_\phi} $.
With the amplitude factored out, the behaviour of $\xi_P(R)$ is 
independent on exact value of ${\cal L}_{\bar\phi}/r_i$ and
is demonstrated in Figure~\ref{fig:asymp_meanF}.

\subsection{Antisymmetric correlations: illustration}
\label{subsec:antisym}

We can illustrate the effect of antisymmetric correlations that arise in the 
presence of regular magnetic field using a simple toy model of weak anisotropy,
$\xi_i(\mathbf{R},\Delta z) = \xi_0(r) 
\left( 1 + A (\mathbf{\hat r} \cdot \hat\lambda)^2 \right) $ 
and power-law
$\xi_0(r) = (R^2 + \Delta z^2)^{-m/2}$.
The odd part of 3D correlations is proportional to the degree of anisotropy
\begin{equation}
\xi_i^o = A \xi_0(r) \frac{2 R \Delta z}{R^2 + \Delta z^2} \sin 2\theta 
\end{equation}
and $\sin2\theta$ of the angle $\phi$ between the direction of the symmetry axis
and the line-of-sight. Despite relying here on a specific model, 
this linear in $\Delta z$ odd behaviour appears naturally in a wider context, 
as the first, linear term in expansion for small anisotropy.
The even part of the correlation, has the main isotropic term,
in addition to anisotropic
correction proportional to $A \cos 2\theta$.
\begin{equation}
\xi_i^e = \xi_0(r) \left( 1 + {\cal O}(A \cos 2\theta) \right)
\end{equation}
For our power-law $\xi_0(r)$ and neglecting boundary effects, 
Eq.~(\ref{eq:QU_meanF_ani}) can be transformed by integration by parts in
\begin{eqnarray}
\left\langle Q({\bf X}_1)U({\bf X}_2) -  U({\bf X}_1)Q({\bf X}_2) \right\rangle
&\approx& A \overline{\phi} \lambda^2 R \sin 2\theta 
\left[2 L \int_{0}^{\infty} \!\!\! d \Delta z 
\cos \left( 2 \overline{\phi}\lambda^2 \Delta z \right)
(R^2 + \Delta z^2)^{-m/2} \right]
\nonumber \\
&\approx& A \overline{\phi} \lambda^2 R \sin 2\theta 
\left\langle Q({\bf X}_1)Q({\bf X}_2) +  U({\bf X}_1)U({\bf X}_2) \right\rangle
\end{eqnarray}
where we have left only the leading in anisotropy term in the last expression.
This result, with details depending on an exact model of turbulence, shows
the main effects that determine the ratio of the imaginary antisymmetric
and the real symmetric terms in the polarization correlation. They are the
degree of anisotropy, amount of Faraday rotation over the distance of
the separation $R$ between the two lines-of-sight, and the dependence on twice
the angle between anisotropy preferred axis and the line-of-sight
that makes the effect vanishing for $\theta=0$ and $\theta=\pi/2$.

Without analyzing more realistic models we may nevertheless see that the analysis of the
imaginary part of the correlations can provide the information about the direction of the
angle between the line-of-sight and the anisotropy preferred axis determined by magnetic field.
The positional angle in the plane of the sky can be obtained either from the polarization direction
or from the anisotropy measurement technique that is described in LP12. Therefore, we may state that
the study of synchrotron fluctuations should provide the 3D orientation of the vector of magnetic field,
which is very advantageous.

\section{Line-of-sight measures}
\label{sec:lineofsight}
The other regime that we would like to study is the multi-wavelength
observations along a fixed line-of-sight.
In this section we focus on the statistical measures of polarization
along a fixed line-of-sight, but at different wavelengths. 
Such approach that we shall call
Polarization Frequency Analysis (PFA) is complimentary to PSA. Compared to the
single wavelength regime, which place demands on the spatial resolution of the
measurements, the regime that we study below is essentially spectroscopic,
requiring sufficient wavelength resolution and coverage. We remind the reader
that the basic $\lambda^{\gamma-1}$ wavelength dependence of synchrotron
intensity is assumed to be scaled out in our measures.

\subsection{Variance}
\label{sec:variance}

Along a fixed line-of-sight the correlation of polarization signal measured at different wavelengths 
(label ${\bf X}$ can be omitted) is
\begin{equation}
\left\langle P(\lambda^2_1)P^*(\lambda^2_2) \right\rangle
= \int_0^L dz_1 \int_0^L dz_2  e^{2 i \overline{\phi}\left( \lambda_1^2 z_1  - \lambda_2^2 z_2 \right)} \left\langle P_i(z_1) P_i^*(z_2) 
e^{2 i \left( \lambda_1^2  \Delta \Phi(z_1) - \lambda_2^2  \Delta \Phi(z_2) \right)}
\right\rangle
\label{eq:corrPx=0}
\end{equation}
The variance is a special case of the correlation function studied
in the previous section taken at $R=0$, but it is useful to be looked at
separately as a function of $\lambda^2$
\begin{equation}
\left\langle P^2(\lambda^2) \right\rangle
= \int_0^L \!\!\! dz_1 \int_0^L \!\!\! dz_2 \;  e^{2 i \overline{\phi} \lambda^2 \left( z_1  - z_2 \right)}
\left\langle P_i(z_1) P_i^*(z_2) 
e^{2 i \lambda^2 \left( \Phi(z_1) - \Phi(z_2) \right)}
\right\rangle
\label{eq:varP}
\end{equation}
and, following the discussion preceding Eq.~(\ref{eq:corrP_lambda})
\begin{equation}
\left\langle P^2(\lambda^2) \right\rangle
\approx \int_0^L \!\!\! dz_1 \int_0^L \!\!\! dz_2 \;  e^{2 i \overline{\phi} \lambda^2 \left( z_1  - z_2 \right)}
\left\langle P_i(z_1) P_i^*(z_2) \right\rangle 
e^{-4 \lambda^4 D_{\Delta\Phi}(0,z_1-z_2)}
\label{eq:varP_noncorr}
\end{equation}

\begin{figure}[ht]
\includegraphics[width=0.45\textwidth]{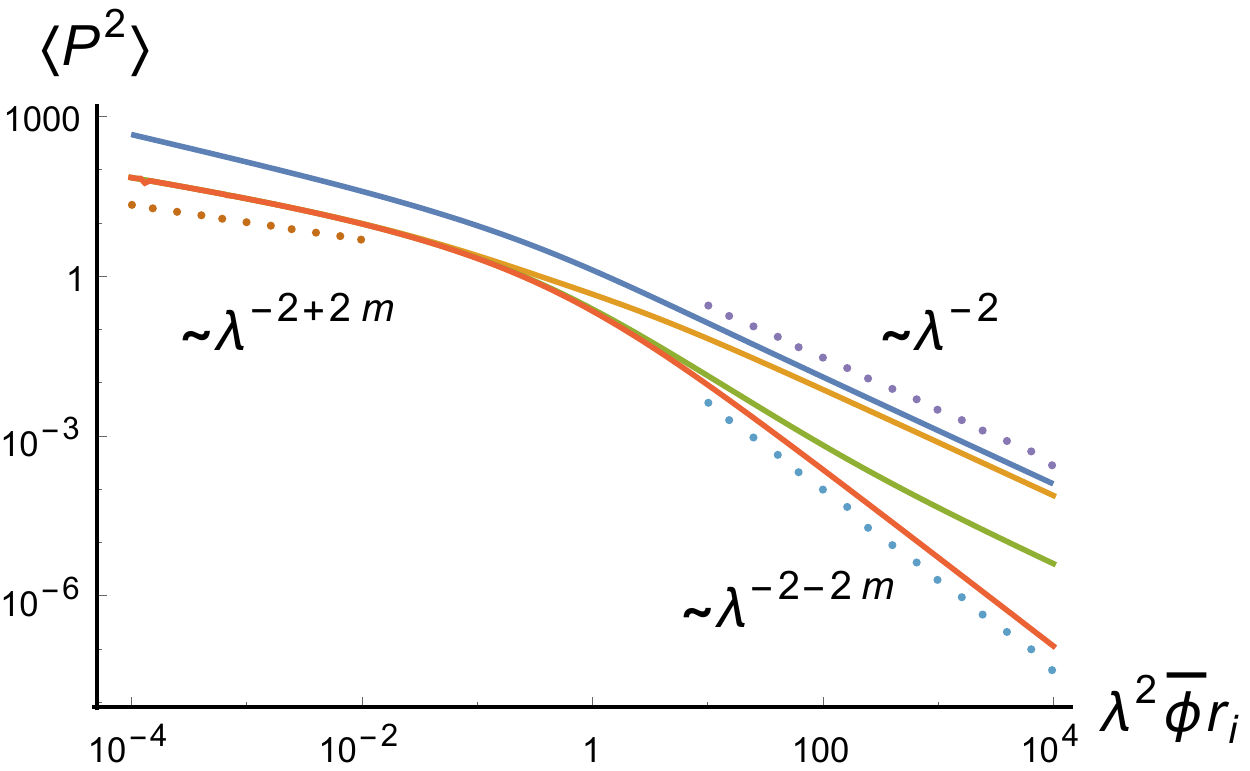}
\includegraphics[width=0.45\textwidth]{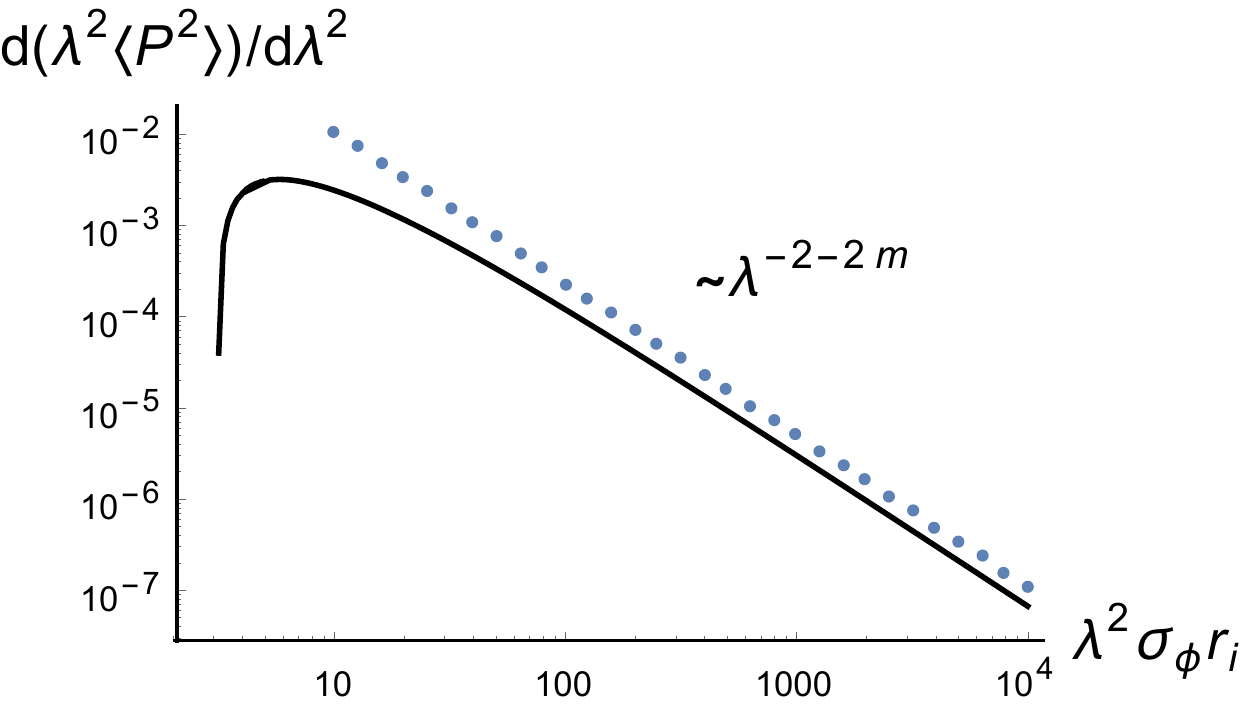}
\caption{Left: Polarization variance $\sigma_P^2(\lambda^2) \equiv
\left\langle P P^* \right\rangle $ as the function of the square of
the wavelength $\lambda^2$. Four curves from top to bottom correspond
to decreasing contribution of the turbulent component of Faraday rotation,
relative to the mean field effect. The sequence is $\sigma_\phi/\bar \phi
= \infty, 1, 0.4, 0.1$. The values $\sigma_\phi/\bar\phi < 0.1$ give
practically the same curve as $0.1$. In case turbulent component of RM
dominates, the variance scales with universal $\lambda^{-2}$ slope
for $\lambda^2 \sigma_\phi r_i > 1$. In case mean RM dominates,
slope of the variance reveals underlying magnetic field scaling via
$\lambda^{-2-2m}$ asymptotical behaviour for $\lambda^2 \bar\phi r_i > 1$.
In the regime of weak, but still present Faraday rotation, 
$r_i < {\cal L}_{\sigma_\phi,\bar\phi} < L$ achieved at small wavelength,
the variance shows scales as $\lambda^{-2+2 m}$ for $m < 1$ and saturates
for $m > 1$.
Right: derivative of the upper curve in the right panel scales as
$\lambda^{-2-2m}$ reflecting the magnetic field scaling. In both plots
amplitude is in arbitrary units and $r_\phi=0.1\;r_i$.
}
\label{fig:variances}
\end{figure}

In Fig.~\ref{fig:variances} we plot the results of numerical analysis
of Eq.~(\ref{eq:varP_noncorr}) for different ratios of $\sigma_\phi/\bar\phi$
that characterizes the transition from the case when RM is predominantly
stochastic to the case when Faraday rotation is mainly due to the uniform
magnetic field and electron distributions.
One could expect that when Faraday window ${\cal L}_{\sigma_\phi}$ or
${\cal L}_{\bar\phi}$ is small enough to resolves $r_i$ 
it may be possible to measure the correlations of the underlying magnetic
field. Our results show that situation, however, is more complicated.

One indeed recovers asymptotically the transverse magnetic field slope,
if Faraday rotation is predominantly uniform $\bar\phi > \sigma_\phi$.
In this case 
\begin{equation}
\left\langle P^2(\lambda^2) \right\rangle \propto \lambda^{-2-2m}~,
\quad \lambda^2 \bar\phi r_i \gg 1~.
\label{eq:var_phi}
\end{equation}
However when turbulent rotation dominates, $\bar\phi < \sigma_\phi $,
$\left\langle P^2(\lambda^2) \right\rangle$ 
follows the universal $\lambda^{-2}$ scaling 
\begin{equation}
\left\langle P^2(\lambda^2) \right\rangle \propto \lambda^{-2}~,
\quad \lambda^2 \sigma_\phi r_i \gg 1 ~,
\label{eq:var_sigma}
\end{equation}
and it is only its weighted derivative
that reflects the turbulent slope of the transverse magnetic field,
as shown in the right panel of Fig~\ref{fig:variances}.
The transition from one regime to another comes at $\sigma_\phi \approx 
\frac{1}{3} \bar\phi$ and is complete as $\sigma_\phi$ changes within the
decade from $\sigma_\phi=0.1\;\bar\phi$ to $\sigma_\phi=\bar\phi$.

Our results can be understood from the following asymptotic considerations.
The dominance of geometrical effects at small line-of-sight 
separations, as reflected in
Eq.~(\ref{eq:DFR0}), allows to argue that even a simplified version of
Eq.~(\ref{eq:varP})
\begin{equation}
\left\langle P^2(\lambda^2) \right\rangle
= \int_0^L \!\!\! dz_1 \int_0^L \!\!\! dz_2 \;  e^{2 i \overline{\phi} \lambda^2 \left( z_1  - z_2 \right)}
\left\langle P_i(z_1) P_i^*(z_2) \right\rangle 
e^{-2 \lambda^4 \sigma^2_{\phi} (z_1 - z_2)^2}
\label{eq:varP_simp}
\end{equation}
provides
a good approximation to full results as long as observations are carried out at sufficiently long
wavelengths that satisfy $\lambda^2 \sigma_{\phi} r_\phi > 1$,
i.e. ${\cal L}_{\sigma_\phi} < r_\phi$.
Given the expectation that $r_\phi < r_i$, 
under this condition one can take 
$\left\langle P_i(z_1) P_i^*(z_2) \right\rangle \propto 
1 - \left(\frac{|\Delta z|}{r_i}\right)^m $,
i.e. the variance sensitively depends on
the spectral index of magnetic field $m$. 

On the contrary, when Faraday rotation is dominated by the random part,
$\sigma_\phi > \bar\phi$,
\begin{eqnarray}
\left\langle P^2(\lambda^2) \right\rangle
& \propto & L \int_0^L  \!\!\! d \Delta z ( 1 - \Delta z/L)
\left(1 -\frac{\Delta z^m}{r_i^m} \right)
e^{-2 \lambda^4 \sigma^2_{\phi} \Delta z^2 } \nonumber \\ 
&\propto& \frac{L}{\sqrt{2} \sigma_{\phi} \lambda^2}
 \left(
1
- 
\left(\sqrt{2} \lambda^2 \sigma_{\phi} r_i \right)^{-m} 
\Gamma\left[\frac{1+m}{2}\right]/\sqrt{\pi}
\right) ~,
\quad \lambda^2 \sigma_\phi r_i > 1~,
\label{eq:varP_scale}
\end{eqnarray}
that shows the universal $\lambda^{-2}$ scaling which masks
the decreasing with
wavelength correction that is sensitive to $m$ slope index.
This information is recovered in the derivative
\begin{equation}
\frac{ \mathrm{d} \left\langle \lambda^2 P^2(\lambda^2) \right\rangle}
{\mathrm{d} \lambda^2} = 
\frac{m \Gamma\left[\frac{1+m}{2}\right]}{\sqrt{\pi}}
L r_i
\left(\sqrt{2} \lambda^2 \sigma_{\phi} r_i \right)^{-1-m} 
\label{derivative}
\end{equation}
which, however, is more difficult to estimate from the observational data. 

If most of the Faraday rotation is due to the mean 
distribution of $\phi=n_e H_\parallel$, $\sigma_\phi < \bar\phi$,
we obtain
\footnote{One can easily see
that $\lim_{L\to \infty} \int_0^L (1-z/L) \cos(A z) = const$} 
\begin{equation}
\left\langle P^2(\lambda^2) \right\rangle
\propto L \int_0^L  \!\!\! d \Delta z ( 1 - \Delta z/L)
\left(1 -\frac{\Delta z^m}{r_0^m} \right)
\cos \left( 2 \overline{\Phi} \lambda^2 \Delta z \right)
\sim L r_i (\lambda^2 \bar\phi r_i)^{-1-m} 
\frac{\sqrt{\pi}\Gamma\left[\frac{1+m}{2}\right]}{2 \Gamma\left[-\frac{m}{2}\right]}
\label{eq:regular}
\end{equation}
in the limit of large emitting area 
$L \gg r_i > {\cal L}_{\bar\phi}$,
so the polarization variance is expected to scale with wavelength reflecting
the transverse magnetic field component scaling.

Figure~\ref{fig:variances} also shows the behaviour of 
$\left\langle P^2(\lambda^2)\right\rangle$ at small wavelength.
In this case, Faraday rotation is small,
${\cal L}_{\sigma_\phi,\bar\phi} > r_i$. Nevertheless,
it still has an effect if $r_i\ll {\cal L}_{\sigma_\phi,\bar\phi} < L$,
resulting in
\begin{eqnarray}
\label{eq:var_short_mean}
\left\langle P^2(\lambda^2) \right\rangle && \propto \lambda^{-2+2m}~, 
\quad\quad\quad ~~ m<1~,
\quad \bar\phi > \sigma_{\phi} \\
\left\langle P^2(\lambda^2) \right\rangle && \propto
\lambda^{-2-2 a m_\phi +2m}~, \quad m<1 ~,
\quad \bar\phi < \sigma_{\phi} 
\label{eq:var_short_sigma}
\end{eqnarray} 
which represents the short wavelength scaling of our expressions.
When Faraday rotation is dominated by the $n_e H_{\parallel}$ fluctuations,
their correlations steepen the shortwave scaling which is reflected in
the negative $-a m_\phi$ contribution to the slope. We have not 
investigated this effect in detail with asymptotic analysis,
but numerical calculations show that $a \approx 1-m$ and from below
the slope is limited to $\lambda^{-4}$. For $m \ge 1$ both
asymptotics saturate at a constant value.

\subsection{Mean polarization}
\label{subsec:mean}

For the sake of completeness, below we provide the expression for the mean
polarization that arises in the presence of the mean magnetic
field $\overline{H_{\perp}}$ in the volume under study. The fluctuating
component of magnetic field is assumed to be present, but the
directional averaging of polarization nullifies its contribution.
Thus the expression for the mean polarization below is quite general: 
\begin{equation}
\left\langle P({\bf X}_1,\lambda^2)\right\rangle
= \int_0^L dz \left\langle  e^{2 i \lambda^2  \Phi({\bf X}_1,z) }
\right\rangle
= \int_0^L dz   e^{2 i \lambda^2  \bar\phi z }
e^{-4 \lambda^4  D_{\Delta\Phi}(0,z)}
\label{eq:Pmean}
\end{equation}
This expression assumes that the $x$ axis of
the coordinate system is aligned with $\overline{H_{\perp}}$ so that
there is a purely $Q$ (set to unity) uniform polarization at the source. 
Separately for observed $Q$ and $U$ components
\begin{eqnarray}
\label{eq:Qmean}
\left\langle P_Q({\bf X}_1,\lambda^2)\right\rangle
&=& \int_0^L dz   \cos\left(2 i \lambda^2  \bar\phi z \right)
e^{-4 \lambda^4  D_{\Delta\Phi}(0,z)}\\
\left\langle P_U({\bf X}_1,\lambda^2)\right\rangle
&=& \int_0^L dz   \sin\left(2 i \lambda^2  \bar\phi z \right)
e^{-4 \lambda^4  D_{\Delta\Phi}(0,z)}
\label{eq:Umean}
\end{eqnarray}

Eq~(\ref{eq:Qmean}) is practically identical
to Eq.~(\ref{eq:varP_noncorr}) for the 
variance if we replace in the latter the intrinsic correlation by a constant
Thus, the mean polarization scaling can be deduced from the results for 
the variance by setting $m=0$ in
Eqs~(\ref{eq:var_phi},\ref{eq:var_sigma},\ref{eq:var_short_mean},\ref{eq:var_short_sigma})
(due to difference between single and double line-of-sight integrals in
Eq~(\ref{eq:Qmean}) and Eq.~(\ref{eq:varP_noncorr}) there is an additional
factor $L$ in resulting expressions for the variance ).
\begin{eqnarray}
\left\langle P_Q(\lambda^2) \right\rangle && \propto \lambda^{-2}~,
\quad\quad\quad~ \bar\phi > \sigma_\phi \\
\left\langle P_Q(\lambda^2) \right\rangle && \propto \lambda^{-2}~,
\quad\quad\quad~ \bar\phi < \sigma_\phi ~,\quad\quad
\quad \lambda^2 \sigma_\phi r_\phi \gg 1 ~,\\
\left\langle P_Q(\lambda^2) \right\rangle && \propto \lambda^{-2-2 m_\phi}~,
\quad \bar\phi < \sigma_\phi ~,\quad\quad
\quad \lambda^2 \sigma_\phi r_\phi \ll 1 ~,
\end{eqnarray}

The appearance of average $U$ polarization
(in the frame oriented with magnetic field)
occurs only if the mean rotation is present. Moreover, 
$Q$ polarization at the source rotates to be almost purely $U$ polarization
at the observer if $\bar\phi \gg \sigma_{\phi}$,
while still scaling as $\lambda^{-2}$
\begin{equation}
\left\langle P_U(\lambda^2) \right\rangle \propto \lambda^{-2} \gg 
\left\langle P_Q(\lambda^2) \right\rangle ~,
\quad\quad\quad~ \bar\phi \gg \sigma_\phi ~.
\end{equation}
When the mean Faraday effect is subdominant, the relative magnitude of the
$U$ polarization is found numerically to change linearly
\begin{equation}
\frac{\left\langle P_U(\lambda^2) \right\rangle}  
{\left\langle P_Q(\lambda^2) \right\rangle} 
\approx A(\lambda) \frac{\bar\phi}{\sigma_\phi} ~\, \quad\quad
A(\lambda) \approx 1,~ \quad \lambda^2 \sigma_\phi r_\phi \gg 1 ~,
\end{equation}
but $A(\lambda)$ being scale dependent at small wavelengths
$ \lambda^2 \sigma_\phi r_\phi < 1 $.
The direction of the average
magnetic field projected on the sky can be determined
by independent techniques, e.g. via studies of the 
anisotropy of the intensity correlation. Then the ratio between Q and U
polarization in its frame may provide the information on $\bar\phi$ versus
$\sigma_\phi$.

\subsection{Effect of finite resolution}
Realistic observations have finite angular resolution while observing the synchrotron emission from a particular direction.
This leads to averaging of the correlation signal over the beam of neighbouring lines-of-sight. Assuming
isotropic sensitivity described by the beam $B$ of width $\Delta B$ in
the approximation of the parallel lines-of-sight
\begin{equation}
P(\mathbf{X}_1,\lambda^2) =
\int \! d \mathbf{X}  \; 
B\left(\left|\mathbf{X}_1 - \mathbf{X}\right|/\Delta B\right) 
P({\bf X},\lambda^2)
\end{equation}

To study turbulence at short scales where scaling laws are established,  the experiment 
should obviously be able to resolve the  $r_i$ scale,  i.e., 
$\Delta B < r_i$.  This, however,  is not sufficient as the study of
the variance of the polarization exemplifies.
The generalization of the approximation
Eq.~(\ref{eq:varP_simp}) to finite resolution gives 
\begin{eqnarray}
\left\langle P^2(\lambda^2) \right\rangle
&=&4  \int_0^\infty \!\!\!R d R B^2(R/\Delta B) 
\int_0^L  \!\!\! d\Delta z ( L - \Delta z)
\xi_i(R,\Delta z)
\cos \left( 2 \overline{\Phi} \lambda^2 \Delta z \right)
e^{-4 \lambda^4 \xi_{\phi}(R) \Delta z^2} ~,
\label{eq:varP_scale_finrez}
\end{eqnarray}
where
$ B^2(R/\Delta B) = 
\int \! d \mathbf{X}  
B\left(\left|\mathbf{X} - \frac{1}{2}\mathbf{R}\right|/\Delta B\right) 
B\left(\left|\mathbf{X} + \frac{1}{2}\mathbf{R}\right|/\Delta B\right) 
$.
Its analysis shows that the scaling regimes in
Eqs.~(\ref{eq:varP_scale},\ref{eq:regular})
are recovered when $\Delta B < {\cal L}_{\sigma_\phi,\bar\phi}$.
So to summarize, the window of opportunity to recover turbulence statistics with PPF is the range of wavelengths
\begin{equation}
 1/r_i <   \lambda^2 \mathrm{max}(\sigma_{\phi},\bar\phi)  < 1/\Delta B ~.
\end{equation}

We remark that finite resolution does not affect the scaling of
the mean polarization with the wavelength.

\subsection{Faraday rotation synthesis: example of usage for turbulence study}

Faraday rotation synthesis is currently becoming more popular with better frequency coverage
available within PPF cubes. Below we consider how to apply our approach using the data subject
to the Faraday rotation synthesis. This presentation here serves mostly to illustrate the applicability
of of the synthesis within our approach to study turbulence with synchrotron fluctuations. More detailed studies of the synthesis will be provided elsewhere. In the spirit of this section we consider the Faraday rotation synthesis asymptotics for the
same line of sight data while in Appendix D we provide a more general formulation of the problem.

In what follows we illustrate the use of Faraday rotation synthesis approach to studying
magnetic turbulence by obtaining the Faraday rotation synthesis expression corresponding to the variance that we studied in
the previous section. For this purpose we use the Eq. (\ref{FRS}) from Appendix D taking coinciding lines of sight, i.e. $R=0$.
Along a fixed line-of-sight $\Psi_1-\Psi_2$ has the mean value 
$\overline{\Phi}(z_1)-\overline{\Phi}(z_2)=\overline{\phi}\Delta z$
and the variance 
$2 D_{\Delta\Phi}(z_1-z_2)$.
Under approximation of independence of $P_i$ and $\Phi$ 
we obtain
\begin{equation}
\xi_F(0,\Psi_1-\Psi_2) 
\approx \frac{1}{2} \int_0^L \!\! d\Delta z(L-\Delta z)
\left\langle P_i(H_{\perp1}) P^*_i(H_{\perp2}) \right\rangle
\frac{1}{\sqrt{4 \pi D_{\Delta\Phi}(\Delta z)}} 
e^{-1/2 \frac{(\Psi_1-\Psi_2 - \overline{\phi}\Delta z)^2}{4 D_{\Delta\Phi}(\Delta z)}}
\label{eq:FPsi1-Psi2}
\end{equation}
which is an integral transform of the  
intrinsic polarization line-of-sight correlations.
Modeling the latter, for instance, as a saturated power law
$\propto \frac{\sigma_i^2 r_i^m}{(r_i^2+z^2)^{m/2}}$ shows that 
the dispersion function correlations recover the scaling
of the underlying turbulence whether the fluctuations in Faraday RM
dominate the mean or the mean is more important
\begin{eqnarray}
\sigma_{\phi} > \bar\phi : &\quad\quad&
\xi_F(0,\Psi_1-\Psi_2) \sim \frac{L \sigma_i^2}{\sigma_{\phi}}
\left(\frac{r_i \sigma_{\phi}}{|\Psi_1 - \Psi_2|} \right)^{m-a m_\phi}
\quad \quad 
\Delta\Psi > r_i \sigma_{\phi}
\\
\label{-1}
\bar\phi > \sigma_{\phi} : &\quad\quad&
\xi_F(0,\Psi_1-\Psi_2) \sim \frac{L \sigma_i^2}{\bar\phi}
\left(\frac{r_i \bar\phi}{|\Psi_1 - \Psi_2|} \right)^m
\quad \quad\quad\quad 
\Delta\Psi > r_i \bar\phi
\label{-2}
\end{eqnarray}
The results that we obtained with the Faraday rotation synthesis are 
reciprocal to the results
that we obtained with the variance for the regime of short wavelengths, see
Eqs.~(\ref{eq:var_short_mean},\ref{eq:var_short_sigma}).\footnote{In the case
of mean RM dominance, the asymptotic behaviour is rigorously
obtained by replacing in Eq.~(\ref{eq:FPsi1-Psi2}) 
the Gaussian with $\delta$-function as $\sigma_\phi \to 0$.
When fluctuative RM dominates, same $m$ scaling is obtained by approximating
$D_{\Delta\Phi}(\Delta z) \approx \sigma_\phi^2 \Delta z^2$. 
We have not derived rigorously, bu conjectured based on reciprocity with
Eq.~(\ref{eq:var_short_sigma}) the presence of the $-a m_\phi,~a \approx 1-m$
correction to the slope due to correlations in RM density.}
Other regimes (see Table~\ref{tab:mainresults}) should also be possible to
obtain through the Faraday rotation synthesis approach, but we do not provide
the corresponding expressions in this paper.

\section{Additional ways of polarimetric studies of magnetic turbulence}
\label{sec:additionalways}
The formalism that we have developed allows introduction of new measures
that can provide additional information about turbulence spectra and can open other
ways of studying magnetic field. Some of them are discussed below.

\subsection{Correlation of polarization derivative wrt $\lambda^2$}
\label{sec:dP}
In this subsection we introduce a measure that is more sensitive to Faraday
rotation.  Combining it with the measures in the earlier subsection appears
very synergetic.

Multi-wavelength Position-Position-Frequency synchrotron polarization datasets
contain more information that just single-frequency sky maps or 
line-of-sight multi-frequency analysis that we have discussed in the previous
sections. Full 3D Position-Position-Frequency analysis is beyond the scope
of this paper, but a step towards utilizing frequency information in 
polarization maps is to study the sky correlation of the derivatives of the
measured polarization \textit{wrt} the (square of) wavelength $\lambda^2$.
\begin{eqnarray}
\left\langle \frac{d P({\bf X}_1)}{d \lambda^2}
\frac{ d P^*({\bf X}_2)}{d \lambda^2} \right\rangle
&=& \int_0^L \!\! dz_1 \int_0^L \!\! dz_2  
e^{2 i \overline{\phi}\lambda^2 ( z_1 - z_2)} \times \nonumber \\
&\times&
\left\langle
P_i({\bf X}_1,z_1) P_i^*({\bf X}_2,z_2) 
\Delta \Phi({\bf X}_1,z_1) \Delta \Phi({\bf X}_2, z_2)
e^{- 2 \lambda^2 i 
\left( \Delta \Phi({\bf X}_1,z_1) - \Delta \Phi({\bf X}_2, z_2) \right)}
\right\rangle
\label{eq:corrPdlambda}
\end{eqnarray}
Again assuming negligible correlation between intrinsic polarization 
$P_i$ and the RM $\Phi$ we have
\begin{eqnarray}
\left\langle \frac{d P({\bf X}_1)}{d \lambda^2}
\frac{ d P^*({\bf X}_2)}{d \lambda^2} \right\rangle
&=& \int_0^L \!\! dz_1 \int_0^L \!\! dz_2  
e^{2 i \overline{\phi}\lambda^2 ( z_1 - z_2)}
\xi_i({\bf R},z_1-z_2) 
\left\langle
\Delta \Phi({\bf X}_1,z_1) \Delta \Phi({\bf X}_2, z_2)
e^{- 2 \lambda^2 i 
\left( \Delta \Phi({\bf X}_1,z_1) - \Delta \Phi({\bf X}_2, z_2) \right)}
\right\rangle
\nonumber\\
&=& \int_0^L \!\! dz_1 \int_0^L \!\! dz_2  
e^{2 i \overline{\phi}\lambda^2 ( z_1 - z_2)}
\xi_i({\bf R},z_1-z_2)  \times  \nonumber \\
&\times&\left(
\xi_{\Delta\Phi}(\mathbf{R},z_1,z_2) +
\lambda^4
\left( D_{\Delta\Phi}^2(\mathbf{R},z_1,z_2) - \frac{1}{4}
\left( \sigma^2_{\Delta\Phi}(z_1) - \sigma^2_{\Delta\Phi}(z_2) \right)^2
\right)
\right)
e^{- 4 \lambda^4 D_{\Delta\Phi}(\mathbf{R},z_1,z_2) }
\label{eq:corrPdlambda2}
\end{eqnarray}
Main complexity of this result is that $\Delta\Phi$ is inhomogeneous,
e.g. its variance depends on $z$ along the line-of-sight, but also its 
correlation functions depend on $z_1+z_2$ as well as $z_1-z_2$.
Several limiting cases are tractable,
but the most advantageous is to correlate the derivatives when Faraday rotation
is weak, $\lambda^4 D_{\Delta\Phi} < 1$, but still non-vanishing. In this case
\begin{eqnarray}
\left\langle \frac{d P({\bf X}_1)}{d \lambda^2}
\frac{ d P^*({\bf X}_2)}{d \lambda^2} \right\rangle
&\approx& \int_0^L \!\! dz_1 \int_0^L \!\! dz_2  
e^{2 i \overline{\phi}\lambda^2 ( z_1 - z_2)}
\xi_i({\bf R},z_1-z_2)
\xi_{\Delta\Phi}(\mathbf{R},z_1,z_2)
\end{eqnarray}
contains information about the RM as well as polarization
at the synchrotron sources. This is in contrast with correlation of the 
polarization itself, which in the limit of weak Faraday rotation given by
Eq.~(\ref{eq:corrP_weakF}) is almost insensitive to Faraday rotation.
In particular, correlating the derivatives in case of large mean component
in distribution of sources, will measure the correlation of RM with 
asymptotical behaviour for structure function
\begin{equation}
\left\langle \frac{d P({\bf X}_1)}{d \lambda^2}
\frac{ d P^*({\bf X}_2)}{d \lambda^2} \right\rangle \propto R^{1+\widetilde{m}_\phi}
 \label{add_1}
 \end{equation}
  that is
analogous to
Eqs.~(\ref{eq:corrP_weakF_scaling}).
In general, by combining polarization and its derivative data it 
is potentially possible to separate the information about sources of
synchrotron and RM.

In the absence of the mean magnetic field the correlation of derivatives 
is manifestly symmetric with respect to intrinsic and Faraday density
correlations 
\begin{eqnarray}
\left\langle \frac{d P({\bf X}_1)}{d \lambda^2}
\frac{ d P^*({\bf X}_2)}{d \lambda^2} \right\rangle
&\approx& \int_0^L \!\! dz_1 \int_0^L \!\! dz_2 \int_0^{z_1} \!\! dz^\prime
\int_0^{z_2} \!\! dz^{\prime\prime} \;
\xi_i({\bf R},z_1-z_2) 
\xi_{\phi}(\mathbf{R},z^\prime-z^{\prime\prime}) \nonumber \\
&=& \int_0^L \!\! dz^\prime \int_0^L \!\! dz^{\prime\prime}
\int_0^{z^\prime} \!\! dz_1 \int_0^{z^{\prime\prime}} \!\! dz_2 \;
\xi_i({\bf R},z_1-z_2) 
\xi_{\phi}(\mathbf{R},z^\prime-z^{\prime\prime})
\label{eq:xidP_weak}
\end{eqnarray}
It is important to note that in this measure the relative importance of
contribution 
from the intrinsic fluctuations of polarization at the source and from turbulent
Faraday rotation does not depend on how relatively strong the 
fluctuations are, i.e on $\sigma_i$ versus $\sigma_\phi$. It depends only
on the interplay of 
correlation lengths $r_i, r_\phi$ and scalings $m, m_\phi$ of the two terms.

\begin{figure}[ht]
\includegraphics[width=0.31\textwidth]{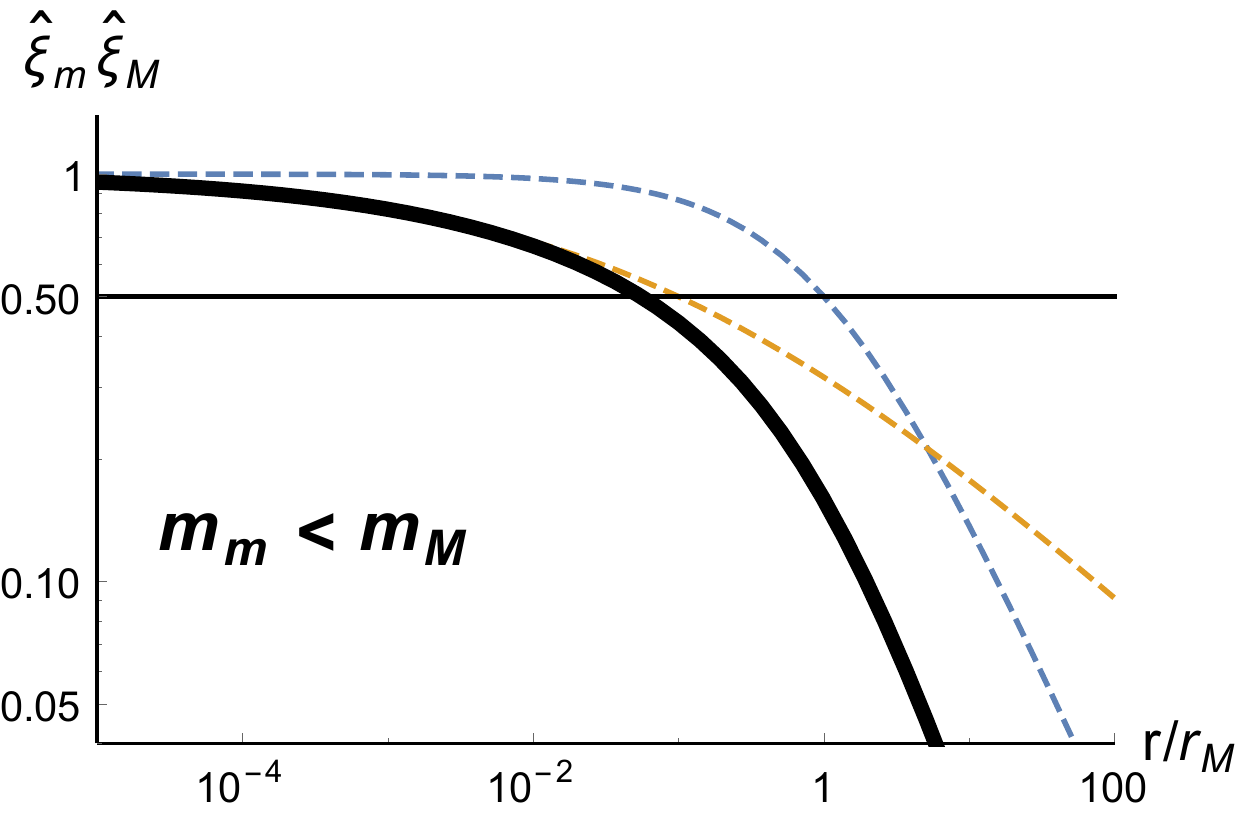}
\includegraphics[width=0.31\textwidth]{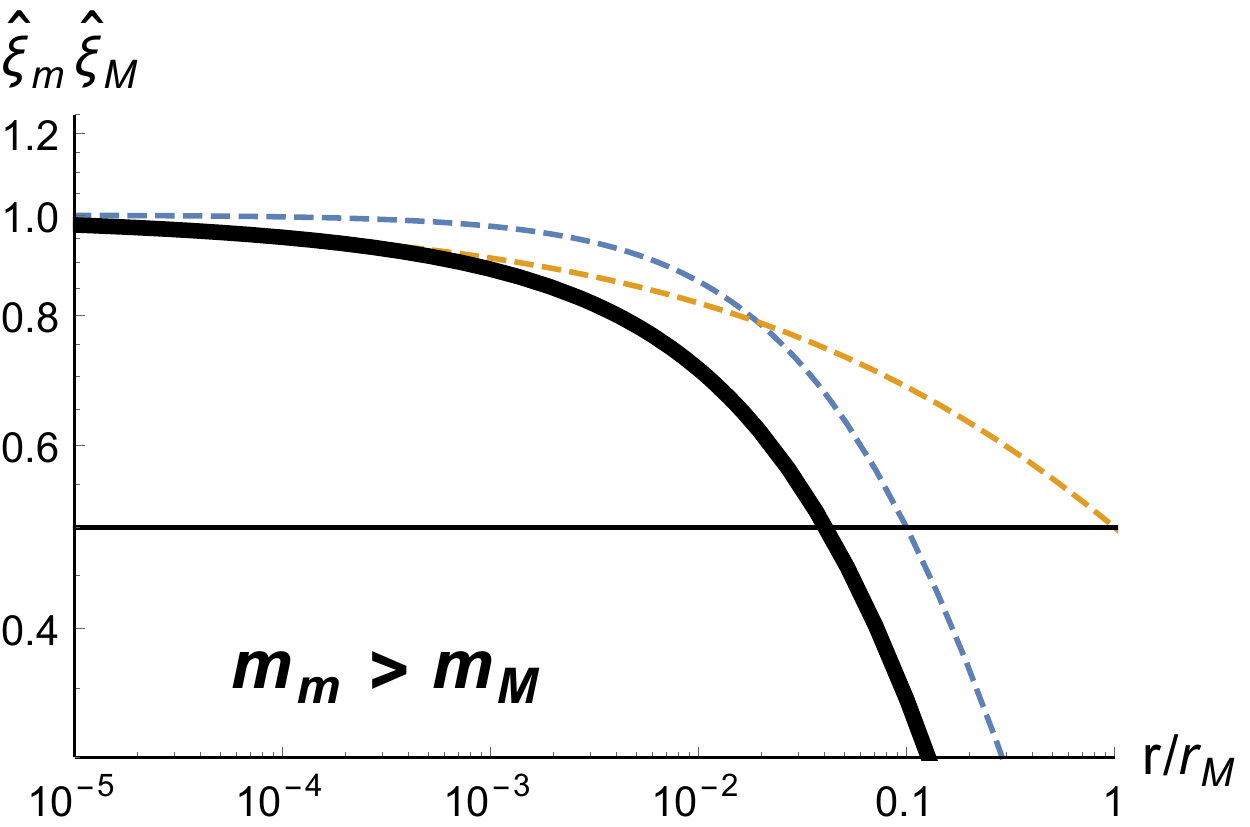}
\includegraphics[width=0.31\textwidth]{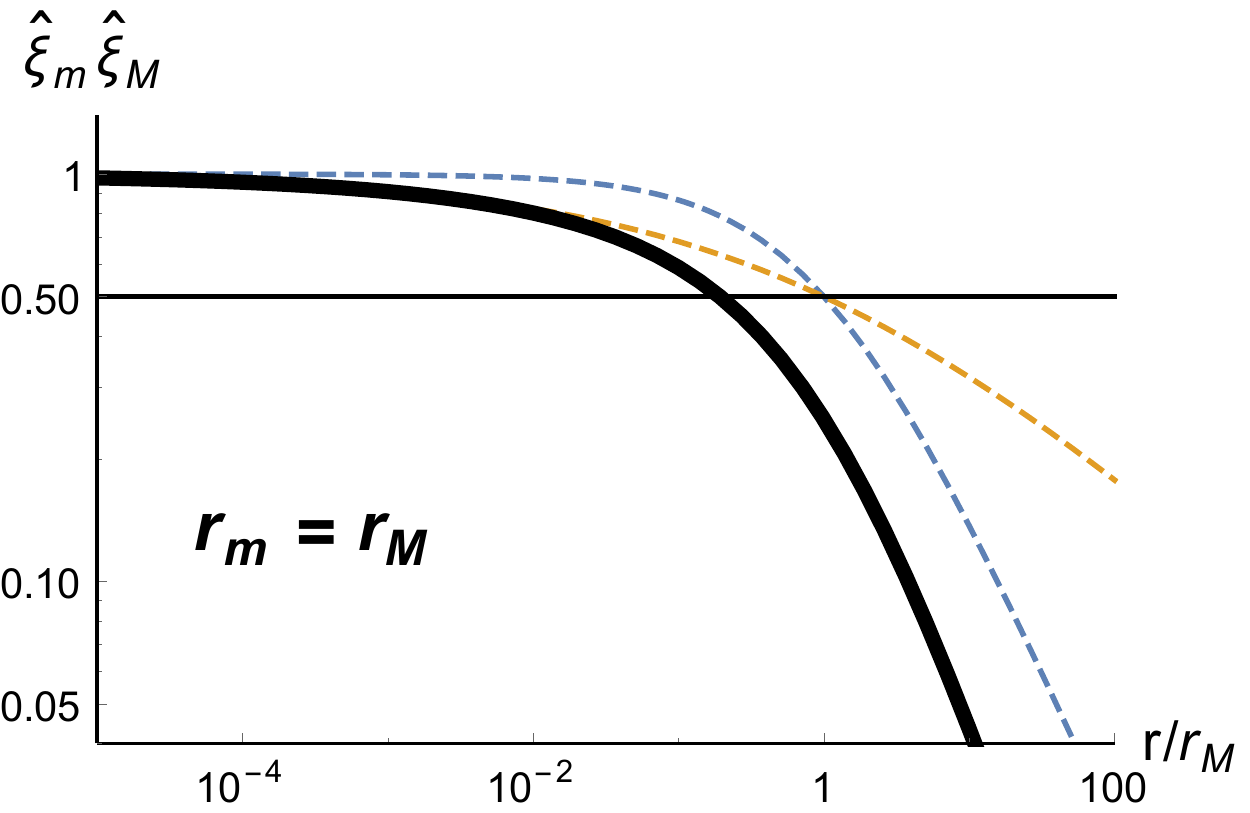}
\caption{Product of the two normalized
correlation functions (bold solid line) versus
individual terms (dashed) lines. Intersections of the dashed lines with the
horizontal at $0.5$ value mark correlation lengths $r_m$ and $r_M$.
Scales are given in the units of $r_M$, thus $r_M=1$.
Left: the case
$r_m \ll r_M$, $m_m < m_M$, here $r_m=0.1~r_M$, $m_m=1/3$ and $m_M=4/5$.
Center: the case
$r_m \ll r_M$, $m_m > m_M$, here $r_m=0.1~r_M$, $m_m=4/5$ and $m_M=1/3$.
This plot is given in a zoomed version.
Right: the degenerate case $r_m = r_M$,  with $m_m=1/3$ and $m_M=4/5$.
}
\label{fig:xii_xiphi}
\end{figure}

While the line-of-sight projections in Eq.~\ref{eq:xidP_weak} complicate 
the discussion, much of the
qualitative result can be understood by considering simply 
the product of the intrinsic and Faraday correlations. 
Let us denote
\begin{equation}
r_m \equiv \mathrm{min}(r_i,r_\phi), \quad r_M \equiv \mathrm{max}(r_i,r_\phi)
\label{eq:r_mM}
\end{equation}
and correspondent to $r_m$ and $r_M$ scaling indexes as $m_m$ and $m_M$
where we can have either $m_m \le m_M$ or $m_m > m_M$
(here and below indexes $m$ and $M$ mark the terms by their
respective correlation length).
The product
of normalized correlations $\widehat\xi_i = \xi_i/\sigma_i^2$ and 
$\widehat\xi_\phi = \xi_\phi/\sigma_\phi^2$ in three principal cases,
(a) $r_m \ll r_M$, $m_m \le m_M$, (b) $r_m \ll r_M$, $m_m > m_M$ and
(c) $ r_m \approx r_M$ is shown in Figure~\ref{fig:xii_xiphi}.
We note two points  important for the discussion:
First, the combined correlation at the shortest separations follows 
the term with the shallowest slope. This behaviour extends until $r_m$
with slope $m_m$ in cases (a) and (c) when the shallowest correlation 
corresponds also to the shortest or
similar correlation length, or until
\begin{equation}
R_* \approx r_M (r_m/r_M)^\frac{m_m}{m_m-m_M} < r_m 
\label{eq:R_*}
\end{equation}
with the slope $m_M$
in case (b) when the shallowest correlation has the longest correlation length.
In the latter case scaling at $r_m > R > R*$ is determined by steeper
$m_m$. Secondly, the presence of two effects decorrelates the signal from
sources separated by large distances and as the result
the overall amplitude of correlations integrated
over the line-of-sight will be diminished relative to the result obtained in
Eq.~\ref{eq:corrP_weakF_scaling}.

Let us now focus at small separations $R < r_m$.
The short scale  
behaviour is best revealed in the measurements of the structure function
\begin{eqnarray}
\label{eq:derivD_full}
\left\langle \left| \frac{d P({\bf X}_1)}{d \lambda^2}
- \frac{ d P({\bf X}_2)}{d \lambda^2}\right|^2 \right\rangle
&\approx& \sigma_i^2 \sigma_\phi^2 
\int_0^L \!\! dz_1 \int_0^L \!\! dz_2 \int_0^{z_1} \!\! dz^\prime
\int_0^{z_2} \!\! dz^{\prime\prime}  \times
\\
&\times & \left( \left[\widehat D_i(R,\Delta z)-\widehat D_i(0,\Delta z)\right]
+ \left[\widehat D_\phi(R,\Delta z^\prime)-\widehat D_\phi(0,\Delta z^\prime)\right]
\right. \nonumber \\
&-&
\left. \left[ \widehat D_i(R,\Delta z) \widehat D_\phi(R,\Delta z^\prime)
- \widehat D_i(0,\Delta z) \widehat D_\phi(0,\Delta z^\prime)\right]
\right)
\nonumber 
\end{eqnarray}
where $\widehat D \equiv D/(2 \sigma^2)$ are structure functions normalized
by the correspondent variances. Since $\widehat D < 1$, to linear
order intrinsic correlations and Faraday rotation contribute additively, thus
we will be comparing the full results with individual linear terms. Quadratic 
correction is important to understand the amplitude of the correlations.

In the first case of $ m_M > m_m$ we can approximate
$D_M(R<r_m,z) \approx D_M(0,z)$ to obtain the estimate
\begin{eqnarray}
\left\langle \left| \frac{d P({\bf X}_1)}{d \lambda^2}
- \frac{ d P({\bf X}_2)}{d \lambda^2}\right|^2 \right\rangle
&\approx& 
\sigma_\phi^2 \sigma_i^2
\int_0^L \!\! d\Delta z \widehat D_m(R,\Delta z) W_M(L,\Delta z) \\
W_M(L,\Delta z) &\equiv& 
\int_{\Delta z/2}^{L-\Delta z/2}\!\! dz_p \int_0^{z_p+\Delta z/2} \!\! dz^\prime
\int_0^{z_p-\Delta z/2} \!\! dz^{\prime\prime}  
\left(1 - \widehat D_M(0,\Delta z^\prime) \right) ~.
\end{eqnarray}
This result demonstrates that the scaling is determined by the 
contribution with
the shortest correlation length, while the effect of the other term, compressed
in $W_M$ window, leads to amplitude suppression relative to the uniform limit 
$r_M \to \infty$ when $W_M=L^3/3$. In detail
\footnote{We are giving only the
leading scaling with $(r/L)$, more accurate study shows that for $m_M < 1$
$W_M/L^3 \sim A (r_M/L)^{m_M} $, where coefficient $A$ depends
on $m_M$ varying from $\sim 2$ for $m_M=1/3$ to $\sim 11$ for $m_M=4/5$.
Next order term in $r/L$ is also not negligible for accurate calculations.},
for $r_M \ll L$ we find
$W_M \propto L^3 (r_M/L)^{m_M} (1-\Delta z/L)^2 $ which leads to
\begin{equation}
\left\langle \left| \frac{d P({\bf X}_1)}{d \lambda^2}
- \frac{ d P({\bf X}_2)}{d \lambda^2}\right|^2 \right\rangle
\propto \sigma_i^2 \sigma_\phi^2 L^3 (r_M/L)^{m_M} R^{1+m_m}/r_m^{m_m}
\quad\quad   R <  r_m, \quad m_m \le m_M~.
\label{add_2}
\end{equation}

\begin{figure}[ht]
\includegraphics[width=0.45\textwidth]{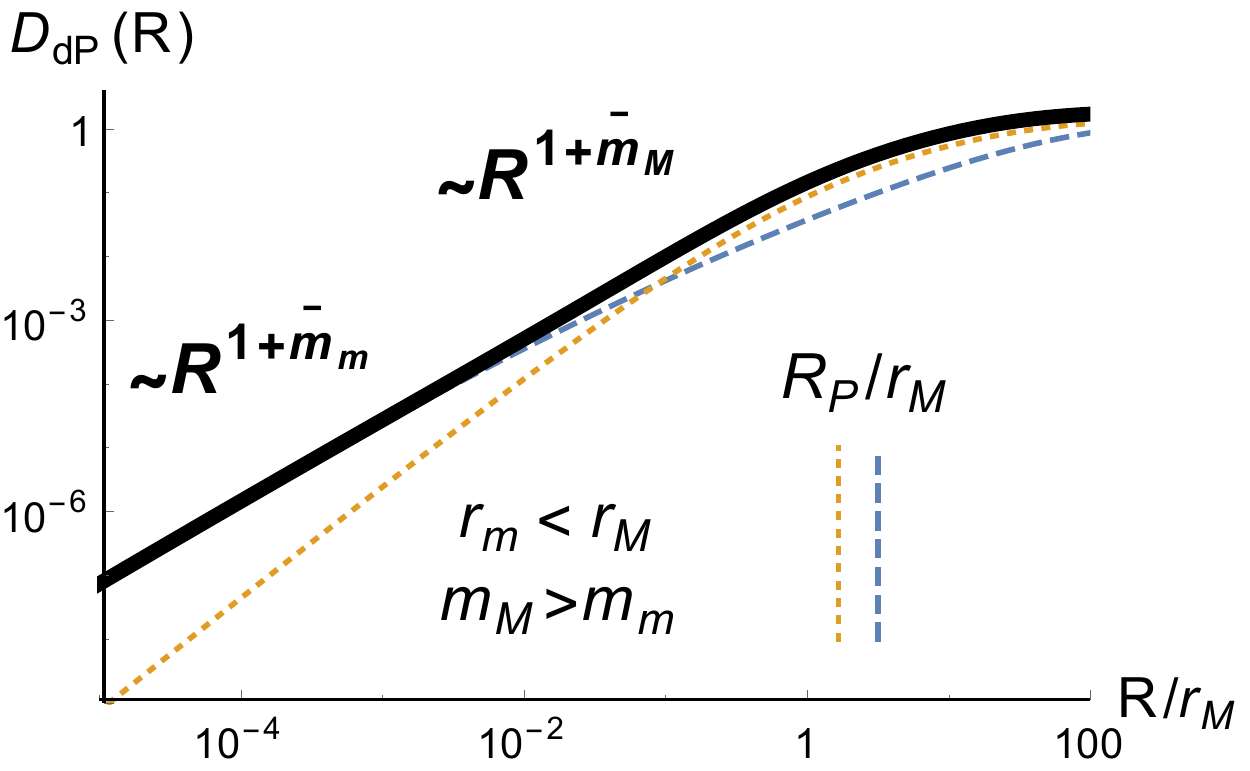}
\includegraphics[width=0.45\textwidth]{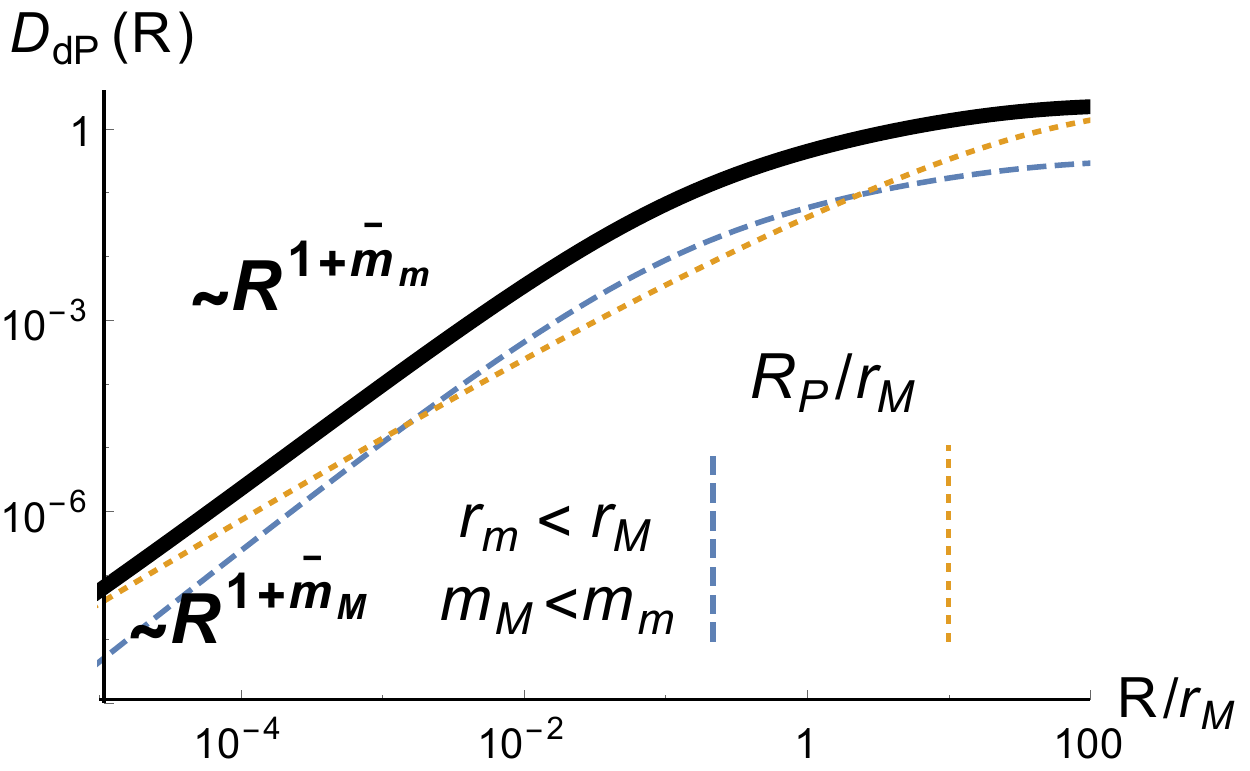}\\
\includegraphics[width=0.45\textwidth]{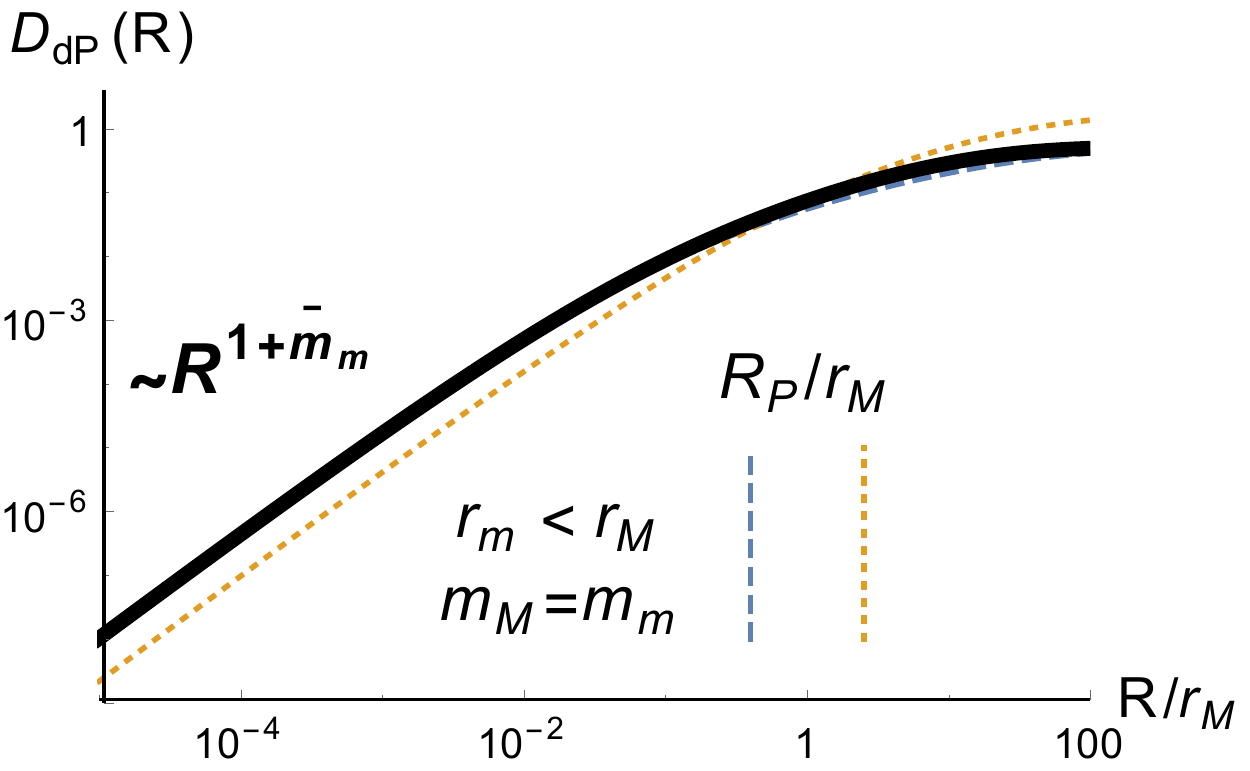}
\includegraphics[width=0.45\textwidth]{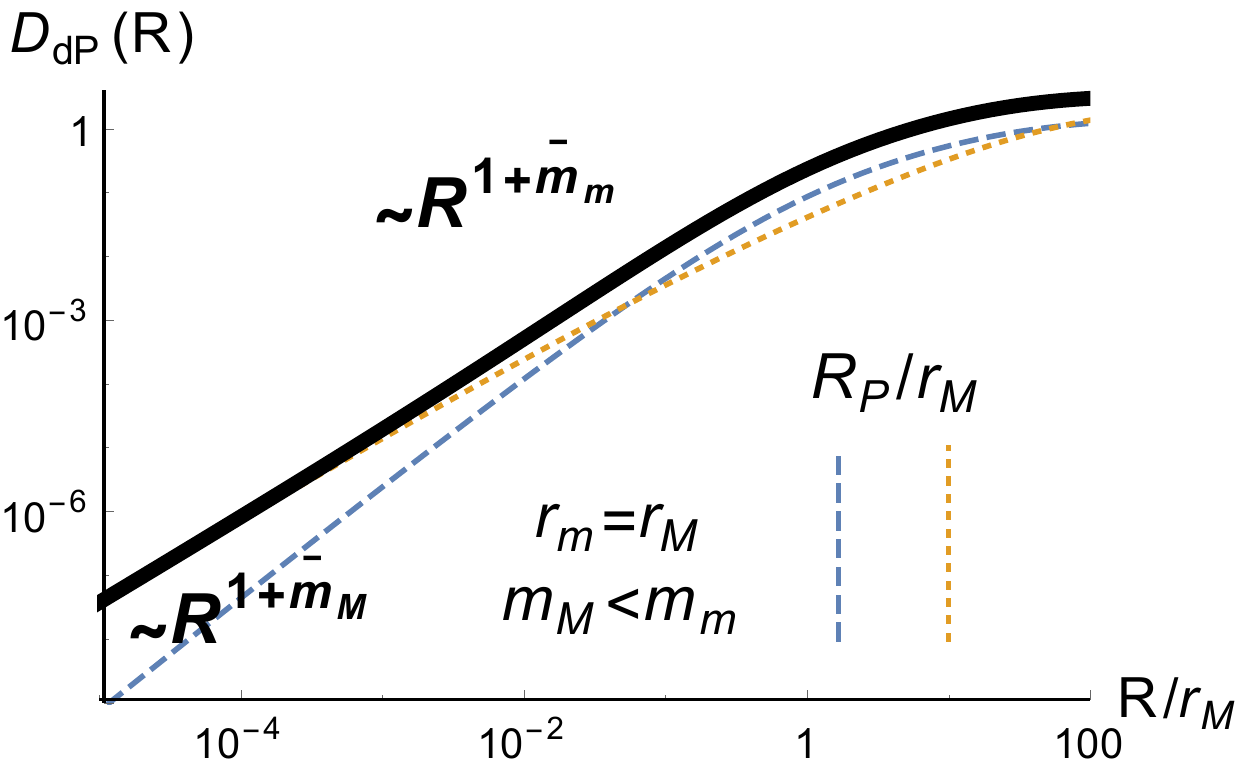}
\caption{Structure function of the derivative of the polarization \textit{wrt}
$\lambda^2$, $D_{dP} \equiv 
\left\langle \left| \frac{d P({\bf X}_1)}{d \lambda^2}
- \frac{ d P({\bf X}_2)}{d \lambda^2}\right|^2 \right\rangle
$, in units of $\sigma_i^2 \sigma_\phi^2 L^3$.
Example parameters are $r_i = r_M$, $r_\phi = r_m = 0.1\; r_i$, $L = 100\; r_i$.
Bold solid line is full result, in the left column scaled up by
$(L/r_M)^{m_M}$ and in the right column scaled up by $(L/r_m)^{m_m}$.
Individual contribution from intrinsic ($r_M$) and Faraday ($r_m$) correlations
are shown by dotted and dashed lines, respectively.
Vertical lines mark projected correlation lengths $R_P$ for the two terms.
Top left: $m_\phi = 1/3$, $m = 4/5$, corresponding to the case $m_M > m_m$.
For $R < r_m=r_\phi$ $D_{dP}$ scaling is determined by 
Faraday correlations, while for $R > r_m$ $D_{dP}$ follows intrinsic correlation
scaling. Projected correlation scales $R_P$ are inverted relative to 3D ones.
Top right: $m_\phi = 4/5$, $m = 1/3$, corresponding to the case $m_M < m_m$.
For $R < R_* \approx 10^{-3} r_i$ $D_{dP}$ scaling is determined by 
intrinsic correlations, while for $R > R_*$ $D_{dP}$ follows Faraday correlation
scaling at an enhanced amplitude.
Bottom row: degenerate cases.
Bottom left: $m_\phi = m = 2/3$ corresponding to the case $m_M = m_m$.
$D_{dP}$ is determined by the $r_m=r_\phi$ contribution everywhere.
The difference with the top-left case is that the projected scales $R_P$
here follow
the order of 3D correlation lengths, which is why $r_M$ contribution scaling
is not picked up at large separations.
Bottom right: $r_i=r_\phi$ case with $m_\phi=4/5$, $m=1/3$, which
can be viewed as a continuation of the top-right case.
}
\label{fig:DdP}
\end{figure}

In the case
$m_M < m_m$, however, the very short scales $R < R_*$ are dominated by the
term with the shallowest slope, which in this case is $m_M$. The transition
scale $R_*$, given above in Eq.~(\ref{eq:R_*}),
is estimated from the condition
$\widehat D_m(R_*)= \widehat D_M(R_*)$,
and is determined by the underlying correlations scales and not the size of
the emitting region.
Thus we have two regimes
\begin{eqnarray}
\label{add_3}
\left\langle \left| \frac{d P({\bf X}_1)}{d \lambda^2}
- \frac{ d P({\bf X}_2)}{d \lambda^2}\right|^2 \right\rangle
&\propto& \sigma_i^2 \sigma_\phi^2 L^3 (r_m/L)^{m_m} R^{1+m_M}/r_M^{m_M}
\quad\quad   R <  R_*, ~\quad\quad\quad m_m > m_M
\\
&\propto& \sigma_i^2 \sigma_\phi^2 L^3 (r_m/L)^{m_m} R^{1+m_m}/r_m^{m_m}
\quad\quad   
R_* < R < r_m, \quad m_m > m_M
\label{add_4}
\end{eqnarray}
where the first asymptotics is obtained by setting 
$D_m(R<R_*,z) \approx D_m(0,z)$ and the second one by matching at $R_*$.
Both scalings, for intrinsic polarization at the source
and Faraday rotation depth, can potentially be determined
in this case.
We note that while using $m$ and $M$ indexes highlights the formal
symmetry between intrinsic and Faraday depth correlations in our measure,
we expect the Faraday depth to have shorter correlation length,
thus $r_m = r_\phi$, 
$r_M = r_i$ and $m_m =m_\phi$, $m_M = m$ is a more probable identification.
In Figure~\ref{fig:DdP} we illustrate the discussed regimes.

The result at $R > r_m$ has a
complicated dependence on $r_M, r_m, m_M, m_m$ and $L$,
but can always be computed numerically. We notice however from
Eq.~(\ref{eq:derivD_full}) that as soon as $R$ is large enough that one of the 
structure functions in the integrand saturates, the
quadratic correction saturates the full result.
Figure~\ref{fig:DdP} confirms that the contribution
with the shortest \textit{projected} correlation length $R_P$ determines
the large separation behaviour and 
the correlation length of the polarization derivative signal
\footnote{Here we remind our discussion 
of Eq.~\ref{eq:corrP_weakF_scaling} in \S\ref{subsec:turbdom} that showed that
the correlation length of the projected signal in the weak Faraday limit
$R_P \approx r (L/r)^\frac{1-m}{1+m}$ exceeds the 3D one $r$, 
as Fig~\ref{fig:DdP} also readily illustrates.
Moreover, it is possible that the contribution with the shorter 3D $r$ scale
has the longer projected $R_P$ scale, as is the case in the upper-left panel
of Figure~\ref{fig:DdP}.}.

In conclusion, measuring the correlation of the derivative of the polarization
with respect to wavelength allows to study weak Faraday rotation effect.
Since Faraday depth correlation can be expected to have shorter correlation
lengths than intrinsic polarization, Faraday effect may be dominant in 
this measure, in contrast to correlating the polarization itself.

\subsection{Studying turbulence with interferometric data}
High resolution synchrotron maps are obtained with interferometers. To get a synthesis of interferometric data
obtained for an extended range of baselines is necessary. Obtaining the interferometric measurement to fill the
entire UV plane of spatial frequencies is time consuming and sometimes is impossible. At the same time, missing
spatial frequencies in the data used to restore the synchrotron polarization maps 
can interfere with the turbulence studies and distort the output of techniques. Fortunately, as we discuss further,
restoring of the synchrotron polarization image is not necessary for turbulence studies. Row interferometric data
is sufficient, as we discuss below. In fact, just a few measurements with different baselines are sufficient for 
finding the turbulence spectra.

In the paper thus far we have discussed the use of spatial correlation functions of polarization $P$ and its derivative $dP/d\lambda^2$. Another way to study fluctuations is to use power spectrum:
\begin{equation}
S_P({\bf K})=\int d{\bf R} e^{\bf i{\bf KR}} \langle P({\bf X}_1) P({\bf X}_2) \rangle
\end{equation}
where $\mathbf{R}= {\bf X_1} -{\bf X}_2$. The advantage of this presentation is
that $S_P$ can be available using raw interferometric data without requiring
the full range of spatial frequencies required for restoring the distribution
of polarized intensities. 

Similarly, we can differentiate the observed visibilities by $\lambda^2$ and obtain 
\begin{equation}
 S_{dP}({\bf K})=\int d{\bf R} e^{\bf i{\bf KR}} \left\langle \frac{d P({\bf X}_1)}{d \lambda^2}
\frac{ d P^*({\bf X}_2)}{d \lambda^2} \right\rangle 
\end{equation}

As the asymptotics of $ \langle P({\bf X}_1) P({\bf X}_2) \rangle$ and
 $\left\langle \frac{d P({\bf X}_1)}{d \lambda^2}
\frac{ d P^*({\bf X}_2)}{d \lambda^2} \right\rangle$ are known, obtaining the 
relations between the interferometric measurements on one hand,
and the underlying magnetic field and Faraday rotation statistics on the other
hand is trivial. Let us assume that $r_\phi \le r_i$.
Using Table~\ref{tab:mainresults} for the case of strong rotation
${\cal L}_{\sigma_\phi,\bar\phi} < r_i $.
we get for the interferometric
response for the polarization fluctuations for a fixed frequency, i.e.
within the PSA approach,
\begin{align}
\label{+1}
\quad {\cal L}_{\sigma_\phi} &<{\cal L}_{\bar\phi} &
S_{P}({\bf K}) & \sim K^{-1+\widetilde{m}_\phi}~, &
K & > r_{\phi}^{-1}~, \\
\label{+2}
&&
S_{P}({\bf K}) & \sim K^{-2+m}~, &
K &< r_{i}^{-1}~. \\
\label{+3}
\quad {\cal L}_{\sigma_\phi} & >{\cal L}_{\bar\phi} &
S_{P}({\bf K}) & \sim K^{-2-m}~, &
K &> r_{i}^{-1}~, \\
&&
S_{P}({\bf K}) &\sim K^{-2+m}~, &
K &< r_{i}^{-1} ~.
\label{+4}
\end{align} 

Similarly using Eqs. (\ref{add_2}), (\ref{add_3}) and (\ref{add_4}) for $S_{dP}$ in the regime of weak Faraday rotation
we find from \S\ref{sec:dP} 
\begin{align}
\label{+5}
m &\ge m_\phi &
S_{dP}({\bf K}) &\sim K^{-3-\widetilde{m}_\phi}~, & K & > r_\phi^{-1}~. \\
\label{+6}
m & < m_\phi &
S_{dP}({\bf K}) &\sim K^{-3-\widetilde{m}_\phi}~, & R_*^{-1} > K &> r_\phi^{-1}~,\\
\label{+7}
&& S_{dP}({\bf K}) &\sim K^{-3-\widetilde{m}}~, & K & > R_*^{-1}~. 
\end{align} 
for the case of dominance of fluctuating component
and a rather trivial result that follows from Eq.~(\ref{add_1})
for the dominance of regular magnetic field
\begin{equation}
S_{dP}({\bf K}) \sim K^{-3-\widetilde{m}_\phi}~. 
\label{+8} 
\end{equation}

We see that $S_P$ in most cases reflects the statistics of underlying magnetic
turbulence that is responsible for the emission, while $S_{dP}$ is more focused
on the statistics of Faraday rotation fluctuations. This corresponds to the measures of 
polarization that $S_P$ and $S_{dP}$ present spatial Fourier transforms. 

We note that the measures $S_P (\mathbf{K})$ and $S_{dP} (\mathbf{K})$ depend
also on the direction of $\mathbf{K}$
in respect to the in-plane direction of magnetic field ${\bf H}_{\bot}$. This opens ways for studying ${\bf H}_{\bot}$
direction using sparsely sampled interferometric data. In addition, as we discuss below, with such a data it is possible
to study more sophisticated properties of turbulence related to its anisotropy. 

\subsection{Anisotropy of fluctuations}

Our analysis in this paper was focused on the spectrum of the synchrotron
fluctuations. The actual spectra of MHD turbulence are anisotropic.
This was shown in LP12 to result in anisotropies of synchrotron fluctuations.
If we compare the expression for intensity in Eq.(\ref{i}) and polarized
intensity Eq.(\ref{p}) we observed that the fluctuations of the perpendicular
component of magnetic field enter the same way in both expressions, i.e.
they are both proportional to $\int dz |{\bf H}_{\bot}|^{\gamma}$.
The difference arises from the fact that the expression for
the polarization has the Faraday phase factor. In terms of correlation
properties, this
factor, as our analysis above shows, in most cases results in introduction of
the window which limits the extent of the region from which the signal
is collected. Thus the analysis of anisotropy of synchrotron fluctuations in
LP12 is applicable to the correlation functions of the polarized radiation
with the caveat that the magnetic field studied this way is not the magnetic
field integrated over the line-of-sight through the entire emitting region, but limited by the zone extent of which determined 
by the Faraday window that we discussed in \S\ref{sec:samelambda}.

This provides a few new effects  which are very valuable for magnetic field
studies. The analysis of anisotropies in LP12 suggested that the mean magnetic
field direction in the entire synchrotron -emitting region can be obtained from
the analysis of the quadrupole anisotropy of the synchrotron fluctuations.
In view of our earlier discussion this means that  determination of the
direction of magnetic field is possible within the sub volume of the
synchrotron emitting region if fluctuations of polarization are studied.  By changing the wavelength of the radiation one can sample the region to a different extent. This allows studying the variations of the direction of the magnetic field along the line-of-sight, which is new valuable information. In addition, LP12 found that the analysis of anisotropies can determine the relative strength of Alfv\'en, fast and slow modes within MHD turbulence. It is possible to see that the same analysis can be extended to obtain the variations of the corresponding strengths along the line-of-sight if polarized synchrotron emission is used for the cases when the Faraday
rotation does not impose the fluctuations of its own.
The latter corresponds to the limited parameter space (see left upper part of
Table~\ref{tab:mainresults}).
In that special situation the analysis in LP12 should be extended. 

A particular case of anisotropy analysis considered in LP12 is the case of studying anisotropies
within the Milky Way when the observer is submerged in the emitting synchrotron volume. It was
shown there that if the angle $\theta$ between the lines-of-sight along which the synchrotron intensity is collected is sufficiently large, i.e. $\theta$ is much larger than the ratio of the correlation radius of magnetic fluctuations $r_i$ to the extent of the emitting volume $L$, then the anisotropy analysis samples correlations of magnetic field only to the extent $r_i \theta$, which allows determining magnetic field direction variations within the emitting volume. It is easy to see that the polarization substantially extends this ability of sampling magnetic field and allows studying variations of magnetic field at scales less than $r_i$. 

The direction of magnetic field that obtained by the correlation anisotropy analysis in LP12 will not coincide, in general, with the direction given by the mean polarization. This arises from the fact that the magnetic eddies tend to get elongated along the mean field. The latter direction is fixed in space and not affected, unlike the synchrotron polarization direction, by Faraday rotation. A comparison of the direction provided by the statistical technique above and the direction of mean polarization can provide an estimate of the mean angle of Faraday rotation for the polarized radiation. 

The suggested in LP12 technique of multipole decomposition of the synchrotron radiation potentially opens ways to disentangling the contributions from compressible and incompressible motions. The spectral analysis of the fluctuations of the multipoles obtained as a result of such a decomposition opens an avenue for studies of spectra of magnetic fluctuations corresponding to compressible and incompressible motions separately. In LP12 we assumed that for our study we use single dish data or, 
equivalently, the interferometric data with the full coverage of spatial frequencies. Within this paper we explored a possibility of studying synchrotron with sparsely sampled interferometric data. We feel that such a data can be used within the multipole decomposition technique. We plan to elaborate the corresponding procedures elsewhere.

\section{Major results and prospects of studying turbulence spectra}
\label{sec:results}

\subsection{Starting point}
This paper is the first attempt, as far as we know, to provide the detailed statistical properties of Position-Position-Frequency (PPF) data cubes of polarization data in order to study properties of underlying magnetic turbulence. Synchrotron emission
is pretty complex and this probably kept theorists off the subject. For instance, its properties depend on the index of cosmic
rays, as a result the intensities of both polarized and unpolarized emission depend on $H_{\bot}^{\gamma}$, where $H_\bot$
is the amplitude of magnetic field component perpendicular to the line of sight and
$\gamma$ is related to the spectral index of cosmic rays as it is discussed in Appendix A. 

This paper continues our study of the statistics of synchrotron radiation that we started in LP12. There we obtained
the two point statistics of
synchrotron intensities for arbitrary $\gamma$. In this paper we utilized this advance to describe the statistics
of polarization in the presence of Faraday rotation. As a result, our results are applicable to a variety of situations
characterized by different $\gamma$.

In LP12 we showed that the spectrum of magnetic turbulence within entire synchrotron emitting volume can be determined from synchrotron fluctuations for an arbitrary $\gamma$. The use of polarized fluctuations gives us a possibility to vary the extent of the volume under study.
Indeed, we have seen in the previous sections that the Faraday rotation effect limits the extent of the region where magnetic fluctuations are being sampled.

An important finding of the present study is that we found the situations when the correlations of the polarization
have imaginary parts. In other words, we found that the real part of the polarization correlations, i.e.
$\left\langle Q({\bf X}_1)Q({\bf X}_2) +  U({\bf X}_1)U({\bf X}_2) \right\rangle$ provides the information about the
symmetric turbulence of magnetic field and electron density fluctuations, while the imaginary part, i.e.
$\left\langle Q({\bf X}_1)U({\bf X}_2) -  U({\bf X}_1)Q({\bf X}_2) \right\rangle$ is may not be zero and can deliver very valuable information about properties of magnetic field and turbulence. For the most part of the paper we
concentrate on the real part of polarization correlations, while we briefly discuss that the non-zero imaginary part can inform us about the 3D direction of the magnetic field.  

We note that for the practical analysis of the data one can use the expression $\left\langle Q({\bf X}_1)Q({\bf X}_2) +  U({\bf X}_1)U({\bf X}_2) \right\rangle$ to make sure that the antisymmetric part does not contribute to the signal. In fact, this is the expression that one should identify with the analysis of the correlation of fluctuations of polarization that we discussed in the paper. Alternatively, as we discussed earlier, one can use structure functions of the polarization rather than the correlation functions. Finding that the combination
 $\left\langle Q({\bf X}_1)U({\bf X}_2) -  U({\bf X}_1)Q({\bf X}_2) \right\rangle$ differs from zero in observations will be an important discovery. Depending on the dominant origin of the antisymmetry one can get insight either into the 3D direction of magnetic field in the emitting volume. 
 
 \subsection{Polarization Spatial analysis (PSA) and Polarization Variance Analysis (PVA)}
 
 Naturally, there are different ways of studying fluctuations of polarization. In the present paper we have provided a detailed 
 description of the study of the spatial correlations or polarization measures in spatially separated points 
 $\left\langle P(X_1) P(X_2) \right\rangle$ for the same wavelength and the frequency variations of the dispersion of the 
 polarization $\left \langle P(\lambda) \right \rangle$. The technique utilizing the former measure we termed Polarization Spatial 
 Analysis (PSA) and utilizing the latter measure Polarization Variance Analysis (PVA). 
Table~\ref{tab:mainresults} illustrates different regimes of turbulence study
that are available with the above techniques. This table also reflects the richness of the PPF data in terms of various scales
that are involved and which can be recovered by the analysis of the polarization data (see also Table 1 where these scales
are listed).

The variety of the scales involved arises from the fact that the resulting polarization is the outcome of both polarization of
synchrotron emission collected along the line of sight and the Faraday rotation of the polarization plane. In the main text of
the paper we assumed that the emission and Faraday rotation happen within the same volume. This is the most complex
from the mathematical point of view situation. We consider another limiting case of separate regions of emission and Faraday rotation in Appendix C. Naturally, the intermediate cases can be presented as a superposition of these two limiting cases.

With astrophysical environments presenting us with a rich variety of different circumstances, we consider both the situation
when most of Faraday rotation arises due to the turbulent field and to the mean field. The former case corresponds to 
turbulent decorrelation scale ${\cal L}_{\sigma \phi}$ given by Eq. (\ref{eq:sigmaphi}) being smaller that the scale of Faraday
decorrelation induced by mean magnetic field, i.e. ${\cal L}_{\bar{\phi}}$ given by Eq. (\ref{eq:barphi}). The analysis of the data
allows distinguishing these two cases.

It is essential to keep in mind that ${\cal L}_{\sigma \phi}$  and 
${\cal L}_{\bar{\phi}}$ are all proportional to $\lambda^2$ and change
with the wavelength\footnote{Naturally, this is also applicable to ${\cal L}_{\sigma\phi, \bar{\phi}}$, which is the minimal of the two lengths, see Eq. (\ref{min}).}. This provides one with an opportunity to sample different volumes along the lines and use synchrotron polarization fluctuations to get unique insight into the distribution of magnetic fields and magnetic turbulence
and different distances from the observer. This is the unique feature of using polarized synchrotron emission. 

We deal at length with PSA and PVA techniques, as we view them as the simplest techniques involving synchrotron polarization
data. At the same time, our approach is not limited to these two techniques. In fact, we also applied our approach to find a few
more useful statistical measure that we discuss in the paper.

If we talk about the PSA we have several regimes. It is evident from
Table~\ref{tab:mainresults} that for sufficiently long wavelengths the Faraday
rotation effect is essential (upper left corner of Table~\ref{tab:mainresults},
corresponding to ${\cal L}_{\sigma_\phi} < {\cal L}_{\bar\phi}$ and the
separation of the lines-of-sight less than the correlation radius of $n_e H_{\|}$ fluctuations, i.e.
$R< r_{\phi}$. In this regime the statistics of synchrotron polarization fluctuations are both affected through the presence of $\xi ({\bf R}, 0)$ by the fluctuations of underlying
perpendicular component of magnetic field $H_{\bot}$ and the fluctuations of $n_e H_{\|}$, which
for the given example are assumed to have the spectral power law index $m_{\phi}$. 

By observing the change of the scaling of the polarization fluctuations from one regime given
by Eq. (\ref{eq:xi_sync_smallR_m<1}) to the other regime given by Eq. (\ref{eq:xi_sync_largeR}) which happens as the line separation $R$
becomes comparable with the correlation scale $r_{\phi}$ we can get the characteristic size of the
Faraday rotation fluctuations. 

In other regimes shown in Table~\ref{tab:mainresults} one can obtain
$\xi({\bf R}, 0)$, i.e. to obtain the statistics
of magnetic field, which is not distorted by the Faraday rotation fluctuations.
Combining that with the result obtained in the regime in the upper left corner
of Table~\ref{tab:mainresults}, one can obtain also the spectral properties of
the Faraday rotation fluctuations, e.g. the spectral index $m_{\phi}$.

Depending on the media the fluctuations of $n_e H_{\|}$ can be dominated either by electron  density fluctuations or the parallel to line-of-sight component of magnetic field fluctuations. In many situations the fluctuations of magnetic field and electron density are not correlated and therefore the correlation is factorized, i.e. can be presented as a 
product of the electron density and magnetic field correlations.
If magnetic field statistics is known through the measurements of
$\xi ({\bf R}, 0)$ as we discussed above, then in the case of no
density-magnetic field correlations one can get the correlation function
of electron density. 

Another feature that is obvious from the analysis of Table~\ref{tab:mainresults}
is that, apart from $r_{\phi}$, measuring of which we discussed above, we get
an opportunity to obtain the characteristic correlation scales of the magnetic
field $r_i$. If within PSA we observe at a given wavelength two regimes
depending on the line-of-sight separation, then we can increase the wavelength
and suppress the regime influenced by the Faraday rotation fluctuations
(see upper left corner of Table~\ref{tab:mainresults}). The transition will
happen at ${\cal L}_{\sigma_\phi,\bar\phi} \sim r_i $, which presents an
interesting way to determine the correlation length of the magnetic field
fluctuations.

In short, when magnetic field fluctuations dominate the Faraday rotation measure and in the case of no assumption of magnetic field and electron density correlation the PSA
provides the correlation function of magnetic field and a correlation function of electron density product with magnetic field. If the magnetic field is not correlated with electron density one gets separately the correlation functions of magnetic field and electron density. The observations of
the change of the polarization fluctuation slopes provide also the scales $r_{\phi}$ and $r_i$. In addition, by exploring the very fact that there is a change of the scalings of the polarization fluctuations (compare upper and lower lines of the left corner of Table~\ref{tab:mainresults}) allows to establish whether the magnetic fluctuations or regular magnetic field dominate the Faraday effect. 
If the Faraday rotation is dominated by the regular magnetic rotation, then we have only one 
regime of study and all the fluctuations naturally arise from the fluctuations of the perpendicular component of magnetic field. 

For the PFA, which is complementary to the PSA technique,
Table~\ref{tab:mainresults} illustrates that the relative effect of the ratio
of the mean Faraday measure to its dispersion results in rather conspicuous
consequences. Indeed, the universal
scaling of the variance as $\lambda^{-2}$ for long $\lambda$ testifies that the Faraday dispersion measure dominates the regular Faraday rotation. To find the underlying magnetic field spectral slope one has to differentiate the variance as it suggested by Eq. (\ref{derivative}). Naturally, this procedure
increases effects of the noise present in the data and makes recovering of the underlying turbulence statistics more involved. 

At the same time, Table~\ref{tab:mainresults} shows that within the PFA
technique one can change the wavelength to observe the change of the spectral
behavior either from
$\lambda^{-2}$ to $\lambda^{-2+2m}$ or from $\lambda^{-2-2m}$ to $\lambda^{-2+2m}$ and this
transfer will happen at  the scale ${\cal L}_{\sigma_\phi,\bar\phi} \sim r_i $, i.e. at the correlation scale of magnetic fluctuations. It is evident that the simultaneous use of PSA and PFA increases the reliability of the results on the turbulence statistics that these techniques deliver. 

\begin{table}[h]
\begin{tabular}{r||lcc|lc||lc|lc} 
& \multicolumn{5}{c||}{PSA, $\xi_P(\mathbf{R}) $}
& \multicolumn{4}{c}{PFA, 
$\left\langle P^2(\lambda^2) \right\rangle$ } \\[5pt]
\cline{2-10}
& \multicolumn{3}{c|}{${\cal L}_{\sigma_\phi,\bar\phi} < r_i $}\Tstrut\Bstrut
& \multicolumn{2}{c||}{${\cal L}_{\sigma_\phi,\bar\phi} > r_i $}
& \multicolumn{2}{c|}{${\cal L}_{\sigma_\phi,\bar\phi} < r_i $}
& \multicolumn{2}{c}{${\cal L}_{\sigma_\phi,\bar\phi} > r_i $} \\[3pt]
\hline
${\cal L}_{\sigma_\phi} < {\cal L}_{\bar\phi}$ \Tstrut\Bstrut
& 
$\sim
{\cal L}_{\sigma_\phi}^2 \xi_i({\bf R},0)
\frac{ {\cal L}_{\sigma_\phi} r_\phi^{\widetilde{m}_\phi}}{R^{1+\widetilde{m}_\phi}} $
& $ R < r_\phi $
& Eq.~(\ref{eq:xi_sync_smallR_m<1})
& reflects spectrum $H_{\bot}$
&
& $\propto \lambda^{-2}$
& Eq.~(\ref{eq:var_sigma})
& $\propto \lambda^{-2-2 a m_\phi+2 m}$
& Eq.~(\ref{eq:var_short_sigma})\\ 
& 
$\sim
{\cal L}_{\sigma_\phi}^2 \; \xi_i({\bf R},0)$ 
& $ R > r_\phi$
& Eq.~(\ref{eq:xi_sync_largeR})
&
&
&
&
& 
& \\[5pt] 
${\cal L}_{\sigma_\phi} > {\cal L}_{\bar\phi}$
& $ \sim L \; {\cal L}_{\bar\phi} \; \xi_i({\bf R},0)$
&
& Eq.~(\ref{eq:xi_sync_meanF})
& reflects spectrum $H_{\bot}$
&
& $\propto \lambda^{-2-2m}$ 
& Eq.~(\ref{eq:var_phi})
& $\propto \lambda^{-2+2m}$
& Eq.~(\ref{eq:var_short_mean}) \\ 
\end{tabular}
\caption{Different regimes of synchrotron polarization correlations in
Position-Position-Frequency studies. For additional statistics see 
\S\ref{sec:additionalways}.}
\label{tab:mainresults}
\end{table}

\subsection{Other new ways to study turbulence}

If one is interested in the Faraday rotation fluctuations, the limitation of PSA technique is that in the case of weak Faraday rotation, the statistics $\left\langle P(X_1) P(X_2) \right\rangle$ is only marginally sensitive to the Faraday
rotation. To deal with this limitation 
we proposed a new measure based on the wavelength derivative of polarization
$dP/d\lambda^2$, namely, the spatial correlations
\footnote{One can also consider the quantity $\left \langle [dP/d\lambda^2]^2 \right \rangle$ as a function of wavelength, which would correspond to the analog
of PVA, but focused on Faraday rotation fluctuation statistics.
The corresponding study is, however, beyond the scope of this paper.} 
of $ dP(X_1)/\lambda^2 $.
This measure was shown to be sensitive to the fluctuations of Faraday rotation.
In the case of weak Faraday rotation the asymptotic solutions are given by
Eqs.~(\ref{add_1}), (\ref{add_2}) (\ref{add_3}) and (\ref{add_4}),
the cases for which are explained in \S~6.1. 

Combining the study of the statistics of $dP/\lambda$ with the statistics of  $P$ in the same weak Faraday rotation regime it is possible to get separately both the statistics of underlying magnetic field and Faraday rotation. In may be worth mentioning that the latter statistics {\it in the regime of strong Faraday rotation} is important for studying the distribution of magnetic turbulence at different distances from the observer. 

For the sake of simplicity, dealing with PVA and PSA techniques as well as with the statistics of
$dP/d\lambda^2$ we did not account for the actual
structure of turbulence, e.g. its anisotropy. From the point of view of power law asymptotics that we
were focused on, the anisotropies of the turbulence (see e.g. \S 2.3) shining via synchrotron do not matter if these
anisotropies do not change the spectral index. The arguments that we provided in LP12 suggest that
the spectral index does not change in the global system of reference which is available through observations.
This can be easily tested observationally by measuring the correlations of
polarization for different fixed positional angles in the sky. If changes, in disagreement with the theory, are observed, the
averaging should be performed for a fixed positional angle and all the machinery of PSA would stay
intact otherwise. 

The limitations for the turbulence studies related to the resolution of telescopes can be substantially mitigated with the use of interferometers. An important claim in this paper is that to get a spectrum of magnetic turbulence the full spatial frequency coverage is not necessary. In fact, measuring for a few spatial frequencies may be sufficient. The corresponding measures
$S_P$ and $S_dP$ present the spatial Fourier transforms of fluctuations of $P$ and $dP/d\lambda^2$, respectively. 
The asymptotics of $S_P$ is given for different regimes by Eqs. (\ref{+1}), (\ref{+2}), (\ref{+3}), (\ref{+4}) and for
$S_dP$ by Eqs. (\ref{+5}), (\ref{+6}), (\ref{+7}) and (\ref{+8}). 

Within our treatment the study of turbulence using the Faraday rotation synthesis (see \S 5.4) has a relatively modest role. We did not see any particular advantage of using the synthesis prior to turbulence study. The obtained asymptotics are given by Eqs. (\ref{-1}) and (\ref{-2}). 

Anisotropies, as we discussed earlier, are not the main focus of this paper.
However, they are very informative and interesting by itself. Our study in LV99 was
focused on their study for the case of unpolarized intensity. Polarized radiation opens new windows
for studying the anisotropy.  In other words, our research opens ways to study anisotropy in the correlation
properties of $H_{\bot}$ statistics. Due to the mathematical similarity of
such a problem with the one of anisotropy of synchrotron intensity fluctuations
that we studied in LP12, we can directly apply the results in the
aforementioned work to polarization fluctuations in the regime when the
fluctuations of of $H_{\bot}$ dominate the signal
(see Table~\ref{tab:mainresults}), which for
the Faraday length being less than the correlation scale of $H_{\bot}$
fluctuations, i.e. ${\cal L}_{\sigma_\phi,\bar\phi} < r_i $, satisfies the
following conditions: either ${\cal L}_{\sigma \phi} <{\cal L}_{\bar \phi}$
and $R> r_{\phi}$ or ${\cal L}_{\sigma_\phi}> {\cal L}_{\bar \phi}$\footnote{In
the case, of ${\cal L}_{\sigma \phi} <{\cal L}_{\bar \phi}$ and $R< r_{\phi}$
our anisotropy analysis should be modified to account for the possible
anisotropy induced by Faraday rotation term. Nevertheless, the anisotropy
should still define the direction of the mean field even in this case.}. In
the opposite regime, i.e.  ${\cal L}_{\sigma_\phi,\bar\phi} < r_i $,
Table~\ref{tab:mainresults} testifies that the LP12 analysis is directly applicable.

As we discussed above,  the change in $\lambda$ allows sampling the synchrotron emitting volume to different depths, which opens ways for the
studies of distribution of magnetic turbulence within the emitting volume.  This is applicable both to
turbulence spectrum and to the anisotropy of magnetic turbulence. We believe that this opens a way to do
tomographic analysis of the statistical properties of magnetic turbulence and distribution of magnetic fields
at different distances from the observer. 

 We would like to
stress that the condition ${\cal L}_{\sigma_\phi,\bar\phi} > r_i $ does not
mean that we return within the anisotropy analysis back to the case of
intensity studies in LP12. While the aforementioned condition is fulfilled,
the Faraday rotation can still determine the volume over which the anisotropies
are studied. 

Most of our studies are devoted to the general case of the synchrotron emitting region and Faraday rotation regions coinciding. This is the case for the synchrotron emission from the galactic disk, for instance. However, there are situations when the synchrotron emitting and Faraday rotating regions are separated in space. This is the case of the galactic synchrotron halo, which emits synchrotron polarized emission, but does not provide much of Faraday rotation due to the low concentration of thermal electrons, and galactic disk, which may provide substantial Faraday rotation. This case is simpler than the one we studied. Its results are presented in Appendix C.  Observations may faces a combination of the two cases above, i.e. some synchrotron emission can be subject to Faraday rotation within the source, some can be coming from the outside and experience Faraday rotation in a localized region. With the two limiting situations quantified in our paper formalism allows modeling such situations. The variety of the measures that we described allows obtaining sufficient constraints for the corresponding numerical
studies.

\section{Discussion}
\label{sec:discussion}
\subsection{Synchrotron statistics}

Our present study and LP12 show that synchrotron statistics is very rich and important information about turbulence can be obtained. LP12 outlined perspectives for studying the direction of the mean magnetic field with synchrotron intensity as well as evaluating of the contribution from different modes. The present study employs polarization fluctuations and enables one to get the statistics of synchrotron emission and Faraday rotation at different distances from the observer as well as to estimate the 3D direction of magnetic field. We expect substantial improvements of our knowledge of the galactic and extragalactic magnetic fields from the use of the technique.

The analytical description of synchrotron fluctuations in LP12 and the present study provides an opportunity to describe synchrotron statistics for complex astrophysical situations when the emitting regions and the regions of Faraday rotation are inhomogeneous in space, magnetic field changes in the direction in respect to the line of sight. This is very important for the analysis of galactic and extragalactic synchrotron sources as well as for predicting the statistical properties of synchrotron foregrounds within CMB studies. 

Our study is focused on obtaining turbulent spectra both of magnetic field and of Faraday rotation. While studying astrophysical turbulence one should take into account that the properties of
turbulence may change along the line of sight and averaged rather than local properties of turbulence are available. This, however, did not prevent substantial progress in the studies of turbulent densities and velocities, as we discuss in the paper. Similarly, magnetic turbulence was also studied successfully using synchrotron intensity fluctuations (see LP12 and ref. therein). We would like to mention, however, that in terms of sampling local properties of magnetic turbulence and getting homogeneous in the statistical sense sample the suggested techniques employing synchrotron polarization have an unquestionable advantage. Indeed, using the Faraday rotation depolarization it is possible to limit the extent of the regions that contribute to the signal. This allows more localized studies of turbulence and also mitigates the problem of the sample inhomogeneity. 

In our derivations we used the assumption of magnetic fields being Gaussian. Our recent numerical study (Herron et al. 2015) confirms that even for high Mach number turbulence the magnetic fields preserve Gaussian properties, which is reflected through the skewness and kurtosis calculations. 

The last comment is that while the full statistical 
description of magnetic fields requires the use of anisotropic tensors
presented in \S\ref{subsec:Bfieldstat}, for the limited purpose of obtaining the
spectral slope of magnetic turbulence spectrum, the simplified description of turbulence that we adopted for the most part of this paper is adequate.

\subsection{Relations between the parameters of the problem}

For the most of the paper we consider a homogeneous volume filled with turbulent magnetic fields, relativistic electrons and turbulent thermal plasmas.  We also consider in the Appendix
a particular case when the thermal plasmas and the volume of synchrotron emission are separated in space. In reality, the actual picture may be a combination of the two cases.

In the present problem of studying magnetic turbulence with synchrotron polarization
there are several parameters, the characteristic length of Faraday rotation arising from the mean field ${\cal L}_{\phi}$, the characteristic scale of Faraday rotation arising from the random field ${\cal L}_{\sigma}$, the correlation scale of emitting synchrotron magnetic structures $r_i$, the correlation scale of the Faraday rotation fluctuations $r_\phi$. separation between the lines-of-sight
${\bf R}$ and the size of the emitting region along the line-of-sight $L$.
Both  
${\cal L}_{\phi}$ and  ${\cal L}_{\sigma}$ are proportional to $\lambda^2$. Therefore, by changing the wavelength $\lambda$ is it possible to make these sizes larger or smaller than other sizes involved in the problem. As we discussed earlier, ability to vary $\lambda$ allows to study separately the statistics of fluctuations that arise from the Faraday rotation and
arising from variations of perpendicular component of magnetic field ${\bf H}_{\bot}$ responsible for synchrotron polarized emissivity. 

To see correlations of synchrotron polarization it is most productive to
make the measurements at $R<r_i$,
as no appreciable correlations is expected when the sampling of independent eddies is involved. We also assumed that the thickness of emitting physical volume along the line-of-sight $L$ is much larger than the separation between the lines-of-sight $R$. This is a natural assumption for studying 3D turbulent volumes. In some particular situations of studying turbulence in thin emitting synchrotron shells, the turbulence can be essentially 2D. For such special cases our analysis can be modified accordingly.

In many cases it is natural to assume that the correlation scale of Faraday rotation fluctuations
$r_{\phi}$ should be smaller than the correlation scale of the magnetic field fluctuations $r_i$.
This is natural if fluctuations of electron density are short correlated. However, there was no particular assumption for the relation between $r_i$ and $r_{\phi}$ involved in our derivations.

In some astrophysical settings another parameter related to the optical depth for self-absorption of synchrotron is appropriate. However, in the present study we disregard the effect of self-absorption. This effect will be studied elsewhere.

\subsection{Comparison with LP12 and relation to earlier works}

All the papers on synchrotron spectra determination
(see \S\ref{sec:introduction}) prior the LP12 study used the simplifying
assumption of synchrotron emissivity being proportional to the perpendicular
component of magnetic field squared. This meant
that the solution was suitable only for a single value of the relativistic electron spectral index. In general,
synchrotron emissivity is proportional to $H_{\bot}^\gamma$ with $\gamma$ varying within astrophysical objects and
also varying with the frequency of synchrotron radiation. This limitation was a serious constraint for the statistical
studies and it was overcome in LP12.  

Our present study employs the foundations established  in LP12. Indeed, studying intensities as in LP12 we demonstrated that one can obtain the 
spectrum of magnetic fluctuations as well as the direction
of the mean magnetic field along the line-of-sight. We also showed that one can find the relative importance of compressible and incompressible modes within magnetic
turbulence. Intensities is the simplest emission property to study through observations and it is remarkable
that a lot of information can be obtained even using synchrotron intensities alone. The polarization that we deal with 
in the present paper is a more sophisticated measure. The expressions for the synchrotron polarization are affected by Faraday rotation
and the resulting statistics reflects both magnetic and electron density fluctuations. Naturally, for sufficiently small 
$\lambda$ we are getting expressions which are not affected by the Faraday rotation and the expressions for polarization become
similar to those of intensity. 

In terms of the analysis of the observational signal we can list a few advantages of using the polarization statistics we employ
here compared to that of intensity. 
The most obvious advantage of the present approach is that we get another parameter that we can vary, namely,
the wavelength $\lambda$. This makes the study much more informative as we discuss in \S\ref{sec:lineofsight} and \S\ref{sec:additionalways}.
Additional information should be available by fitting of the model of turbulence with its injection and dissipation scales as it is done in Chepurnov et al. (2010, 2015) for the case of spectral line data.  For obtaining the spectral slope of turbulence the wavelength information can increase the reliability of the measurement.

The predicted dependence on $\lambda$ is not the only advantage of using polarimetric data compared to pure intensity.
The synchrotron radiation is strongly polarized and therefore polarization 
provides additional way to separate it from unpolarized
diffuse radiation, e.g. from the free-free plasma emission. This is important,
for instance, in the context of separating foregrounds in Cosmic Microwave
Background studies.

In addition, as we discussed above, the advantage of using PPF data cubes is the ability of tomographic study of turbulence that the use of polarization
allows. As we have demonstrated in the paper, the polarization signal is being collected from the distances limited
by the Faraday depolarization effect. This allows studying turbulence at different distances from the observer. 

Last, but not the least is that our study reveals the possibility of studying magnetic turbulence using wavelength  
information for the measurements taken along the same line-of-sight. This ability is absent for absent for the intensity
data.

Our work shows the dependence
on the cosmic ray spectral index affects the the polarization measures the same way that it depends intensity\footnote{The numerical
studies of synchrotron fluctuations with cosmic rays corresponding to different $\gamma$
in En{\ss}lin et al. (2010) are consistent with our theoretical finding.}. We believe that 
our approach presents a way to extend the results to the measures that other authors (e.g. En{\ss}lin et al. 2010) employed in
the assumption of $\gamma=2$,
e.g. the way to determine the ``power spectrum of the magnetic tension force'' 
(En{\ss}lin et al. 2010) and other magnetic field statistical
measures that can be obtained with synchrotron Stokes measures. Exploring these possibilities
deserves a separate study.

This study provides a new approach to studies of turbulent magnetic fields. We should stress that attempts  to recover statistical information about the structure of turbulent magnetic  fields in the ICM from the Faraday rotation measure (RM) data (En{\ss}lin \& Vogt 2003; Vogt \& En{\ss}lin 2003, 2005; Govoni et al. 2006; Guidetti et al. 2008), as well as studies of the ISM (Haverkorn et al. 2006, 2008), and the works  based on polarized synchrotron emission data (Spangler 1982, 1983,  Eilek 1989a,b) can be repeated on the basis of the detailed statistical description of synchrotron polarization that we presented in LP12 and this study.

\subsection{Statistics of PPV and PPF data cubes}

Our present paper presents the description of the statistics of synchrotron fluctuations in the PPF space. The corresponding description of
the statistics of spectral line fluctuations in the Position-Position-Velocity (PPV) space was provided in Lazarian \& Pogosyan (2000, 2006). We
see both similarities and differences of the aforementioned statistics. We can observe that while the PPV statistics is strongly
influenced by both the fluctuations of density and velocity, the statistics of fluctuation in the PPF arise from the fluctuations of magnetic
field and those in Faraday rotation. The latter is affected by the product of electron density and parallel component of magnetic field. Thus in both cases we deal with a signal that is affected by
different turbulent fields. 

At the same time, the properties of PPF and PPV data cubes are different. In PPV cubes we could separate
the contributions from velocity and density by changing the thickness of the frequency intervals over which we
sampled the cube, i.e. by changing the thickness of velocity slices. In PPF cubes, for a given spatial separation of the correlated points, the low frequency part of PPF depending on the regime (see Table~\ref{tab:mainresults}) is dominated
by both the Faraday rotation fluctuations and fluctuations of magnetic field or just fluctuations of magnetic field, while the higher frequency part of PPF cube arises from only magnetic fluctuations. In other words, the ways of extracting statistical information from PPV and PPF are different. By studying PPF at sufficiently high frequencies the information about $H_{\bot}$ can be obtained, which if combined with the statistics at low frequencies provides the information about $nH_{\|}$.
Important difference between the PPV and PPF statistics is that PPV statistics is homogeneous in the velocity direction, while PPF statistics is strongly inhomogeneous.
This provides additional information about turbulent volumes. For instance,
changing the wavelength it is possible to explore regions at different distances
from the observer. This is a capability not available with PPV data cubes.
In addition, anisotropy in PPF results in additional effects unique for this
space, e.g. in the imaginary part of correlations as it is discussed in
\S\ref{subsec:antisym}.

Integration over a range of frequencies, or frequency resolution play an important role for VCA, but it is not so important within the PSA if the range of frequencies is within one of the distant regimes of $nH_{\|}$ or
$H_{\bot}$ integration (see Figure \ref{fig:Wl}). If, however, the power is integrated over the frequencies includes the range of the transition from one asymptotic regime to another, the resulting behavior will not reflect any of
the asymptotics.

Our PFA technique can be considered as the analog of the VCS technique
that we developed for the PPV space. Similarly to VCS, this technique does not use spatial information to get the statistics
of turbulence. Therefore, we expect that, similar to the VCS, the new technique can get the statistics of turbulence
using measurements just along 5-10 lines-of-sight. The latter was established numerically in Chepurnov \& Lazarian (2010). 

In the paper we were focused on the asymptotic regimes. In this respect our study follows the theoretical approach in 
our papers dealing with obtaining velocity turbulence using Doppler shifted spectroscopic data (Lazarian \& Pogosyan 2000, 2006). 
However, we also obtained general expressions for the synchrotron polarization in the presence of Faraday rotation.
Our corresponding general expressions obtained for the spectroscopic data were used in the study of Chepurnov
et al. (2010) to provide a detailed description of turbulence, including 
In particular, studying HI emission with GALFA data Chepurnov et al. (2010) provided not only slopes of the underlying
velocity and density fluctuations, but also the injection scale and other parameters. The fruitfulness of such an approach 
was shown in the subsequent study by 
Chepurnov et al. (2015)  that revealed that the corresponding properties of turbulence in Small Magellanic Cloud are rather
different from those in our galaxy. We feel that a similar approach can
be fruitful within our new technique synchrotron polarization technique described in the present paper.

\subsection{Variations of cosmic rays distribution}

The fact that the expression for synchrotron intensity depends on the spectral
index of cosmic rays has been an impediment for the quantitative studies of the
statistics of synchrotron variations. Our study shows that cosmic ray spectral
variations change only the amplitude of the polarization fluctuation that we
discussed. As varying $\lambda$ allows to sample synchrotron emitting regions
to different depths, we may encounter the variations of spectral index in the sampled
regions. Then the use the correction factors from LP12 can be employed, if necessary.

Our study has been performed in the assumption of no magnetic field correlation
with the density of relativistic electrons. This assumption corresponds to
observation of much smoother distribution of relativistic electrons compared to
the distribution of magnetic fields in the Milky Way. Our formalism, however,
may be extended for the case of relativistic electrons correlated with magnetic
fields. We do not expect substantial changes in our present conclusions, e.g. in
the dependence of synchrotron intensity spectrum on $\gamma$.

\subsection{Galaxy and extragalactic sources}

With the enhanced resolution of the synchrotron data available it is possible to
apply our techniques not only to Milky Way, but also to extragalactic sources, e.g. to nearby galaxies and radio lobes of radio galaxies. For such extragalactic applications, the synchrotron radiation and Faraday rotation of Milky Way acts as a foreground. Incidentally, the usage of the raw interferometric data in order to obtain turbulent spectra is most advantageous for this application as getting turbulence spectra would not require full spectral frequency coverage. 

For the two techniques the Milky Way foreground acts differently. It is influence is strongest to the PFA one as the information is collected along a single line-of-sight. For the PSA, if the angular extent of the synchrotron source is relatively small, then the influence of the Milky Way amounts to essentially constant along both lines-of-sight Faraday rotation and synchrotron polarization intensity. Those should not interfere with the analysis. As a result we expect that the technique can study the distribution of turbulence in extragalactic sources.

For the turbulence anisotropy analysis the variations of the Milky way polarization for the beams
separated at distances much smaller than $r_i$ are expected to be small. Therefore the variations
are expected to arise mostly from an extragalactic source and the anisotropy analysis should reveal the magnetic field and its variations within the source.

\subsection{Synergetic use of different techniques}

Galactic data synchrotron emission presents a natural object for studying with the new technique. However, the
technique can be used to measure the statistics of distant objects, e.g. external galaxies and clusters of galaxies as the resolution of instruments increases. 

Magnetic fluctuations that are available through the study of PSA and PSF techniques
described above are very important for understanding both the fundamental properties
of MHD turbulence and its influence on key astrophysical processes, e.g. star formation
(see McKee \& Ostriker 2007) and cosmic ray propagation (see Schlickeiser 2002, Yan \& Lazarian 2008), as well as for the mapping of the sources and sinks of turbulent energy. The synchrotron fluctuations deliver the information on magnetic fluctuations, which 
is complementary to the velocity statistics that is available by other techniques. The 
currently alternative way of studying magnetic fluctuations is through using dust polarization signal, which is a technique that is limited by our the ability of dust grains
to follow magnetic fields and the uncertainties of the grain alignment theory (see Lazarian 2007 for a review).  

The VCA and the VCS techniques that we developed for the analysis of the velocity fluctuations provide the statistics
of velocity fluctuations. In MHD turbulence, the velocity and magnetic field statistics are related and therefore the
simultaneous use of the two measures is advantageous. In addition, the measures may be very complementary, as
spectral lines and velocities may sample different regions along the line-of-sight.

In general, in the presence of the complex, inhomogeneous ISM it is really advantageous to use complementary statistical
measures which can provide a better insight into the complexity of the turbulence. Thus combining velocity and magnetic
field is synergetic.

In addition to velocity that we can obtain with e.g. VCA and VCS techniques we can use the statistics of density. Those can be obtained
using some of the new techniques that have been developed recently. Those include the analysis of moments of
the density probabilities (Kowal et al. 2007, Burkhart et al. 2009, 2010), Tsallis statistics (Esquivel \& Lazarian 2010,
Tauffmire et al. 2011), bispectra and genus (Lazarian 1999, Chepurnov et al. 2008, Burkhart et al. 2009) etc.

Apart for analyzing the density, the aforementioned techniques can be applied to the synchrotron emission distributions.
For instance, in Burkhart et al.  (2012) used kurtosis, skewness and genus of the polarized intensity
gradient distribution to establish the Mach number of turbulence. Such work provides complementary information
about the properties of magnetic turbulence under study. We note that the general formalism describing fluctuations of
polarization that we developed in this paper can be modified for study spectra of gradients of polarization and the anisotropy
of the gradients of polarization. The advantages of using of gradients, e.g. utilizing interferometric data which does not
contain single dish observations, are described in Gaensler et al. (2011). 

\subsection{Utility of the information available}

The techniques described above outline new ways of studying anisotropic astrophysical magnetic turbulence both in Milky Way and beyond. The injection and dissipation scales as well as spectral slope of magnetic turbulence provide keys to addressing many problems of turbulence. These properties are necessary for studying cosmic ray transport and cosmic ray acceleration, as well as studying heat and mass transport in astrophysical environments. In addition, studies of turbulence anisotropy open ways for studying turbulence compressibility. Combined with statistical studies of density and velocity, the new techniques allow ways to identify sources of turbulence in different phases of interstellar media over the range of scales from kiloparsecs to astronomical units. The corresponding knowledge is necessary for a wide range of subjects from star formation to interstellar chemistry. 

We note, that the interest to properties of magnetic turbulence in Milky Way also arises from its role of inducing fluctuations of foreground radiation. In particular, we discussed potential ways of removing foregrounds if the spectrum of turbulence is known.
The application of such an approach for CMB data is discussed in Cho \& Lazarian (2010) and for high redshifted atomic
hydrogen emission in Cho, Lazarian \& Timbie (2012). Last but not the least, the information about turbulence obtained through observations is important for understanding of the enigmatic nature of MHD turbulence. 

\subsection{Future work}

The present paper presents the first to our knowledge detailed statistical description of the synchrotron polarization fluctuations. It capitalizes on our study in LP12, which provided the description of fluctuations for an arbitrary index of cosmic rays. However, this work does not address all the questions. While
it provides analytical expressions of the different regimes of the PSA and PFA studies, it explores analytically only one regime of turbulence study within the Faraday rotation synthesis. Naturally, the corresponding work should be extended.

We have noticed in the present paper that the helical correlations should
result in distinct signature, making the correlation function of polarization
(see Eq. (\ref{eq:corrPfluct}) complex. The second term in Eq. (\ref{eq:corrPfluct}) contains the valuable information about 3D direction of magnetic field and therefore this term deserves a separate theoretical study.

The effects of self-adsorption is neglected within our treatment. This effect may be important for some data sets and also should be accounted for in future. Our experience with PPV data cubes in Lazarian \& Pogosyan (2004) show that absorption can substantially change the observed statistics. Fortunately, effects of self absorption are less common in synchrotron compared to the spectroscopic data.

\section{Summary}
\label{sec:summary}
Motivated by the richness of the synchrotron data available with operating telescopes and the
prospects of higher resolution and sensitivity that is going to be available for synchrotron observations in the near future, we proceed with our studies of synchrotron fluctuations Our present study capitalizes on our previous investigation in LP12, which established the way
to describe synchrotron fluctuations for the arbitrary index of cosmic ray spectrum. Unlike the LP12
it is aimed on using synchrotron polarization rather than intensity. Synchrotron polarization fluctuations carry information both about the statistics of random magnetic field perpendicular to the line-of-sight as well as of the fluctuations of the product of the electron density and the parallel component of magnetic field. Obtaining these statistics from the statistics of synchrotron polarization fluctuations was our goal in the paper. Within the assumptions discussed 
throughout the paper we have obtained analytical results 
which can be briefly summarized in the following way:
\begin{itemize}

\item  We provided an analytical description of the statistics of synchrotron polarization fluctuations within Position-Position-Frequency (PPF) data cubes and employed this description to develop two complementary techniques to study the spectra of the underlying fluctuations of magnetic field and electron density. 

\item The first technique that we termed Polarization Spatial Analysis (PSA) employs the correlations of  the polarization intensities measured at the same frequency. At high frequencies it samples $H_{\bot}$ fluctuation, while at lower frequencies the technique is also sensitive to  the fluctuations of $n_e H_{\|}$ measure, provided that these fluctuations dominate Faraday rotation arising from the regular field. We find that the PSA is a powerful technique that can provide the spectrum of magnetic fluctuations, the slope of Faraday rotation fluctuations, the correlation scale of Faraday rotation fluctuations, the correlations scale or magnetic turbulence fluctuations as well as to give insight into the ratio of the regular and random magnetic fields.

\item The second technique that we termed  Polarization Frequency Analysis (PFA) employs the change of the variance of polarization with frequency. The properties of the variance as a function of frequency were found to depend on whether the rotation measure is dominated by the regular Faraday rotation arising from the mean
field or the Faraday rotation is dominated by $n_e H_{\|}$ fluctuations. In the former case the
dependence of the variance of polarization on frequency reflects the statistics of $H_{\bot}$.
In the latter case the variance of polarization exhibits a universal asymptotic but we suggest a
measure that uses the derivative of the variance and which reflects the $H_{\bot}$ statistics. By 
studying the change of the slope one also can determine the correlation length of magnetic field
fluctuations.

\item Polarization fluctuation are predicted to exhibit anisotropy that arises from the MHD turbulence anisotropy. The technique for such a study is similar to the one in LP12 but has the
advantage that the sampling distance along the lines-of-sight can be controlled by changing the
wavelength of the radiation.  This allows to study the variations of the regular magnetic field along the line-of-sight. 

\item The imaginary part of the correlation functions of polarization contains the information about the 3D direction of mean magnetic field,
which is a very valuable information to be obtained from the PPF studies. 

\item Our analysis of the effects of the spatial and frequency resolution of the telescopes provides
the criteria for the averaging that the finite resolution involves not to distort the results obtained with the techniques. 
\item We introduced a new measure of polarization statistics
$\left\langle \left| \frac{d P({\bf X}_1)}{d \lambda^2}
- \frac{ d P({\bf X}_2)}{d \lambda^2}\right|^2 \right\rangle $
which we found is more sensitive to Faraday rotation and therefore is complementary to the polarization correlations
used with PSA technique.

\item We showed that interferometers can be used directly to study magnetic turbulence and provided the corresponding asymptotics. This extends the possibilities of using synchrotron for studying turbulence in different astrophysical objects, including external galaxies.
\end{itemize}

\acknowledgments
A.~L. acknowledges the NSF grant AST 1212096 and Center for Magnetic Self Organization (CMSO). D.~P. thank 
the Department of Physics, 
Universitade Federal do Rio Grande do Norte (Natal, Brazil) for hospitality.
AL acknowledges a distinguished visitor PVE/CAPES appointment at the Physics Graduate Program of the Federal University of Rio Grande do Norte and thanks the INCT INEspaço and Physics Graduate Program/UFRN. 
We thank Ann Mao for the discussion of the Faraday rotation synthesis technique with us. Comments 
and suggestions by Christopher Herron, Torsten En{\ss}lin and Rainer Beck are acknowledged. We thank the anonymous referee
for suggestions improving our paper. 

\appendix
\section{Synchrotron intensity and synchrotron polarization}
\subsection{Basics}
\label{app:basics}
Synchrotron emission arises from relativistic electron spiraling about magnetic
fields. The emission has been discussed in many monographs (see Pacholczyk 1970, Fleishman 2008 and references therein). Careful study of the formation of the synchrotron signal
(see Westfold 1959) revealed that the
signal is \textit{essentially} non-linear in the magnetic field $H$ ($B$).
The origin of nonlinearity is in relativistic effects.
Nonlinearity comes from the fact that the signal is formed only over the narrow
fraction of the electron cycle, and the two leading orders in the
deviation of the trajectory from the straight line give the same contribution
in terms of the inverse of the Lorentz factor $1/\gamma_L$. Summation of the result
over ``flashes'' is also non-straightforward, and produces spread frequency
spectrum (Westfold 1959).  Situation is remarkably different from cyclotron
(non relativistic) emission where emission is monochromatic and has intensity
just quadratic in the magnetic field.

If the distribution of relativistic electrons in terms of the Lorentz factor is
\begin{equation}
N_{re}(\gamma_L)d{\gamma_L}=N_0 {\gamma_L}^{-p} d{\gamma_L}
\label{N}
\end{equation}
where $N_0$ is a normalization constant proportional
to the density of relativistic electrons
and $p$ is the spectral index of the electron distribution,
then the observer sees intensity of the synchrotron emission
\begin{equation}
I_{sync}({\bf X}) \propto \int dz H_{\perp}^\gamma({\bf x})
\end{equation}
where ${\bf X} = (x,y)$ is the 2D position vector on the sky and
$H_{\perp} = \sqrt{H_x^2 + H_y^2} $ is the magnitude of the magnetic
field perpendicular to the line-of-sight $z$. Note that
$\gamma=\onehalf(p+1)$ is, generally, a fractional power.

The linearly polarized radiation can be characterized by two directional
intensities $I^x$, $I^y$ and the angle of polarization $\xi^{xy}$
that are connected to the Stokes parameters as (see
Pacholczyk 1970):
\begin{equation}
I=I^x+I^y ~,\quad
Q=I^x-I^y ~,\quad
U=(I^x-I^y) \mathrm{tan} 2\xi^{xy}
\end{equation}
The subject of the LP12 study was the properties of intensity $I$, while the present study
is focused on the polarization $Q$ and $U$. In the frame aligned with the
direction of $\vec{H}_\perp$ only Q is present.
It is convenient to subdivide the emissivity into two components, respectively
perpendicular and parallel to $\vec{H}_\perp$ (see Rybicki \& Lightman 1979)
\begin{eqnarray}
j_\perp (\omega,\vec{x}) &=& \left[F(p) + G(p)\right]\omega^{(1-p)/2} 
|\vec{H}_\perp(\vec{x})|^{\gamma}
\label{jpe}
\\
j_\parallel (\omega,\vec{x}) &=& \left[F(p) - G(p)\right]\omega^{(1-p)/2} 
|\vec{H}_\perp(\vec{x})|^{\gamma},
\label{jpa}
\end{eqnarray}
where $\omega = 2\pi c/\lambda$, $\lambda$ is the observation wavelength,
$c$ is the speed of light, 
and 
\begin{eqnarray}
F(p) &=& \frac{\sqrt{3 \pi}\,e^3}{32\pi^2 m_ec^2} 
\left(\frac{2m_ec}{3e}\right)^{(1-p)/2} N_0
\, \Gamma\left(\frac{p}{4}-\frac{1}{12}\right) 
\frac{2^{(p+1)/2}}{p+1} \Gamma\left(\frac{p}{4}+\frac{19}{12}\right)
\frac{\Gamma\left(\frac{p}{4}+\frac{5}{4}\right)}{\Gamma\left(\frac{p}{4}+\frac{7}{4}\right)}, 
\\
G(p) &=& \frac{\sqrt{3 \pi}\,e^3}{32\pi^2 m_ec^2} 
\left(\frac{2m_ec}{3e}\right)^{(1-p)/2} N_0
\, \Gamma\left(\frac{p}{4}-\frac{1}{12}\right) 
2^{(p-3)/2} \Gamma\left(\frac{p}{4}+\frac{7}{12}\right)
\frac{\Gamma\left(\frac{p}{4}+\frac{5}{4}\right)}{\Gamma\left(\frac{p}{4}+\frac{7}{4}\right)},
\end{eqnarray}
where $m_e$ is the electron mass, $e$ is its charge,
and $N_0$ is the prefactor of the electron distribution (Eq.~(\ref{N})).
Then, the specific intensity ${\rm I}$ and the polarized specific intensity
${\rm PI}=Q+iU$ are (see Waelkens et al. 2009)
\begin{eqnarray}
{\rm I}(\omega,\vec{X}) &=& \int_{b}^{a} dz 
\left[j_\perp(\omega,\vec{x}) + j_\parallel(\omega,\vec{x})\right],
\label{i}
\\
{\rm PI}(\omega,\vec{X}) &=& \int_{b}^{a} dz 
\left[j_\perp(\omega,\vec{x}) - j_\parallel(\omega,\vec{x})\right]
e^{2i\chi(\vec{x})},
\label{p}
\end{eqnarray}
where $a$ and $b$ provide the boundaries of the emitting region
along the line-of-sight and 
\begin{equation}
\chi(\vec{x}) = \xi_0^{xy}(\vec{x}) + \lambda^2 \Phi(\vec{x}), 
\label{eq::chi}
\end{equation}
where $\xi_0^{xy}(\vec{x})$ is
the polarization angle at the source and $\Phi(\vec{x})$ (often denoted in
the literature by ${\rm RM}(\vec{x})$) is the Faraday rotation measure
\begin{equation}
\Phi(\vec{x}) = \frac{e^3}{2\pi m_e^2 c^4} 
\int_{b}^{z} dz^\prime\, n_{e}(\vec{X},z^\prime) H_z(\vec{X},z^\prime),
\end{equation}
where $n_{e}$ is the density of thermal electrons and $H_z$ is the 
projection of the magnetic field on the line-of-sight. Taking into
account  Eqs.(\ref{jpe}), (\ref{jpa}) it  is easy to see that expressions
for intensity given by Eq. (\ref{i}) and polarization given by Eq. (\ref{p})
have the same dependence on the power of $H_{\bot}$. This relates the 
problem of the statistics of emissivity for arbitrary $p$ to the statistic of polarization.

\subsection{Emission for fractional p}

The relativistic electron power law index $p$ changes from object to object and also varies with energy of electrons.
For galactic radio halo tested at meter wavelengths, observations indicate that
$p\approx 2.7$ (see Pohl 1996). The index $p$
and  therefore $\gamma$ may vary due to processes of acceleration and loses.  In reality,
the acceleration of particles is a more sophisticated process which in case of a
shock includes the formation of the precursor and its interaction with the media
(see  Beresnyak, Jones \& Lazarian 2009) as well as various feedback processes. As a result
of these effects and effects of propagation $\gamma$ will vary\footnote{In LP12 we considered
variations of $\gamma$ in the range from 1 to  4, which covers the
astrophysically important cases we are aware of.}.

To correlate the measures of polarized intensity, one has to correlate magnetic fields in
the power of $\gamma$. This problem was addressed in LP12, where the  correlation function of the synchrotron intensity
\begin{equation}
\xi_{sync}({\bf R}) \equiv \left\langle I_{sync}({\bf X_1}) I_{sync}({\bf X_2}) \right\rangle
\propto \int_0^L dz_1 \int_0^L dz_2 \left\langle
H_{\perp}^\gamma \left({\bf x}_1 \right)
H_{\perp}^\gamma \left({\bf x}_2 \right)
\right\rangle
\label{eq:synchro_gen}
\end{equation}
were obtained. There the averaging over an ensemble of the pairs of the
sky measurements at fixed two dimensional separation ${\bf R}=\bf {X_1} - {\bf X_2} $ is performed.
This function is the projection of the three-dimensional correlation of emissivity
\begin{equation}
\xi_{H_\perp^\gamma}({\bf R},z) = \left\langle
H_{\perp}^\gamma \left({\bf x}_1 \right)
H_{\perp}^\gamma \left({\bf x}_2 \right)
\right\rangle
\label{eq:3Dcorr_emissivity}
\end{equation}
which for homogeneous turbulence depends only on ${\bf x_1} - {\bf x_2} = ({\bf R}, z)$.

The structure function is formally related to the correlation one
\begin{equation}
D_{sync}({\bf R}) \equiv  \left\langle \left(I_{sync}({\bf X_1}) -  I_{sync}({\bf X_2})\right)^2 \right\rangle
\propto 2 \int_0^L dz_1  \int_0^L dz_2
\left( \xi_{H_\perp^\gamma}(0,z_1-z_2) - \xi_{H_\perp^\gamma}({\bf R},z_1-z_2) \right)
\label{eq:synchro_strucut}
\end{equation}
but it can been sometimes defined even when the correlation function itself is 
divergent.

This presents one with the problem of describing the correlation  between
the fractional powers of the magnitudes of the orthogonal projections of
a vector field,  $\xi_{H_\perp^\gamma}$. In LP12 where 
expression for emissivity correlations was obtained:
\begin{eqnarray}
\xi_{H_\perp^\gamma}({\bf r}) \approx  {\cal P}(\gamma) \xi_{H_\perp^2}({\bf
r}), \quad\quad
D_{H_\perp^\gamma}({\bf r}) \approx {\cal A}(\gamma) {\cal P}(\gamma)
 D_{H_\perp^2}({\bf r}),
\end{eqnarray}
where the strongly dependent on $\gamma$ amplitude ${\cal P}(\gamma) \equiv
\frac{\left\langle (H_\perp^\gamma)^2 \right\rangle - \left\langle
H_\perp^\gamma \right\rangle^2}{\left\langle H_\perp^4 \right\rangle -
\left\langle H_\perp^2 \right\rangle^2}$ is factorized from
the scaling and angular dependences described by $\gamma=2$ term.
For the structure function, adjusting the amplitude in a weakly
$\gamma$-dependent fashion with ${\cal A}(\gamma) \sim 1$ makes the
approximation even more accurate at small scales.

Results in LP12 translate into the observable structure function of synchrotron
polarization
\begin{eqnarray}
D_{sync, \gamma}({\bf R}) \approx p(\gamma)^2 {\cal A}(\gamma) {\cal P}(\gamma)
D_{sync, \gamma=2}({\bf R})
\end{eqnarray}
according to which the scaling and angular dependence of the
polarization structure function can be understood from studying $\gamma=2$ case.
For isotropic magnetic fields the above expression can be written as (see LP12)
\begin{eqnarray}
D_{sync, \gamma}({\bf R}) \approx p(\gamma)^2{\cal A}(\gamma)
2^{\gamma-2} H^{2\gamma-4} \left(
\Gamma\left[1+\gamma\right]
-\Gamma\left[1+\onehalf\gamma\right]^2 \right)
D_{sync, \gamma=2}({\bf R})
\end{eqnarray}

In case of axisymmetric turbulence one can choose the $x$ coordinate to be
aligned with the sky projection of the symmetry axis. In this frame 
the covariance of the components of the magnetic field is diagonal,
$\sigma_{xy}=0$ and the ratio of the variances $P(\gamma)$ can be expressed
via $\sigma^2=\sigma_{xx}+\sigma_{yy}$ and $\epsilon \equiv
\frac{\sigma_{xx}-\sigma_{yy}}{ \sigma_{xx} + \sigma_{yy}}$ to give
\begin{eqnarray}
D_{sync, \gamma}({\bf R}) &\approx& {\cal A}(\gamma)
\sigma^{2\gamma-4} D_{sync, \gamma=2}({\bf R}) \times  \\
&\times& \frac{\left(1 - \epsilon^2\right)^{\gamma}}{1+\epsilon^2}
\left[\left(1-\epsilon^2\right)^{\onehalf} 
\Gamma(1+\gamma) 
\mbox{}_{2}F_1 \left(\frac{1}{2}+\frac{\gamma}{2},1+\frac{\gamma}{2},1,
\epsilon^2\right) 
-\Gamma\left(1+\frac{\gamma}{2}\right)^2
\mbox{}_{2}F_1 \left(\frac{1}{2}+\frac{\gamma}{4},1+\frac{\gamma}{4},1,
\epsilon^2\right)^2
\right] \nonumber
\label{impor}
\end{eqnarray}

Eq. (\ref{impor}) relates the structure function of synchrotron intensity for arbitrary index of relativistic
electrons, i.e. for arbitrary $\gamma$, with the 
structure function for synchrotron intensity fluctuations corresponding to $\gamma=2$. The additional factors
do not depend on the distance between points for which the correlation is thought, but uniformly change the amplitude of the structure function. Both $\gamma$ and the $\epsilon$ can be obtained independently as (see LP12). It is important to us that Eq. (\ref{impor}) allows us to generalize the results obtained for $\gamma=2$ for an arbitrary index $\gamma$\footnote{For the mean intensity of fluctuations the approximation for
arbitrary $\gamma$ was obtained by Burn (1966).}. 

\subsection{Antisymmetric correlations of the magnetic field and polarization}
\label{app:antisym}

General expression for statistically axisymmetric correlations of the 
solenoidal vector field, e.g. magnetic field, 
has been given by Oughton et al. 1997
\begin{eqnarray}
\left\langle H_i(\mathbf{x_1}) H_j(\mathbf{x_2}) \right\rangle
= \frac{1}{(2\pi)^3} \int d^3{\bf k} 
\; e^{i {\bf k}\cdot (\mathbf{x_1}-\mathbf{x_2})} 
&&  \left[
 E({\bf k}) \left( \delta_{ij} - \hat k_i \hat k_j
\right) \right. \nonumber \\
&& +  F({\bf k}) \frac{ ({\bf \hat k} \cdot {\bf \hat \lambda})^2 \hat k_i \hat
k_j + \hat \lambda_i \hat \lambda_j - ({\bf \hat k} \cdot {\bf \hat \lambda})
(\hat k_i \hat \lambda_j+\hat k_j \hat \lambda_i)}{1 - ({\bf \hat k} \cdot {\bf
\hat \lambda})^2} \nonumber \\
&& - i C(\mathbf{k}) \left(\delta_{i\mu} \epsilon_{j\alpha\beta} 
+ \delta_{j\mu} \epsilon_{i\alpha\beta} \right)
{\hat \lambda}^\alpha {\hat k}^\beta
\left( {\hat \lambda}^\mu - {\hat k}^\mu  ({\bf \hat k} \cdot {\bf \hat \lambda}) \right)
\nonumber \\
&& + \left. i H({\bf k}) \epsilon_{ij\alpha} {\hat k}^\alpha
\right] ~.
\label{eq:axisym_turb}
\end{eqnarray}
where $\hat\lambda$ is the direction of the symmetry axis. 
The correlation is described by four spectral terms, $E, F, C$ and $H$, 
each defined by its specific  tensor structure. In particular, for 
statistically isotropic field, only $E$ and $H$ spectra are present, 
with $E$ term describing the energy spectrum and $H$-term the helicity
spectrum of the magnetic field. $F$ and $C$ term appear 
when isotropy is broken, although the anisotropy can also manifest itself
in direction dependence of $E$ and $H$ spectral functions.
In LP12 we have shown that Alfv\'en, fast and slow modes of MHD turbulence
give naturally rise to $E$ and $F$ type correlations of the magnetic field.

The property important for our discussion here is that $E$, $F$ and $C$ terms
lead to index symmetric correlation tensor 
$\langle H_i(\mathbf{x_1})H_j(\mathbf{x_2})\rangle = 
\langle H_j(\mathbf{x_1})H_i(\mathbf{x_2})\rangle$
while the helical term $H$ is the only one that gives antisymmetric
contribution to the correlation tensor of the magnetic field. Let us see
how it is reflected in the correlations of the polarization of the 
synchrotron radiation.

The general correlation matrix of the polarization of two emitters
in terms of $Q$ and $U$
Stokes parameters defined in a frame with line-of-sight along z coordinate
is
\begin{equation}
\left(
\begin{array}{c c}
 \left\langle Q_1 Q_2 \right\rangle &
\left\langle Q_1 U_2 \right\rangle \\
 \left\langle U_1 Q_2\right\rangle &
 \left\langle U_1 U_2\right\rangle
\end{array}
\right)
\end{equation}
where we have used the short-hand index notation $1$
(and, correspondingly, $2$)  
to designate the 3D position of emitter $\mathbf{x_1}=(\mathbf{X_1},z_1)$
at distance $z_1$ along the line-of-sight $\mathbf{X_1}=(x_1,y_1)$. It 
has two linear in the correlation invariants with respect to rotation of the
observer frame in the $(x,y)$ plane. These are the trace 
$ \left\langle Q_1 Q_2 \right\rangle + \left\langle U_1 U_2 \right\rangle $ and
the possible antisymmetric part of the matrix
$\left\langle Q_1 U_2 \right\rangle - \left\langle U_1 Q_2 \right\rangle $.
These invariants are conveniently encoded as the real and the imaginary parts
of the correlation of the complex polarization $P = Q + i U$ given by
Eq.~(\ref{eq:Pcorr_inQU}).

In the quadratic approximation to synchrotron emissivity, $\gamma=2$,
(Waelkens et al. 2009, LP12).
\begin{eqnarray}
Q &\propto& \left( H_x H_x - H_y H_y \right) \\ 
U &\propto& \left( 2 H_x H_y \right)
\end{eqnarray}
which gives for the correlations
\begin{eqnarray}
\left\langle Q_1 Q_2 \right\rangle + \left\langle U_1 U_2 \right\rangle 
&\propto& \left( \sigma_{xx} + \sigma_{yy} \right)^2 + 
2 \left( \left\langle H_{x1} H_{x2} \right\rangle + 
\left\langle H_{y1} H_{y2} \right\rangle \right)^2
- 
2 \left( \left\langle H_{y1} H_{x2} \right\rangle - 
\left\langle H_{x1} H_{y2} \right\rangle \right)^2
\\
\left\langle Q_1 U_2 \right\rangle - \left\langle U_1 Q_2 \right\rangle 
&\propto&
4 \left( \left\langle H_{x1} H_{x2} \right\rangle + 
\left\langle H_{y1} H_{y2} \right\rangle \right)
\left( \left\langle H_{y1} H_{x2} \right\rangle - 
\left\langle H_{x1} H_{y2} \right\rangle \right)
\end{eqnarray}
We see that invariants of the polarization correlation at the source
are constructed from the
respective invariants of the magnetic field correlation tensor. Importantly,
antisymmetric term in polarization is only present if the corresponding
antisymmetric term 
$\left\langle H_{y1} H_{x2} \right\rangle - 
\left\langle H_{x1} H_{y2} \right\rangle 
$
is non-zero for the magnetic field, i.e. helical correlations are present.
For isotropic turbulence, antisymmetric correlation induced by helicity in
the magnetic field  has the form
$\langle H_i H_j \rangle = \epsilon_{ij\alpha} \xi_H(r) r^\alpha$, as
can be obtained by Fourier transformation of the $H$ term in 
Eq.~(\ref{eq:axisym_turb}). Therefore
$
\left\langle H_{y1} H_{x2} \right\rangle - 
\left\langle H_{x1} H_{y2} \right\rangle
\propto \xi_H(r) \Delta z
$
and $
\left\langle Q_1 U_2 \right\rangle - \left\langle U_1 Q_2 \right\rangle 
$
is an odd function of the line-of-sight separation $\Delta z$. Helical term
also contributes an even in $\Delta z$ correction to the trace
$\left\langle Q_1 Q_2 \right\rangle + \left\langle U_1 U_2 \right\rangle$.

\section{Quadratic approximation to RM structure function}
\label{app:DPhi}
Here we show that in the general case the RM structure function given by
Eq.~(\ref{eq:DPhi_def}) 
has a valley along the $z_1=z_2$ line and
can be approximated as a quadratic in $\Delta z = z_1-z_2$ away from it.

Differentiating Eq.~(\ref{eq:DPhi_def}) we find
\begin{eqnarray}
\frac{\partial D_{\Delta \Phi}({\bf R},z_1,z_2)}{\partial z_1}
&=&
\int_0^{z_1} \!\!\!\! dz' \xi_{\phi}(0,z_1-z') 
-\int_0^{z_2}\!\!\!\! dz' \xi_{\phi}({\bf R},z_1-z') \\
\frac{\partial D_{\Delta \Phi}({\bf R},z_1,z_2)}{\partial z_2}
&=&
\int_0^{z_2} \!\!\!\! dz' \xi_{\phi}(0,z'-z_2) 
-\int_0^{z_1}\!\!\!\! dz' \xi_{\phi}({\bf R},z'-z_2)
\end{eqnarray}
i.e. 
\begin{equation}
\left. \frac{\partial D_{\Delta \Phi}({\bf R},z_1,z_2)}
{\partial (z_1-z_2) } \right|_{z_2=z_1} =0
\end{equation}
which signifies the bottom of the valley along $z_1=z_2$ line. 
Parameterizing the coordinate along this line with 
$z_+ = \frac{1}{2}(z_1+z_2)$, the slope
along the bottom  becomes
\begin{equation}
\left. 
\frac{\partial D_{\Delta \Phi}({\bf R},z_1,z_2)}
{\partial (z_1+z_2)/2} \right|_{z_2=z_1}
= 2 \int_0^{z_+} \!\!\!\! dz' \left( \xi_{\phi}(0,z_+-z')
- \xi_{\phi}({\bf R},z_+-z')  \right)
\end{equation}

The curvature of $D_{\Delta\Phi}(\textbf{X},z_1,z_2)$ surface
at the $z_1=z_2$ line is determined by the Hessian
which general expression 
\begin{equation}
\frac{\partial^2 D_{\Delta \Phi}({\bf R},z_1,z_2)}
{\partial z_i \partial z_j}
= \left(
\begingroup
\renewcommand\arraystretch{1.5}
\begin{array}{cc}
\xi_{\phi}(0,z_1)
-\xi_{\phi}({\bf R},z_1)
+\xi_{\phi}({\bf R},z_1-z_2)
& -\xi_{\phi}({\bf R},z_1-z_2) \\
-\xi_{\phi}({\bf R},z_1-z_2)
& 
\xi_{\phi}(0,z_2)
-\xi_{\phi}({\bf R},z_2)
+\xi_{\phi}({\bf R},z_1-z_2)
\\
\end{array}
\endgroup
\right)
\end{equation}
is reduced at $z_1=z_2$ to
\begin{equation}
\left. 
\frac{\partial^2 D_{\Delta \Phi}({\bf R},z_1,z_2)}
{\partial z_i \partial z_j} \right|_{z_2=z_1}
= \left(
\begingroup
\renewcommand\arraystretch{1.5}
\begin{array}{cc}
\xi_{\phi}(0,z_+)
-\xi_{\phi}({\bf R},z_+)
+\xi_{\phi}({\bf R},0)
&
-\xi_{\phi}({\bf R},0)
\\
-\xi_{\phi}({\bf R},0)
& 
\xi_{\phi}(0,z_+)
-\xi_{\phi}({\bf R},z_+)
+\xi_{\phi}({\bf R},0)
\\
\end{array}
\endgroup
\right)
\end{equation}
The eigenvalues of this Hessian are 
\begin{eqnarray}
\Lambda_{-}(R,z_+) &=&
\xi_{\phi}(0,z_+) -\xi_{\phi}({\bf R},z_+) +2 \xi_{\phi}({\bf R},0) \\
\Lambda_+(R,z_+) &=& \xi_{\phi}(0,z_+) -\xi_{\phi}({\bf R},z_+) \quad ,
\label{eq:Lambda+}
\end{eqnarray}
that correspond, respectively, to the eigendirections orthogonal to and
parallel to $z_1=z_2$ line.
Direction orthogonal to $z_1=z_2$ is the direction of
the largest curvature, as is attested by the $R=0$ case when the 
$z_1=z_2$ valley is exactly
flat, $\Lambda_+ = 0$ while $\Lambda_-=2 \xi_\phi(0,0) = 2 \sigma_\phi^2$.

\section{Separated regions of synchrotron emissivity and Faraday rotation } 

The focus on this paper was on the most complex situation when the sources of
the synchrotron radiation and thermal electrons causing Faraday rotation occupy
the same volume.  Much simpler, but also astrophysically important is the
situation when synchrotron originates in one distinct region while Faraday
rotation acts on synchrotron radiation in another region. For instance, in the studies of synchrotron emission from high galactic latitudes one may consider that the synchrotron emission is generated in the galactic halo, while the regions of the galactic disk close to the observer are responsible for the Faraday rotation.
 Below we cover this special case of synchrotron turbulence study.  To do this let us assume that Faraday rotation occurs close to the observer up to distance $L$, while synchrotron is produced in remote region at line-of-sight distances from $L_s$ to $L_f$, $L_s, L_f > L$.   Then, the observed polarization is
\begin{equation}
P({\bf X},\lambda^2) = \int_{L_s}^{L_f} dz P_i({\bf X},z) e^{2 i \lambda^2 \Phi({\bf X},L) }
\label{eq:Pscreen}
\end{equation}
Coming from different regions, polarization at the source and the rotation measure are uncorrelated, thus
\begin{equation}
\left\langle P({\bf X_1},\lambda^2_1)P^*({\bf X_2},\lambda^2_2) \right\rangle
= \Xi_i({\bf X_1}-{\bf X_2}) 
\left\langle 
e^{2 i \left( \lambda_1^2  \Phi({\bf X_1},L)  - \lambda_2^2  \Phi({\bf X_2},L) \right)}
\right\rangle
\label{eq:corrPscreen}
\end{equation}
where
\begin{equation}
\Xi_i({\bf X_1}-{\bf X_2})  \equiv 
\int_{0}^{L_f-L_s} d\Delta z (L_f-L_s-\Delta z) \xi_i(\mathbf{R},\Delta z)
\label{eq:bigXi_def}
\end{equation}
is the correlation function of the synchrotron integrated over the region
of its emission, assumed to be statistically homogeneous. This function
has been discussed in LP12 with the focus on anisotropy dependence
on the separation vector $\mathbf{R}$. In this paper it has appeared in the
discussion of the weak turbulent Faraday limit in \S~\ref{subsec:turbdom}.
In this section we shall consider $\Xi_i(\mathbf{R})$ as a given function 
that describes the 2D sky correlation of synchrotron sources. To be specific, 
we'll characterize it by the correlation scale $R_i$ and the slope $M_i$,
so that at $ R < R_i$
\begin{equation}
\Xi_i(\mathbf{R}) \approx \Xi_i(0)\left(1- (R/R_i)^{M_i}\right).
\label{bb}
\end{equation}

Correlating two line-of-sight measurements at the same wavelength we obtain
\begin{equation}
\left\langle P({\bf X_1})P^*({\bf X_2}) \right\rangle
= \Xi_i({\bf R}) 
e^{- 4 \lambda^4  D_{\Delta \Phi}(\mathbf{R}, L,L) } 
\label{eq:corrPscreen1}
\end{equation}
where from Eq.~(\ref{eq:DPhi_valley})
\begin{equation}
D_{\Delta \Phi}({\bf R},L,L)
= 2 \int_0^L d\Delta z (L - \Delta z)
\left[ \xi_{\phi}(0,\Delta z) - \xi_{\phi}({\bf R},\Delta z) \right]
\end{equation}
i.e $D_{\Delta \Phi}({\bf R},L,L)$ is $z_+=L$ limit of 
$D_{\Delta \Phi}^+({\bf R},z_+)$ 
described in Figure~\ref{fig:DDF} (right panel)
and Eqs.~(\ref{eq:Dz+_smallRsmallz}-\ref{eq:Dz+_largeRlargez}).
Thus, in terms of the correlation function, observed 
polarization correlation is determined by the projected onto the sky
correlation of the sources $\Xi(\mathbf{R})$,
suppressed by random foreground Faraday rotation, more so
larger the separation $R$ is.
Naturally, for a negligible Faraday rotation, e.g. in the case of high
frequency CMB polarization studies,
$\left\langle P({\bf X_1})P^*({\bf X_2})\right\rangle\approx \Xi_i(\mathbf{R})$,
where $\Xi_i(\mathbf{R})$ is given by Eq.~(\ref{bb}). This is the simplest case.
We note that the mean foreground Faraday rotation has no effect as
long as the depth
of the Faraday rotating medium is the same along the two line of sights. 
and that Faraday rotation does not affect the variance
\begin{equation}
\left\langle P({\bf X})P^*({\bf X}) \right\rangle = \Xi_i(0) ~ .
\end{equation}

The behaviour of the Faraday term can be assembled from 
$D_{\Delta \Phi}^+({\bf R},z_+)$ asymptotics in 
Eqs.~(\ref{eq:Dz+_smallRsmallz}-\ref{eq:Dz+_largeRlargez}), viewed from
the point of view of $R$ dependence at fixed $z_+ = L$. In the case Faraday
screen is \textit{thick}, in the sense of $ L > r_\phi$,
the function $D_{\Delta \Phi}({\bf R},L,L)$ grows with $R$ through
the sequence of the regimes
\begin{eqnarray}
\label{eq:DRLL_thicksmallR}
D_{\Delta\Phi}(\mathbf{R},L,L) &\sim& 
\sigma_\phi^2 L R \left(R/r_\phi \right)^{\widetilde{m}_\phi}
~, \quad\quad\quad\quad\quad R < r_\phi ~, \\
D_{\Delta \Phi}({\bf R},L,L) &\sim& 
\sigma_\phi^2 L R \left(R/r_\phi \right)^{-\widetilde{m}_\phi}
~, ~ ~\quad\quad\quad\quad
r_\phi < R < L ~, \\ 
D_{\Delta \Phi}({\bf R},L,L) &\sim& \sigma_\phi^2 L^2
\left(L/r_\phi \right)^{-\widetilde{m}_\phi}
~, \quad\quad\quad\quad\quad
R > L  ~,
\end{eqnarray}
while in the case the screen is \textit{thin}, $ L < r_\phi $, it grows as
\begin{eqnarray}
D_{\Delta\Phi}(\mathbf{R},L,L) &\sim& 
\sigma_\phi^2 L R \left(R/r_\phi \right)^{\widetilde{m}_\phi}
~, \quad\quad\quad\quad\quad R < L ~, 
\\
D_{\Delta\Phi}(\mathbf{R},L,L) &\sim& 
\sigma_\phi^2 L^2 \left(R/r_\phi \right)^{\widetilde{m}_\phi}
~, \quad\quad\quad\quad\quad ~\; L < R < r_\phi ~, \\
D_{\Delta \Phi}({\bf R},L,L) &\sim& \sigma_\phi^2 L^2
~, \quad\quad\quad\quad\quad\quad\quad\quad\quad\quad
R > r_\phi ~. 
\end{eqnarray}

The characteristic scale $R_s$ above which Faraday rotation suppresses
the correlations is determined by the condition 
$ \lambda^4 D_{\Delta \Phi}(R_s,L,L) \approx 1$.
In particular, when the screen is \textit{thick} and Faraday rotation is
sufficiently strong, namely
$ (\sqrt{2} \lambda^2 \sigma_\phi)^{-1}={\cal L}_{\sigma_\phi} < \sqrt{ L r_\phi}$,
it can be estimated from small $R$ behaviour given by 
Eq.~(\ref{eq:DRLL_thicksmallR})
\begin{equation}
R_s \approx r_\phi \left( {\cal L}_{\sigma_\phi} /\sqrt{L r_\phi}\right)^{\frac{2}{1+\widetilde{m}_\phi}}
\label{eq:Rs_screen}
\end{equation}
If the RM is uncorrelated, suppression factor is constant and of order
$e^{-4 \lambda^4 \sigma_\phi^2 L^2} = e^{-2 (L/{\cal L}_{\sigma_\phi})^2}$.

Let us now turn to the small $R < r_\phi$ behaviour,
for which the structure function is a more appropriate measure
With the focus on the case of weak Faraday rotation,
$\lambda^4 D_{\Delta \Phi}(R,L,L) < 1$, we can expand the exponent to obtain
\begin{equation}
\left\langle \left|P({\bf X_1})-P({\bf X_2})\right|^2 \right\rangle
\approx 2 \Xi_i(0)\left(1 - \frac{\Xi_i({\bf R})}{\Xi_i(0)}
+ 4 \lambda^4  D_{\Delta \Phi}(\mathbf{R}, L,L) \right)
~, \quad\quad \lambda^4 D_{\Delta \Phi}(R,L,L) < 1 ~.
\label{eq:DPscreen1}
\end{equation}
which gives
\begin{eqnarray}
\left\langle \left|P({\bf X_1})-P({\bf X_2})\right|^2 \right\rangle
&\approx& \Xi_i(0)\left( \left(\frac{R}{R_i} \right)^{M_i}
+ 2 \frac{L r_\phi}{ {\cal L}_\phi^2}\left(\frac{R}{r_\phi}\right)^{1+\widetilde{m}_\phi}
\right)
~, \quad R < \mathrm{min}(r_\phi,L)~, 
\quad {\cal L}_{\sigma_\phi}^2 > L \mathrm{min}(r_\phi,L)
\\
\left\langle \left|P({\bf X_1})-P({\bf X_2})\right|^2 \right\rangle
&\approx& \Xi_i(0)\left( \left(\frac{R}{R_i} \right)^{M_i}
+ 2 \left(\frac{L}{ {\cal L}_\phi}\right)^2
\left(\frac{R}{r_\phi}\right)^{\widetilde{m}_\phi}
\right)
~, \quad L < R < r_\phi~, 
\quad\quad {\cal L}_{\sigma_\phi}^2 > L^2 ~.
\end{eqnarray}
where the second regime is only present for the \textit{thin} Faraday screen.
Thus, both the source and Faraday correlations contribute to the observed
structure function of polarization,
however Faraday contribution is suppressed proportionally
to $\lambda^4$, which is small in the regime of the weak Faraday rotation.

Let us now consider what information can be extracted from the measurements
of the derivative of the polarization \textit{wrt} $\lambda^2$ 
\begin{equation}
\frac{\mathrm{d} P({\bf X},\lambda^2)}{\mathrm{d} \lambda^2}
= 2 i \int_{L_s}^{L_f} dz P_i({\bf X},z) \Phi(\mathbf{X},L) 
e^{2 i \lambda^2 \Phi({\bf X},L) }
\label{eq:Pdlambdascreen}
\end{equation}
The correlation function of these derivatives is
\begin{eqnarray}
\left\langle \frac{d P({\bf X}_1)}{d \lambda^2}
\frac{ d P^*({\bf X}_2)}{d \lambda^2} \right\rangle
= \Xi_i({\bf R})
\left(
\xi_{\Delta\Phi}(\mathbf{R},L,L) +
\lambda^4 D_{\Delta\Phi}^2(\mathbf{R},L,L)
\right)
e^{- 4 \lambda^4 D_{\Delta\Phi}(\mathbf{R},L,L) }
\label{eq:corrPdlambda2_screen}
\end{eqnarray}
Perhaps the most interesting regime is when decorrelation due to Faraday effect
is small, $\lambda^4 D_{\Delta\Phi}(\mathbf{R},L,L) \ll 1$, in which case
correlating the derivative gives us a product of the source synchrotron and
RM correlations
\begin{eqnarray}
\left\langle \frac{d P({\bf X}_1)}{d \lambda^2}
\frac{ d P^*({\bf X}_2)}{d \lambda^2} \right\rangle
\approx \Xi_i({\bf R}) \xi_{\Delta\Phi}(\mathbf{R},L,L)
\label{eq:corrPdlambda2_weakscreen}
\end{eqnarray}
Further approximation at small $R < R_i, r_\phi$ gives for the structure function
\begin{eqnarray}
\left\langle \left| \frac{d P({\bf X}_1)}{d \lambda^2} - 
\frac{ d P({\bf X}_2)}{d \lambda^2} \right|^2 \right\rangle
\approx \Xi_i(0) D_{\Delta\Phi}(\infty,L,L)
\left( 1-\frac{\Xi_i({\bf R})}{\Xi_i(0)}
+ \frac{D_{\Delta\Phi}(\mathbf{R},L,L)}{D_{\Delta\Phi}(\infty,L,L)}
\right)
\label{eq:strucPdlambda2_weakscreen}
\end{eqnarray}
This expression is seemingly similar to Eq.~(\ref{eq:DPscreen1}) giving
the sum of the source and Faraday structure functions with one
important difference -- Faraday contribution is not suppressed by the small
factor $\propto {\cal L}_{\sigma_\phi}^{-2} \sim \lambda^4 $. We have
for the \textit{thick}, $L > r_\phi$, Faraday screen
\begin{eqnarray}
\left\langle \left| \frac{d P({\bf X}_1)}{d \lambda^2} - 
\frac{ d P({\bf X}_2)}{d \lambda^2} \right|^2 \right\rangle
&\propto& \left(\frac{R}{R_i} \right)^{M_i}
+ \left(\frac{r_\phi}{L}\right)^{1-\widetilde{m}_\phi}
\left(\frac{R}{r_\phi}\right)^{1+\widetilde{m}_\phi}
~, \quad R < r_\phi~, \quad {\cal L}_{\sigma_\phi}^2 > L r_\phi
\end{eqnarray}
and for the \textit{thin}, $ L < r_\phi$, one
\begin{eqnarray}
\left\langle \left| \frac{d P({\bf X}_1)}{d \lambda^2} - 
\frac{ d P({\bf X}_2)}{d \lambda^2} \right|^2 \right\rangle
&\propto& \left(\frac{R}{R_i} \right)^{M_i}
+ \frac{r_\phi}{L}
\left(\frac{R}{r_\phi}\right)^{1+\widetilde{m}_\phi}
~, \quad\quad R < L~, 
\quad\quad\quad\quad {\cal L}_{\sigma_\phi}^2 > L^2 ~,
\\
\left\langle \left| \frac{d P({\bf X}_1)}{d \lambda^2} - 
\frac{ d P({\bf X}_2)}{d \lambda^2} \right|^2 \right\rangle
&\propto& \left(\frac{R}{R_i} \right)^{M_i}
+\left(\frac{R}{r_\phi}\right)^{\widetilde{m}_\phi}
~ ~\quad\quad, \quad\quad L < R < r_\phi~, 
\quad\quad {\cal L}_{\sigma_\phi}^2 > L^2 ~.
\end{eqnarray}
where the balance between the source and Faraday contributions is determined by
their respective correlation lengths.

Thus, correlation of the polarization derivatives \textit{wrt} the wavelength
is more sensitive to Faraday rotation. By measuring it together with the 
correlation of the polarization itself allows in principle to 
learn separately about correlation of the sources and the Faraday rotation
measure in the foreground.
\section{Studying Turbulence using Faraday Rotation Synthesis: general approach}

One defines the Faraday dispersion function as a Fourier transform of the
polarization surface brightness \textit{wrt} 
the $\lambda^2$ variable\footnote{the fact that $\lambda^2$ is always positive
is not significant for the following derivation}
\begin{equation}
F({\bf X},\Psi) \equiv \int P({\bf X}, \lambda^2) e^{-2 i \lambda^2 \Psi} d \lambda^2
\end{equation}
which for the model Eq.~(\ref{eq:Plambda2}) becomes
\begin{equation}
F({\bf X},\Psi) = \frac{1}{2} \int_0^L \! dz \; P({\bf X}, z) \;
\delta_D\left(\Phi({\bf X, z}) - \Psi \right)
\end{equation}

The correlation of the dispersion function is
\begin{equation}
\left\langle F({\bf X_1},\Psi_1) F^*({\bf X_2},\Psi_2) \right\rangle
=\frac{1}{4} \int_0^L dz_2 \int_0^L dz_1
\left\langle P_i({\bf X_1}, z_1) P^*_i({\bf X_2}, z_2)
\delta_D \left(\Phi({\bf X_1, z_1}) - \Psi_1 \right)
\delta_D \left(\Phi({\bf X_2, z_2}) - \Psi_2 \right) \right\rangle
\end{equation}
This expression is very useful if, as in our case, one is able to construct the joint distribution function 
${\cal P}(P_1,P_2,\Phi_1,\Phi_2)$ (here indices $1$ and $2$ refer to two positions in 3D space), In this case, after
performing averaging integrations over $\Phi_1$ and $\Phi_2$ using the $\delta$-functions,we obtain a general 
form for the statistical average involved in the correlation that is, in principle, tractable.
\begin{equation}
 \left\langle F({\bf X_1},\Psi_1) F^*({\bf X_2},\Psi_2) \right\rangle
=\frac{1}{4} \int_0^L dz_2 \int_0^L dz_1 \int d P_1 dP_2 {\cal P}(P_1,P_2,\Psi_1,\Psi_2) P_1 P^*_2
\label{eq:Fcorr}
\end{equation}
To obtain statistics that is homogeneous in $\Psi$ coordinate and to decrease
the noise in the estimate, we can average over all values of the sum 
$\Psi_1+\Psi_2$, while keeping the difference $\Psi_1-\Psi_2$ fixed.
\begin{equation}
\xi_F(\mathbf{R},\Psi_1-\Psi_2) \equiv \int d(\Psi_1+\Psi_2)
\left\langle F({\bf X_1},\Psi_1) F^*({\bf X_2},\Psi_2) \right\rangle
=\frac{1}{4} \int_0^L dz_2 \int_0^L dz_1 \int d P_1 dP_2 {\cal P}(P_1,P_2,\Psi_1-\Psi_2) P_1 P^*_2
\label{eq:Psi1+Psi2_average}
\end{equation}
Evaluation of $\xi_P$ depends on knowing the joint distribution function of
only three variables, namely, $P_i(1), P_i(2)$ and
$\Delta\Phi = \alpha \int_{z_2}^{z_1} n_e H_z dz'$.
The correlation function $\xi_F(\mathbf{R},\Psi_1-\Psi_2)$ obtained
via averaging procedure of Eq.~(\ref{eq:Psi1+Psi2_average})
has a simple relation to the same-wavelength polarization correlations
studied in the previous sections. It is the Fourier
transform of them with respect to $\lambda^2$
\begin{equation}
\xi_F(\mathbf{R},\Psi_1-\Psi_2) =
\int \left\langle P({\bf X}_1,\lambda^2)P^*({\bf X}_2,\lambda^2) \right\rangle
e^{-2 i \lambda^2 (\Psi_1 - \Psi_2) } d \lambda^2
\label{FRS}
\end{equation}
The use of the expressions is illustrated in the main text.

\end{document}